\documentclass[%
    aps,
    prd,
    letterpaper,
    superscriptaddress,
    preprintnumbers,
    floatfix,
    twocolumn,
    nofootinbib,
    showpacs,
    hyphens
]{revtex4-1}  

\usepackage{subfigure}
\usepackage{graphicx}
\usepackage{dcolumn}
\usepackage{bm}
\usepackage{hyperref}
\usepackage{xcolor}
\usepackage{amsmath,amsfonts,amssymb,amsthm}
\newcommand{\ket}[1]{\left| #1 \right>} 
\newcommand{\bra}[1]{\left< #1 \right|} 
\newcommand{\braket}[2]{\left< #1 \vphantom{#2} \right|
 \left. #2 \vphantom{#1} \right>} 
\newcommand{\mbraket}[3]{\left< #1 \vphantom{#2#3} \right|
 #2 \left| #3 \vphantom{#1#2} \right>} 

\usepackage{slashed}

\newcommand{\bs}{\boldsymbol}
\newcommand{\eul}{\gamma_{\text{E}}}

\DeclareMathOperator{\tr}{tr}

\begin{document}

\preprint{FERMILAB-PUB-23-127-T}

\title{Baryons, multi-hadron systems, and composite dark matter in non-relativistic QCD}
\author{Beno\^{i}t Assi}
\affiliation{Fermi National Accelerator Laboratory, Batavia, IL, 60510}
\author{Michael~L.~Wagman}
\affiliation{Fermi National Accelerator Laboratory, Batavia, IL, 60510}

\date{\today}

\begin{abstract}
We provide a formulation of potential non-relativistic quantum chromodynamics (pNRQCD) suitable for calculating binding energies and matrix elements of generic hadron and multi-hadron states made of heavy quarks in $SU(N_c)$ gauge theory using quantum Monte Carlo techniques. We compute masses of quarkonium and triply-heavy baryons in order to study the perturbative convergence of pNRQCD and validate our numerical methods. Further, we study $SU(N_c)$ models of composite dark matter and provide simple power series fits to our pNRQCD results that can be used to relate dark meson and baryon masses to the fundamental parameters of these models. 
For many systems comprised entirely of heavy quarks, the quantum Monte Carlo methods employed here are less computationally demanding than lattice field theory methods, although they introduce additional perturbative approximations. The formalism  presented here may therefore be particularly useful for predicting composite dark matter properties for a wide range of $N_c$ and heavy fermion masses.
\end{abstract}

\maketitle

\tableofcontents

\section{Introduction}

Heavy quark systems provide a theoretically clean laboratory for studying quantum chromodynamics (QCD) because of the large separation of scales between the heavy quark mass and the confinement scale.
Spurred initially by the discovery of doubly-heavy mesons $J/\psi$~\cite{E598:1974sol,SLAC-SP-017:1974ind} and $\Upsilon$~\cite{E288:1977xhf}, the
use of non-relativistic (NR) effective field theory (EFT) to study heavy quarkonium in QCD~\cite{Bodwin:1994jh,Brambilla:2004jw,Pineda:1997bj,Thacker:1990bm}, analogous to the previous treatment of positronium in NR quantum electrodynamics (NRQED)~\cite{Caswell:1985ui}, has been investigated extensively~\cite{Brambilla:1999xf, Brambilla:2004jw,Kniehl:2002br,Kniehl:1999ud,Georgi:1990um,Pineda:1997bj,Pineda:2011dg}. Prior to this first principles treatment of quarkonia with EFTs derived from QCD, studies mainly relied on potential quark models~\cite{Wilson:1974sk,Lucha:1991vn,Eichten:1979pu,Gromes:1984ma,Barchielli:1986zs,Brambilla:1997vg}. Such models rely on phenomenological input whose connection with QCD parameters is obscure and thus cannot be systematically improved.

Beyond quarkonium, there has been recent excitement about understanding the properties of baryons and exotic hadrons containing heavy quarks including tetraquarks, pentaquarks, hadronic molecules, hybrid states containing explicit gluon degrees of freedom, and more~\cite{Berwein:2015vca,Brambilla:2017uyf,Brambilla:2018pyn,Brambilla:2019esw,Chen:2022asf}.
Theoretically calculating the spectra of baryons and exotic states experimentally observed so far and predicting the presence of other states provide tests of our understanding of QCD in more complex systems than quarkonium.
In particular, doubly-heavy baryons have recently been experimentally observed~\cite{LHCb:2019gqy,Engelfried:2006eki,SELEX:2002xhn}, and triply-heavy baryons, although not yet observed experimentally, have long been of theoretical interest as probes of confining QCD dynamics that are free from light quark degrees of freedom requiring relativistic treatment~\cite{Bjorken:1985ei}.

Additionally, one can consider generic composite states analogous to QCD, bound under a confining $SU(N_c)$ gauge theory. Such states have received particular attention recently as attractive dark matter (DM) candidates~\cite{Gudnason:2006yj,Kribs:2009fy,Hambye:2009fg,Lewis:2011zb,Buckley:2012ky,Hietanen:2012sz,Hietanen:2013fya,LatticeStrongDynamicsLSD:2013elk,LatticeStrongDynamicsLSD:2014osp,Hochberg:2014kqa,Boddy:2014yra,Antipin:2015xia,Appelquist:2015yfa,Soni:2016gzf,Cline:2016nab,Kribs:2016cew,Mitridate:2017oky,Geller:2018biy,DeLuca:2018mzn,DeGrand:2019vbx,Cline:2021itd,Asadi:2021yml,Asadi:2021pwo}. Motivated by the stability of the proton in the Standard Model (SM), a dark sector with non-Abelian gauge interactions can give rise to a stable, neutral dark matter candidate. Simple models of an $SU(N_c)$ dark sector with one heavy quark can provide UV-complete and phenomenologically viable models of composite DM~\cite{Asadi:2021yml,Asadi:2021pwo}.
It would therefore be interesting to probe masses, lifetimes, and self-interactions in composite DM theory to make predictions for experiments.

In this work, we study the description of generic hadronic bound states composed entirely of heavy quarks that are well-described by the EFT of potential NRQCD (pNRQCD)~\cite{Bodwin:1994jh,Caswell:1985ui,Brambilla:1999xf,Pineda:1997bj,Pineda:1998kj}. This EFT takes advantage of the experimental evidence that heavy quark bound state splittings are smaller than the quark mass, $m_Q$. Thus, all dynamical scales are small relative to $m_Q$. Assuming quark velocity is therefore small, $v\ll 1$, one can exploit the hierarchy of scales $m_Q\gg p_Q\sim m_Qv\gg E_Q\sim m_Q v^2$ in the system ~\cite{Caswell:1985ui}. NRQCD is obtained from QCD by integrating out the hard scale, $m_Q$, and pNRQCD is obtained from integrating out the soft scale $p_Q\sim m_Q v$. The inverse of the soft scale gives the typical size of the bound state, analogous to the Bohr radius in the Hydrogen atom. In QCD, one has to consider the confinement scale $\sim \Lambda_{\rm QCD}$, below which non-perturbative effects other than resummation of potential gluons must be included.
Here, we will work in the so-called weak coupling regime~\cite{Pineda:2011dg}, $m_Qv\gg \Lambda_{\rm QCD}$, which is valid for treatment of top and bottom bound states and starts to become less reliable for charm-like masses and below.
Both the weak- and strong-coupling regimes can be studied using lattice QCD (LQCD), and in particular lattice calculations of NRQCD are useful for studying heavy quark systems.
The advantage of using pNRQCD to study the weak-coupling regime is that precise results can be obtained using modest computational resources: the quantum Monte Carlo (QMC) calculations below use ensembles of 5,000 configurations with $3N_Q$ degrees of freedom representing the spatial coordinates of $N_Q$ heavy quarks in contrast to LQCD calculations that commonly use ensembles of hundreds or thousands of configurations with $10^8$ or more degrees of freedom representing the quark and gluon fields at each lattice site.

In many previous studies of pNRQCD, the main focus was heavy quarkonia in QCD~\cite{Brambilla:1999xj,Kniehl:2002br,Brambilla:1999xf,Pineda:2011dg,Pineda:1997hz}. The heavy quarkonium spectrum, as well as other properties such as decay widths, were studied in detail to $\rm{N^3LO}$. Ultrasoft effects were also considered as they play a role beyond NNLO~\cite{Kniehl:1999ud}. Additionally, pNRQCD was extended for doubly- and triply-heavy baryons in QCD~\cite{Brambilla:2005yk}.  The three-quark potential was also recently determined for baryon states and was shown to contribute at NNLO~\cite{Brambilla:2009cd,Brambilla:2013vx}. 

In this work, we employ a pNRQCD formalism previously developed for the case of heavy quarkonia~\cite{Brambilla:1999xf,Pineda:2011dg}, in which we take the operators to be dependent on heavy quark and antiquark fields. In particular, we generalize this formalism to apply to arbitrary hadronic systems comprised totally of heavy quarks. Thus, we can probe exotic states and multi-hadron systems such as tetra-quarks, meson-meson molecules, and the deuteron in the heavy quark limit. Moreover, we generalize all the components of the EFT to treat arbitrary bound systems of heavy fermions charged under $SU(N_c)$.  We determine the operators and matching coefficients describing the action of two- and three-quark potentials on arbitrary hadronic states up to NNLO for general $N_c$. 

Our formalism is then applicable to extract properties of the bound states such as binding energies and matrix elements with the use of variational Monte-Carlo (VMC) and Green's function Monte-Carlo (GFMC) methods ~\cite{Carlson:2014vla,Yan_2017,Gandolfi:2020pbj}. Both VMC and GFMC are state-of-the-art in nuclear physics simulations, and we apply them to study heavy-quark bound states in QCD and $SU(N_c)$ gauge theories in general. Recently, VMC was employed to determine the binding energy and mass spectra of triply-heavy bottom and charm baryons in QCD~\cite{Jia:2006gw,Llanes-Estrada:2011gwu}. The results are mass-scheme dependent, and in this work, we tie our heavy quark mass to the spin-averaged mass of the measured $1S$ state of the associated quarkonia.
After tuning the charm and bottom quark masses to reproduce the quarkonia masses, we predict the mass spectrum of triply-heavy bottom and charmed baryons and compare with previous LQCD results for the same masses.

As for the dark sector, we study the spectra of heavy dark mesons and baryons in $SU(N_c)$ gauge theory for $N_c \in \{3,\ldots,6\}$ and extrapolate to large $N_c$. We demonstrate that QMC calculations using pNRQCD can provide predictions for composite DM observables that enable efficient scanning over a wide range of mass scales. The computational simplicity of this approach is beneficial for studying composite DM, in which the fundamental parameters of the underlying theory are not yet known.
Further, we fit our QMC pNRQCD results for dark meson and baryon masses to power series in the dark strong coupling constant and $1/N_c$ that provide analytic approximations that can be used straightforwardly in phenomenological studies of composite DM.

The remainder of this work is organized as follows. Section~\ref{sec:pnrqcd} introduces pNRQCD in a formulation suitable for studying multi-hadron systems. Section~\ref{sec:manyb} reviews QMC methods that can be used to compute matrix elements of the Hamiltonian and other operators.
In Section~\ref{sec:trial}, we describe and justify the choice of initial trial wavefunctions used as inputs for VMC and GFMC calculations of heavy quarkonium and triply-heavy baryons. Results of these calculations for heavy mesons and baryons in QCD are described in Section~\ref{sec:qcd}, and results for $SU(N_c)$ dark mesons and baryons are described in Section~\ref{sec:bsm}. We discuss some prospects for future investigations in Section~\ref{sec:out}.

\section{pNRQCD for multi-hadron systems}
\label{sec:pnrqcd}

$SU(N_c)$ gauge theory with $n_f$ light fermions and $n_h$ heavy fermions is a straightforward generalization of QCD at the perturbative level.
In this section, this theory will be referred to as QCD with ``quark'' and ``gluon'' degrees of freedom; however, all the formalism we present is relevant for the more general case of $SU(N_c)$ gauge theory discussed for dark hadrons in Sec.~\ref{sec:bsm}.

\subsection{pNRQCD formalism}

The QCD Lagrange density is given by
\begin{equation}
\begin{split}
    \mathcal{L}_{\rm QCD} &= \mathcal{L}_g + \mathcal{L}_l + \mathcal{L}_h,
    \end{split}
\end{equation}
where the gluon, light quark, and heavy quark terms are
\begin{align}
    \mathcal{L}_{g}(x) &= -\frac{1}{2}\tr\left[ G_{\mu\nu}(x) G^{\mu\nu}(x) \right], \\
    \mathcal{L}_l(x) &= \sum_{f=1}^{n_f} \overline{q}_{f}(x)[i\slashed{D}+m_f]q_{f}(x), \\
    \mathcal{L}_h(x) &= \sum_{h=1}^{n_h} \overline{Q}_{h}(x)[i\slashed{D}+m_Q]Q_{h}(x), \label{eq:qcdlagrangian}
\end{align}
where $G_{\mu\nu} = [D_\mu, D_\nu] = G_{\mu\nu}^a T^a$ is the gluon field-strength tensor, $D_\mu = \partial_\mu + i g_s A_\mu^a T^a$ is the  gauge-covariant derivative, $A_\mu^a$ is the gluon field, $g_s$ is the strong coupling, $m_f$ and $m_h$ are light and heavy quark masses respectively, and the $T^a$ are generators of $\mathfrak{su}(3)$ normalized as $\tr[T^a T^b] = \frac{1}{2}\delta^{ab}$.
Light quarks with $m_f \ll \Lambda_{\rm QCD}$ contribute to the  renormalization group (RG) evolution of $\alpha_s(\mu) = g(\mu)^2/(4\pi)$ and will be approximated as massless below.
Heavy quarks with $m_h \gg \Lambda_{\rm QCD}$ have negligible effects on the RG evolution of $\alpha_s$ for $\mu \lesssim m_h$, and in systems where heavy quarks are nonrelativistic EFT methods can be used to expand observables in power series of $\Lambda_{\rm QCD}/m_h$.
The $\overline{\text{MS}}$ renormalization scheme is used throughout this work for simplicity.
In the $\overline{\text{MS}}$ scheme, effective interactions between heavy quarks only depend on $n_f$ and $n_h$ through the number of flavors with mass less than $\mu$ in the RG evolution of $\alpha_s(\mu)$ and in the values of other EFT couplings, defining this number to be $n_f$ leads to a decoupling of heavy quarks from one another, and we, therefore, omit heavy flavor indices and denote the heavy quark mass by $m_Q$ below.

NRQCD is the EFT employed to study systems of two or more heavy quarks. The Lagrangian is determined by integrating out degrees of freedom with the energy of the order of the heavy-quark masses~\cite{Bauer:1997gs,Bodwin:1994jh,Caswell:1985ui,Manohar:1997qy}. The Lagrangian operators are determined by QCD symmetries and are organized as a power series in inverse quark mass, $m_Q$, with $m_Q \gg \Lambda_{\text{QCD}}$. The NRQCD Lagrangian including light quarks reads~\cite{Caswell:1985ui,Georgi:1990um},
\begin{equation}
  \mathcal{L}_{\rm NRQCD} =\mathcal{L}_{\psi}+\mathcal{L}_{\chi}+\mathcal{L}_{\psi\chi}+\mathcal{L}_{\psi\psi}+\mathcal{L}_{\chi\chi}+\mathcal{L}_g +\mathcal{L}_l,
  \label{eq:nrqcdlagrangian}
\end{equation}
where $\psi$ and $\chi_c = -i\sigma_2 \chi^*$ are the Pauli spinors that annihilate a quark and create an antiquark, respectively, which are related to the QCD heavy quark field by
\begin{equation}
  Q(x) = \sqrt{Z}\begin{pmatrix} e^{-im_Qt}\psi(x) \\ e^{im_Qt} \chi(x) \end{pmatrix}, \label{eq:NRpsi}
\end{equation}
in the Dirac basis in which $\gamma^0 = \text{diag}(1,1,-1,-1)$; for further discussion see Ref.~\cite{Pineda:2011dg}. 
In Eq.~\eqref{eq:nrqcdlagrangian}, the NRQCD gauge and light quark terms $\mathcal{L}_g^{\rm NRQCD}$ and $\mathcal{L}_l^{\rm NRQCD}$ are identical to their QCD counterparts $\mathcal{L}_g$ and $\mathcal{L}_l$ in Eq.~\eqref{eq:qcdlagrangian} up to $O(1/m_Q^2)$ corrections.
Interaction terms with light degrees of freedom are suppressed by $\mathcal{O}(\alpha_s/m_Q^2)$ and are given in Ref.~\cite{Brambilla:2004jw}.
The effects of these heavy-light interactions on heavy-heavy interactions are suppressed by the square of this factor and are therefore $\mathcal{O}(\alpha_s^2/m_Q^4)$ and neglected below; however, these interactions could be relevant for studies of heavy-light systems.
The heavy quark one-body term is given by
\begin{equation}
\begin{split}
  \mathcal{L}_{\psi}&=\psi^{\dagger}\left\lbrace i D_0 + c_2\frac{\bs{D}^2}{2m_Q} + c_4 \frac{\bm{D}^4}{8m_Q^3}  + c_F g_s\frac{\bs{\sigma}\cdot\bs{B}}{2m_Q}
  \right. \\ 
  &\hspace{10pt} + c_D g_s\frac{[\bs{D}\cdot\bs{E}]}{8m_Q^2} +\left. ic_S g_s\frac{\bs{\sigma}\cdot(\bs{D}\times\bs{E}-\bs{E}\times\bs{D})}{8m_Q^2}  \right \rbrace\psi,
    \label{eqn:L2pt}
\end{split}
\end{equation}
 where $c_2 = c_4 = 1$ is guaranteed by reparameterization invariance~\cite{Luke:1992cs}, and the remaining Wilson coefficients to $\mathcal{O}(1/m_Q^3)$ are given in~\cite{Manohar:1997qy}. The corresponding heavy antiquark one-body terms $\mathcal{L}_\chi$ are equal to $\mathcal{L}_\psi$ with $\psi\rightarrow \chi$.
There are also four-quark operators involving the heavy quark and antiquark~\cite{Pineda:1998kj, Brambilla:2004jw},
\begin{equation}
    \begin{split}
  \mathcal{L}_{\psi\chi}&=\frac{d_{\mathbf{1}\mathbf{1}}}{m_Q^2}\psi_i^{\dagger}\psi_j\chi_k^{\dagger}\chi_l\delta_{ij}\delta_{kl}+\frac{d_{\mathbf{1}\mathbf{3}}}{m_Q^2}\psi_i^{\dagger}\bs{\sigma}\psi_j\chi_k^{\dagger}\bs{\sigma}\chi_l\delta_{ij}\delta_{kl} \\
  &\hspace{10pt}  +\frac{d_{\mathbf{8}\mathbf{1}}}{m_Q^2}\psi_i^{\dagger}T^a_{ij}\psi_j\chi_k^{\dagger}T^a_{kl}\psi_l
  \\
  &\hspace{10pt}  +\frac{d_{\mathbf{8}\mathbf{3}}}{m_Q^2}\psi_i^{\dagger}T^a_{ij}\bs{\sigma}\psi_j\chi_k^{\dagger}T^a_{kl}\bs{\sigma}\chi_l. 
    \label{eqn:L4pt}
    \end{split}
\end{equation}
as well as operators involving either quarks or antiquarks,
\begin{equation}
    \begin{split}
  \mathcal{L}_{\psi\psi}&=\frac{d_{\mathbf{\bar{3}}\mathbf{1}}}{m_Q^2}\psi_i^{\dagger}\psi_j^{\dagger}\psi_k\psi_l\epsilon_{ijm}\epsilon_{klm}+\frac{d_{\mathbf{\bar{3}}\mathbf{3}}}{m_Q^2}\psi_i^{\dagger}\bs{\sigma}\psi_j^{\dagger}\psi_k^{\dagger}\bs{\sigma}\psi_l\epsilon_{ijm}\epsilon_{klm} \\
  &\hspace{10pt}  +\frac{d_{\mathbf{6}\mathbf{1}}}{m_Q^2}\psi_i^{\dagger}\psi_j\psi_k^{\dagger}\psi_l(\delta_{il}\delta_{jk}+\delta_{jl}\delta_{ik})
  \\
  &\hspace{10pt}  +\frac{d_{\mathbf{6}\mathbf{3}}}{m_Q^2}\psi_i^{\dagger}\bs{\sigma}\psi_j\psi_k^{\dagger}\bs{\sigma}\psi_l(\delta_{il}\delta_{jk}+\delta_{jl}\delta_{ik}),
   \\
  \hspace{10pt} 
  \mathcal{L}_{\chi\chi}&= \mathcal{L}_{\psi\psi}(\psi\leftrightarrow\chi_c),
    \label{eqn:L4ptq}
    \end{split}
\end{equation}
The Wilson coefficients, $d_{rr'}$, sub-scripted by color and spin representations, are given for both equal and unequal mass cases in Refs.~\cite{Pineda:1998kj}. 
The covariant derivative is $D^{\mu}=\partial^{\mu}+ig_sA^{\mu}_aT^a\equiv(D_t,-\bs{D})$, such that $iD_t=i\partial_t -g_sA_0$ and $i\bs{D}=i\bs{\partial}+g_s\bs{A}$. The chromo-electric and magnetic fields are defined as $B^i=\frac{i}{2g_s}\epsilon^{ijk}[D_j,D_k]$ and $\bs{E}=-\frac{i}{g_s}[D_t,\bs{D}]$, respectively. The matching coefficients $c_i$, and $d_{rr'}$ for the equal and unequal mass cases are known to two- and one-loop order in QCD and the SM, respectively \cite{Pineda:1998kj,Gerlach:2019kfo,Assi:2020piz}. Note that Eqs.~\eqref{eqn:L2pt} and~\eqref{eqn:L4pt} are constructed by including all parity-preserving, rotationally invariant,
Hermitian combinations of $iD_t$, $\bs{D}$, $\bs{E}$, $i\bs{B}$, and $i\bs{\sigma}$.

Although NRQCD is a powerful tool for studying heavy quarkonium, it fails to exploit the entire hierarchy of scales in such a system, namely momentum, $|\bs{p}|\sim m_Q |\bs{v}|\ll m_Q$ and binding energy, $E\sim m_Q |\bs{v}|^2\ll |\bs{p}|$. As we are interested in physics at the scale of the binding energies, we can further expand NRQCD in $|\bs{p}|\gg E$. The resulting EFT is an expansion in powers of $E/|\bs{p}|$ known as potential NRQCD (pNRQCD)~\cite{Pineda:1998kj,Brambilla:1999xf}.
Interactions in the pNRQCD Lagrangian that are suppressed by powers of $E/|\bs{p}|$ are local in time but non-local in space and are therefore equivalent to nonrelativistic (two- or more-body) potentials.
Non-potential quark-gluon interactions are also present in pNRQCD but are suppressed by powers of $\alpha_s(\mu)$.
The renormalization scale $\mu$ should ideally be chosen in the range $|\bm{p}| < \mu < m_Q$ for typical momentum scales since logarithms of $\bm{p}/\mu$ arise in matching NRQCD to pNRQCD and logarithms of $m_Q/\mu$ are present from matching QCD to NRQCD.

There are two different kinematic regions with different pNRQCD descriptions: the weak ($|\bs{p}|\gg\Lambda_{\text{QCD}}$) and strong ($|\bs{p}|\sim\Lambda_{\text{QCD}}$) coupling regimes.
In the strong-coupling regime, matching between NRQCD and pNRQCD must be performed non-perturbatively and has been studied by using lattice QCD results in matching calculations to determine pNRQCD potentials; for a review see~\cite{Brambilla:2004jw}.
In this work, we will consider only the weak-coupling regime, where matching between NRQCD and pNRQCD can be performed perturbatively in a dual expansion in $\alpha_s$ and $1/m_Q$, as reviewed in Ref.~\cite{Pineda:2011dg}. Weak-coupling pNRQCD has been used extensively to study heavy quarkonium with the degrees of freedom typically taken to be a composite field describing the heavy $Q\overline{Q}$ system, light quarks, and gluons. Analogous composite $QQQ$ fields have been used in pNRQCD studies of baryons~\cite{Brambilla:2005yk,Brambilla:2009cd}. It is also possible to use the nonrelativistic quark spinor degrees of freedom of NRQCD as the heavy quark degrees of freedom of pNRQCD~\cite{Pineda:2011dg}.
This latter choice of degrees of freedom is not commonly used. However, it permits a unified construction of the pNRQCD operators relevant for describing arbitrary multi-hadron states composed of heavy quarks, and the construction of the pNRQCD Lagrangian with explicit heavy quark degrees of freedom is therefore pursued below.

With this choice of degrees of freedom, the fields of pNRQCD are identical to those of NRQCD. 
The theories differ in that pNRQCD includes spatially non-local heavy quark ``potential'' interactions in its Lagrangian,
\begin{equation}
    L_{\text{pNRQCD}} = L_{\text{NRQCD}}^{\text{us}} + L_{\text{pot}}.
\end{equation}
The potential piece, $L_{\rm pot}$, is given by a sum of a quark-antiquark potential as well as quark-quark, three-quark, and higher-body potentials relevant for baryon and multi-hadron systems composed of heavy quarks,
\begin{equation}
    L_{\text{pot}} = L^{\text{pot}}_{\psi\chi} + L^{\text{pot}}_{\psi\psi} + L^{\text{pot}}_{3\psi} + \ldots. \label{eq:Lpot_parts}.
\end{equation}
The different terms in $L_{\rm pot}$ will be discussed below.
The remaining term $L_{\text{NRQCD}}^{\text{us}}(t)$ corresponds to $L_{\text{NRQCD}}(t) \equiv \int d^3x\ \mathcal{L}(t,\bs{x})$ with only ultra-soft gluon modes included: in other words a multipole expansion of the quark-gluon vertices is performed, and contributions which are not suppressed by $E/|\bs{p}|$ are explicitly removed since they correspond to the soft modes whose effects are described by $L_{\text{pot}}$~\cite{Pineda:1997bj,Pineda:1998kn,Kniehl:1999ud}.
The remaining subleading multipole contributions correspond to ultra-soft modes, and since they do not include infrared singular contributions by construction, they can be included perturbatively.
Ultra-soft contributions to meson and baryon masses in pNRQCD have been studied and found to be N${}^3$LO effects suppressed by $O(\alpha_s^3)$ compared to the LO binding energies~\cite{Kniehl:1999ud,Pineda:1997bj}.
The state-dependence of ultra-soft gluon effects arises through integrals over coordinate space involving the initial- and final-state wavefunctions and are therefore  N${}^3$LO for arbitrary { color-singlet} hadron or multi-hadron systems. Ultrasoft gluon effects will be neglected below since we work to NNLO accuracy. We note, however, that they could be included as perturbative corrections to the binding energies computed in Sec.~\ref{sec:qcd}-\ref{sec:bsm} by determining the baryonic analogs of ultrasoft gluon corrections to quarkonium energy levels, as discussed in Refs.~\cite{Kniehl:1999ud,Pineda:2011dg}.
 { The same construction could be applied in the alternative velocity NRQCD (vNRQCD) power counting~\cite{Hoang:2002ae,Hoang:2002yy}. Differences between the power countings first appear in the N${}^3$LO ultrasoft contributions that are neglected here, and therefore all results of this work are immediately applicable to vNRQCD~\cite{Brambilla:2004jw,Hoang:2002ae,Hoang:2002yy,Hoang:2011gy,Pineda:2001et,Pineda:2011dg} }

\subsection{Quark-antiquark potential}

A sum of color-singlet and color-adjoint terms gives the quark-antiquark potential for arbitrary $N_c$,
\begin{equation}
    \begin{split}
        L^{\text{pot}}_{\psi\chi} =& - \int d^3\bs{r}_1d^3\bs{r}_2\,  \psi^{\dagger}_i(t,\bs{r}_1)\chi_j(t,\bs{r}_2) \chi^{\dagger}_k(t,\bs{r}_2)\psi_l(t,\bs{r}_1) \\
       &\times \left[ \frac{1}{N_c}\delta_{ij}\delta_{kl} V^{\psi\chi}_{\mathbf{1}}(\bs{r}_{12})  + \frac{1}{T_F}T^a_{ij} T^a_{kl} V^{\psi\chi}_{\text{Ad}}(\bs{r}_{12}) \right], \label{eq:psichi}
    \end{split}
\end{equation}
where $\bs{r}_{12}\equiv\bs{r}_1-\bs{r}_2$,  $T_F = 1/2$, and the fermion spin indices are implicitly contracted with indices of the potential.
The potential depends on the renormalization scale $\mu$ as well as $\bs{p}_a = -i\bs{\nabla}_a$ and $\bs{S}_{a}=\bs{\sigma}_a/2$, but these dependencies will generally be kept implicit except where otherwise noted.
Here and below, $i,j,k,\ldots$ represent fundamental color indices, and $a,b,c,\ldots$ index adjoint color indices.
The color-singlet potential is expanded as a power series in $1/m_Q$,
\begin{equation}
\begin{split}
    V^{\psi\chi}_{\mathbf{1}}(r)={}& V_{\mathbf{1}}^{\psi\chi,(0)}(r)+\frac{V_{\mathbf{1}}^{\psi\chi,(1)}(r)}{m_Q} \\
    &\hspace{10pt} +\frac{V_{\mathbf{1}}^{\psi\chi,(2)}(r)}{m_Q^2}+\mathcal{O}(1/m_Q^3).
    \end{split}
\end{equation}
The $\mathcal{O}(1/m_Q^0)$ and $\mathcal{O}(1/m_Q)$ potentials are given by
\begin{align}
    V_{\mathbf{1}}^{\psi\chi,(0)}(r,\mu) ={}&-C_F\frac{\alpha_{V_{\mathbf{1}}}(r,\mu)}{r},
    \\
    V_{\mathbf{1}}^{\psi\chi,(1)}(r,\mu) ={}& -\frac{C_FC_A}{2m_Qr^2}D^{(1)}(\mu),
   \label{eqn:singpots}
\end{align}
where $\mu$ dependence is shown explicitly, the perturbative expansion of $\alpha_{V_{\mathbf{1}}}(r,\mu)$ is discussed below, $C_A = N_c$, $C_F = (N_c^2-1)/(2N_c)$, and $D^{(1)}(\mu) = \alpha_s(\mu)^2 + \mathcal{O}(\alpha_s^3)$ in Coulomb gauge as discussed in in Refs.~\cite{Kniehl:2001ju,Brambilla:2004jw}.
At $\mathcal{O}(1/m_Q^2)$ there are spin-independent and spin-dependent potentials that arise,
\begin{align}
   \label{eqn:singpotsOLO}
    V_{\mathbf{1}}^{\psi\chi,(2)}(r) ={}& V^{\psi\chi,(2)}_{\mathbf{1},\text{SI}}(r) +V^{\psi\chi,(2)}_{\mathbf{1},\text{SD}}(r) ,
    \\
    V^{\psi\chi,(2)}_{\mathbf{1},\text{SI}}(r) ={}& -\frac{C_FD_{1}^{(2)}}{2m_Q^2}\left\lbrace\frac{1}{r},\bs{p}^2\right\rbrace+\frac{C_FD_2^{(2)}}{2m_Q^2r^3}\bs{L}^2 \\\nonumber
    {}& +\frac{\pi  C_F D_{\delta}^{(2)}}{m_Q^2}\delta^{(3)}(\bs{r}),\\
   V^{\psi\chi,(2)}_{\mathbf{1},\text{SD}}(r) ={}& \frac{4\pi C_FD_{S^2}^{(2)}}{3m_Q^2}\bs{S}^2\delta^{(3)}(\bs{r})+ \frac{3 C_FD_{LS}^{(2)}}{2m_Q^2r^3}\bs{L}\cdot\bs{S} \\\nonumber
   {}& + \frac{C_F D_{S_{12}}^{(2)}}{4m_Q^2r^3}\bs{S}_{12}(\hat{\bs{r}}),
\end{align}
where $\bs{S}=\bs{S}_1+\bs{S}_2$, $\bs{S}_{12}(\hat{\bs{r}})= 3\hat{\bs{r}}\cdot\bs{\sigma}_1\hat{\bs{r}}\cdot\bs{\sigma}_2-\bs{\sigma}_1\cdot\bs{\sigma}_2$, $\bs{L}=\bs{r}\times\bs{p}$, and the $D^{(2)}$ coefficients are given in~\cite{Brambilla:2004jw,Pineda:2000sz}. The singlet potential is known to $\rm{N^3LO}$ in QCD and $\rm{NLO}$ in the SM~\cite{Kniehl:2002br,Assi:2020srd}. The adjoint (octet for $N_c=3$) potential is also known to $\rm{N^3LO}$ and is given in Ref.~\cite{Anzai:2013tja}.

The potentials such as $V_{\bm 1}^{(0)}(r)$ appearing in Eq.~\eqref{eqn:singpots} are Wilson coefficients in pNRQCD, which can be obtained by matching with NRQCD.
{ By considering matching with a color-singlet quarkonium state, it can be seen that $V_{\bm 1}^{(0)}(r)$ is identical to the color-singlet potential present in traditional formulations of pNRQCD with a Langrangian including hadron-level interpolating operators. }
The color-singlet potential has been computed to $\text{N}^3\text{LO}$ in the $\overline{\text{MS}}$ scheme for the case of heavy quarks with equal masses (the unequal mass case is not fully known to $\mathcal{O}(1/m_Q^2)$, although various pieces have been computed~\cite{Peset:2015vvi}) 
and has the perturbative expansion
\begin{align}
   \alpha_{V_{\mathbf{1}}}(r,\mu) = \alpha_s(\mu)\left(1+\sum_{n=1}^3\left(\frac{\alpha_s(\mu)}{4\pi}\right)^n \tilde{a}_n(r;\mu)\right),
   \label{eqn:alphav1}
\end{align}
where
\begin{align}
    \tilde{a}_1(r;\mu)={}&a_1+2\beta_0\ln(r\mu e^{\eul}), \nonumber\\
    \tilde{a}_2(r;\mu)={}&a_2+\frac{\pi^2}{3}\beta_0^2+(4a_1\beta_0+2\beta_1)\ln(r\mu e^{\eul})
    \nonumber\\
    {}&+4\beta_0^2\ln^2(r\mu e^{\eul}), \nonumber\\
    \tilde{a}_3(r;\mu)={}&a_3+a_1\beta_0^2\pi^2+\frac{5\pi^2}{6}\beta_0\beta_1+16\zeta_3\beta_0^3+\nonumber\\    &+\left(2\pi^2\beta_0^3+6a_2\beta_0+4a_1\beta_1+2\beta_2 \right.\nonumber\\
    &\left.+\frac{16}{3}C_A^3\pi^2\right)\ln(r\mu e^{\eul}) + \left(12a_1\beta_0^2\right.\nonumber\\   &\left.+10\beta_0\beta_1\right)\ln^2(r\mu e^{\eul})+8\beta_0^3\ln^3(r\mu e^{\eul}).
\end{align}
The coefficients up to $\text{N}^3\text{LO}$ are given in Ref.~\cite{Kniehl:2002br}.
The numerical calculations presented below are carried out to NNLO accuracy and therefore require the coefficients
\begin{align}
\beta_0&=\frac{11}{3}C_A-\frac{4}{3}T_Fn_f, \\
\beta_1&=\frac{34}{3}C_A^2-4 C_FT_Fn_f-\frac{20}{3}C_AT_Fn_f,
\end{align}
and
\begin{align}
    a_1={}& \frac{31}{9}C_A-\frac{20}{9}T_Fn_f, \\
    a_2={}& \left(\frac{4343}{162}+4\pi^2-\frac{\pi^4 }{4}+\frac{22}{3}\zeta_3\right)C_A^2 \nonumber\\ {}& -\left(\frac{55}{3}-16\zeta_3\right)C_FT_Fn_f+\frac{400}{81}T_F^2n_f^2 
    \nonumber\\ {}&-\left(\frac{1798}{81}+\frac{56}{3}\zeta_3\right)C_AT_Fn_f.
\end{align}
Note that in obtaining the pNRQCD Lagrangian presented here, a single fixed renormalization scale $\mu$ is assumed to be used during matching from QCD to NRQCD and NRQCD to pNRQCD. This renormalization scale, therefore, acts as an effective cutoff for both heavy-quark momenta $\bm{p}$ satisfying $|\bm{p}|\ll m_Q$ as well as the four-momenta $\ell^\mu$ of the light degrees of freedom satisfying $\ell^\mu \sim \bm{p}^2/m_Q \ll |\bm{p}|$.
Further refinements to pNRQCD can be achieved by RG evolving the NRQCD Wilson coefficients to resum logarithms of $m_Q / \mu$ or performing renormalization-group improvement of the pNRQCD potentials to resum logarithms of $|\bm{p}|/\mu$~\cite{Pineda:2011dg}. However, such improvement is not straightforward to implement in the QMC approaches to studying multi-quark systems in pNRQCD discussed below, and it is not pursued in this work.

\subsection{Quark-quark potential}

The quark-quark potential appearing in Eq.~\eqref{eq:Lpot_parts} is given by a sum of color-antisymmetric and color-symmetric terms,
\begin{equation}
    \begin{split} 
        L^{\text{pot}}_{\psi\psi} =& -\int d^3\bs{r}_1d^3\bs{r}_2\,  \psi^{\dagger}_i(t,\bs{r}_1)\psi^{\dagger}_j(t,\bs{r}_2) \psi_k(t,\bs{r}_2)\psi_l(t,\bs{r}_1) \\
    &\times \left[ \frac{N_c-1}{4}  \left( \mathcal{F}_{ij}^{\text{A}m} 
    \right)^* \mathcal{F}_{kl}^{\text{A}m}  V^{\psi\psi}_{\text{A}}(\bs{r}_{12}) \right. \\
    &\hspace{20pt} \left. + { \frac{1}{2}}  \left( \mathcal{F}_{ij}^{\text{S}\eta } \right)^* \mathcal{F}_{kl}^{\text{S}\eta }  V^{\psi\psi}_{\text{S}}(\bs{r}_{12}) \right],
    \end{split}   \label{eq:Lpsipsi}
\end{equation}
where  $V^{\psi\psi}_{\bm{\rho}}(r)$ with $\rho = \text{A}$ and $\rho = \text{S}$ denote the potentials for quark-quark states in antisymmetric and symmetric representations, respectively, which are presented explicitly below.
The  antisymmetric and symmetric color tensors $\mathcal{F}_{ij}^{\text{A}m} = -\mathcal{F}_{ji}^{\text{A}m }$  and $\mathcal{F}_{ij}^{\text{S}\eta }=\mathcal{F}_{ji}^{\text{S}\eta }$ are orthogonal and satisfy $\mathcal{F}_{ij}^{\text{A}m } \mathcal{F}_{ij}^{\text{A} m' } = \delta^{mm'}$ and $\mathcal{F}_{ij}^{\text{S}\eta } \mathcal{F}_{ij}^{\text{S}\eta' } = \delta^{\eta \eta'}$ where $\eta \in \{1,\ldots, N_c(N_c+1)/2\}$.
Explicit representations for $\mathcal{F}_{ij}^{\text{A}m} $ and $\mathcal{F}_{ij}^{\text{S}\eta }$ can be found in Appendix B of Ref.~\cite{Brambilla:2005yk} but will not be needed below; the products appearing in Eq.~\eqref{eq:Lpsipsi} are given by
\begin{align} \mathcal{F}_{ij}^{\text{A}m}\mathcal{F}_{kl}^{\text{A}m}  &= \frac{1}{(N_c-1)!}\epsilon_{ijo_1\ldots o_{N_c-2}}\epsilon_{klo_1\ldots o_{N_c-2}}, \\
 \mathcal{F}_{ij}^{S\eta }\mathcal{F}_{kl}^{S\eta}  &= \frac{1}{2}\left(  \delta_{il}\delta_{jk}+\delta_{jl}\delta_{ik}\right). \label{eq:colorfacdef}
\end{align}
The coefficients of the operators appearing in Eq.~\eqref{eq:Lpsipsi} are chosen so that the action of $L^{\text{pot}}_{\psi\psi}$ on a quark-quark state in either the antisymmetric or symmetric representation, $\ket{\psi_i(\bs{x}_1)\psi_j(\bs{x}_2)}\mathcal{F}_{ij}^{\text{A}m}$ or $\ket{\psi_i(\bs{x}_1)\psi_j(\bs{x}_2)}\mathcal{F}_{ij}^{\text{S}\delta}$, is equivalent to multiplying that state by $ V^{\psi\psi}_{\text{A}}(\bs{r}_{12}) $ or $ V^{\psi\psi}_{\text{S}}(\bs{r}_{12}) $ respectively, as detailed in Sec.~\ref{sec:hammy} below.

The pNRQCD quark-quark potentials $V^{\psi\psi}_{\bm{\rho}}(r)$ have the same shape as the quark-antiquark potential up to NLO and differ only in the color factors governing the sign and normalization of the potential.
To determine the appropriate color factors, the tensors associated with the two quark and antiquark fields in each operator, $\mathcal{F}_{ij}^{\bs{\rho}\zeta}$ and $\mathcal{F}_{kl}^{\bs{\rho}\zeta}$, can be used as creation and annihilation operators for initial and final states in particular representations (here $\zeta$ denotes a generic irrep row index).
The color factor for this representation is obtained by contracting these initial- and final-state color tensors with the color structure resulting from a given NRQCD Feynman diagram, denoted $\mathcal{D}_{ijkl}^{2\psi,d}$, where the superscript $d$ labels the particular diagram and normalizing the result~\cite{Nadkarni:1986as},
\begin{equation}
     \mathcal{C}_{\bs{\rho}}^{\psi\psi,d} = \frac{ \mathcal{F}_{ij}^{\bs{\rho}\zeta} \mathcal{D}_{ijkl}^{2\psi,d} \left( \mathcal{F}_{kl}^{\bs{\rho}\zeta}  \right)^* }{ \sqrt{ \left( \mathcal{F}_{i'j'}^{\bs{\rho}\zeta'} \right)^* \mathcal{F}_{i'j'}^{\bs{\rho}\zeta'} \left( \mathcal{F}_{k'l'}^{\bs{\rho}\zeta''} \right)^*\mathcal{F}_{k'l'}^{\bs{\rho}\zeta''}} } . \label{eq:colorfac2}
\end{equation}
Summing over all relevant diagrams gives
\begin{equation}
\mathcal{C}_{\bs{\rho}}^{\psi\psi} = \sum_d \mathcal{C}_{\bs{\rho}}^{\psi\psi,d}.
\end{equation}
This color factor can be determined by applying Eq.~\eqref{eq:colorfac2} to the tree-level diagram
\begin{equation}
  \mathcal{D}^{\psi\psi, \rm tree}_{ijkl} = (T^a)_{ij}(T^a)_{kl},
\end{equation}
to give~\cite{Brambilla:2005yk,Brambilla:2009cd}
\begin{align}
C_{\text{A}}^{\psi\psi,\rm tree} &= - \frac{C_F}{N_c -1}, \\
C_{\text{S}}^{\psi\psi,\rm tree} &= \frac{C_F}{N_c + 1}.
\end{align}
The antisymmetric quark-quark potential is therefore attractive, while the symmetric quark-quark potential is repulsive.
No further representation-dependence arises in the potential at NLO, and so, for example, the antisymmetric quark-quark potential is related to the quark-antiquark potential by~\cite{Brambilla:2009cd},
\begin{equation}
  V^{\psi\psi}_{\text{A}} = \frac{1}{N_c-1} V^{\psi\chi}_{\mathbf{1}} + \mathcal{O}(\alpha_s^3). \label{eq:Vpsipsi}
\end{equation}
The same proportionality holds at NLO for generic color representations,
\begin{equation}
 V^{\psi\psi}_{\bs{\rho}} = -\frac{C_{\bs{\rho}}^{\psi\psi,\rm tree}}{C_F} V^{\psi\chi}_{\mathbf{1}} + \mathcal{O}(\alpha_s^3).
\end{equation}

\begin{figure}[t]
    \includegraphics[width=.4\textwidth]{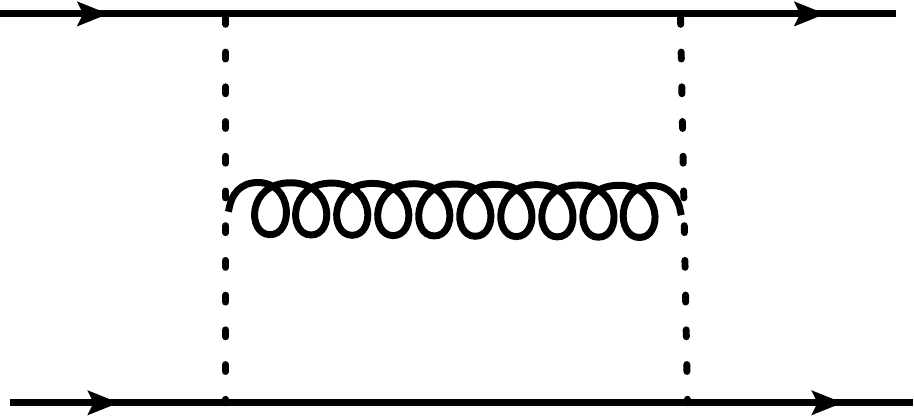}
    \caption{NRQCD Feynman diagram that contributes to the representation-dependent potential $\delta a_{\bm{\rho}}$ when matching to pNRQCD. Dotted and curly lines correspond to longitudinal and transverse gluons in Coulomb gauge. }
    \label{fig:2bodypot}
\end{figure}

At NNLO, the correction to the two body potential of a general color representation $\bm{\rho}$ is known to have the form~\cite{Collet:2011kq},
\begin{equation}
    V^{\psi\psi}_{\bm{\rho}}(r) = -\mathcal{C}^{\psi\psi,\rm tree}_{\bm{\rho}}\left(\frac{1}{C_F}V^{\psi\chi}_{\bm{1}}(r) - \frac{\alpha_s^3}{(4\pi)^2}\frac{\delta a_{\bm{\rho}}}{r}\right). \label{eq:1A}
\end{equation}
The NNLO correction, $\delta a_{\bm{\rho}}$ has been determined for various color representations~\cite{Collet:2011kq,Kniehl:2004rk}, and varies based on the color factor of the H-diagram in Fig.~\ref{fig:2bodypot}, first computed in Ref.~\cite{Kummer:1996jz}. The value of this diagram, modulo coupling and color structure, is $1/r$ times $\mathcal{H} = 2\pi^2(\pi^2-12)$. The color tensor of the H-diagram shown in Fig.~\ref{fig:2bodypot} is
\begin{equation}
\begin{split}
     \mathcal{D}_{ijkl}^{2\psi,H} = (T^aT^c)_{ij}(T^eT^b)_{kl}f^{abd}f^{ced}.
     \end{split}
\end{equation}
The color factors $\mathcal{C}_{\bs{\rho}}^{\psi\psi,H}$ can be determined by projecting into the color symmetric and antisymmetric representations using Eq.~\eqref{eq:colorfac2}.
The NNLO correction factor is then given by $\delta a_{\bm{\rho}} = \mathcal{H} \mathcal{C}_{\bs{\rho}}^{\psi\psi,H} / \mathcal{C}_{\bs{\rho}}^{\psi\psi,\rm tree} $ as 
\begin{align}
   \delta a_{\text{A}}={}&\frac{N_c(N_c-2)}{2}\pi^2(\pi^2-12), \\
   \delta a_{\text{S}}={}&\frac{N_c(N_c+2)}{2}\pi^2(\pi^2-12).
    \label{eqn:delta2}
\end{align}
This completes the two-body potentials needed to study generic multi-hadron systems in pNRQCD at NNLO.
What remains are higher-body potentials, which, as discussed in Ref.~\cite{Brambilla:2009cd} and below, also arise at NNLO.

\subsection{Three-quark potentials}
\label{scn:3higherpots}

\begin{figure*}[t]
    \includegraphics[width=.9\textwidth]{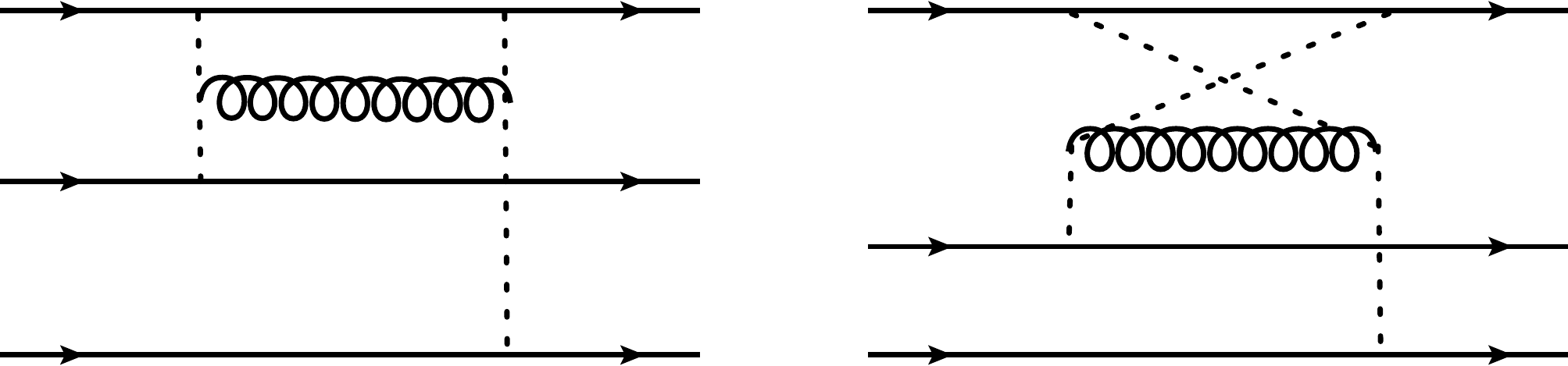}
    \caption{Leading NRQCD Feynman diagrams in Coulomb gauge that lead to non-vanishing contributions to the $3$-body potential when matching to pNRQCD. Dotted and curly lines correspond to longitudinal and transverse gluons, respectively. }
    \label{fig:3bodypot}
\end{figure*}

Three-quark forces first appear in NRQCD at $\mathcal{O}(\alpha_s^3)$. Non-zero contributions in Coulomb gauge arise from the two diagrams shown in Fig.~\ref{fig:3bodypot} and their permutations as discussed in Ref.~\cite{Brambilla:2009cd}.
Specializing first to $N_c=3$, the three-quark potential for a generic representation $\bs{\rho}$ arising in $\mathbf{3}\otimes\mathbf{3}\otimes\mathbf{3} = (\overline{\mathbf{3}}\oplus \mathbf{6})\otimes\mathbf{3} = \mathbf{1}\oplus\mathbf{8}_{\text{A}} \oplus\mathbf{8}_{\text{S}} \oplus\mathbf{10}$ is given by
\begin{equation}
\begin{split}
V_{\bs{\rho}uv}^{3\psi} ={}&\alpha\left(\frac{\alpha}{4\pi}\right)^2 \left[ \mathcal{C}_{\bs{\rho}uv}^{3\psi,1} v_3(\bs{r}_{12},\bs{r}_{13})
    \right. \\ {}&\left.
    +\mathcal{C}_{\bs{\rho}uv}^{3\psi,2} v_3(\bs{r}_{12},\bs{r}_{23})  + \mathcal{C}_{\bs{\rho}uv}^{3\psi,3} v_3(\bs{r}_{13},\bs{r}_{23}) \right],
    \end{split} \label{eq:V3psidef}
\end{equation}
where $\mathbf{r}_{IJ} \equiv \mathbf{r}_I - \mathbf{r}_J$, the indices $u,v \in \{\text{A},\text{S}\}$ label the octet color tensors, which are antisymmetric or symmetric respectively in their first two indices, and should be neglected for $\bs{\rho} \in \{ \mathbf{1}, \mathbf{10} \}$ where only one operator appears, and $\mathcal{C}_{\bs{\rho}uv}^{3\psi,q}$ is the color factor for the permutation of the three-quark diagrams shown in Fig.~\ref{fig:3bodypot} and discussed further below.
Here, $v_3(\bs{r},\bs{r}')$ describes the  spatial structure of the 3-quark potential diagrams, which takes a universal form given by~\cite{Brambilla:2009cd}
\begin{equation}
     v_3(\bs{r},\bs{r}') = 16\pi\int_0^1dxdy \left[ \hat{\bs{r}}\cdot\hat{\bs{r}}' \mathcal{I}_1 + \hat{\bs{r}}^i\hat{\bs{r}}'^j \mathcal{I}_2 \right],
\end{equation}
where $\mathcal{I}_1$ and $\mathcal{I}_2$ are defined in terms of $\bs{R}=x\bs{r}-y\bs{r}'$, $R=|\bs{R}|$, and $A=|\bs{r}|\sqrt{x(1-x)}+|\bs{r}'|\sqrt{y(1-y)}$ by
\begin{align}
     \mathcal{I}_1 &= \frac{1}{R}\left[\left(1-\frac{A^2}{R^2}\right)\arctan{\frac{R}{A}}+\frac{A}{R}\right], \\
     \mathcal{I}_2 &= \frac{\hat{\bs{R}}^i\hat{\bs{R}}^j}{R}\left[\left(1+\frac{3A^2}{R^2}\right)\arctan{\frac{R}{A}}-3\frac{A}{R}\right].
    \label{eqn:3qpot}
\end{align}
The color factors in Eq.~\eqref{eq:V3psidef} can be expressed as contractions of the tensors associated with the three quark and antiquark fields in each operator, $\mathcal{F}_{ijk}^{\bs{\rho}u\zeta}$ and $\mathcal{F}_{lmn}^{\bs{\rho}v\zeta}$, with the color tensor relevant for a particular diagram
\begin{equation}
     \mathcal{C}_{\bs{\rho}uv}^{3\psi,q} = \frac{ \left( \mathcal{F}_{ijk}^{\bs{\rho}u\zeta} \right)^* \mathcal{D}_{ijklmn}^{3\psi,q} \mathcal{F}_{lmn}^{\bs{\rho}v\zeta} }{ \sqrt{ \left( \mathcal{F}_{i'j'k'}^{\bs{\rho}u\zeta'} \right)^* \mathcal{F}_{i'j'k'}^{\bs{\rho}u\zeta'} \left( \mathcal{F}_{l'm'n'}^{\bs{\rho}v\zeta''} \right)^*\mathcal{F}_{l'm'n'}^{\bs{\rho}v\zeta''}} } . \label{eq:colorfac}
\end{equation}
Octet color tensors that are antisymmetric or symmetric respectively in their first two indices are defined by 
\begin{align}
   \mathcal{F}_{ijk}^{\mathbf{8}\text{A}a} &= \frac{1}{\sqrt{2T_F}} \epsilon_{ijp}T^a_{pk},  \\
    \mathcal{F}_{ijk}^{\mathbf{8}\text{S}a} &= \frac{1}{ \sqrt{6 T_F}} \left( \epsilon_{ikp} T^a_{pj} + \epsilon_{jkp} T^a_{pi} \right),
\end{align}
and satisfy $\mathcal{F}_{ijk}^{\mathbf{8}u a} \mathcal{F}_{ijk}^{\mathbf{8}v b} = \delta^{uv} \delta^{ab}$.
Totally antisymmetric and totally symmetric color tensors $\mathcal{F}_{ijk}^{\mathbf{1}}$ and $\mathcal{F}_{ijk}^{\mathbf{10}\delta}$ satisfying $\mathcal{F}_{ijk}^{\mathbf{1}} \mathcal{F}_{ijk}^{\mathbf{1}} = 1$ and $\mathcal{F}_{ijk}^{\mathbf{10}\delta} \mathcal{F}_{ijk}^{\mathbf{10}\delta'} = \delta^{\delta \delta'}$ with $\delta \in \{1,\ldots,10\}$ are explicitly presented in Appendix B of Ref.~\cite{Brambilla:2005yk}; below we only need the products
\begin{align}
 \mathcal{F}_{ijk}^{\mathbf{1}}\mathcal{F}_{lmn}^{\mathbf{1}}  &= \frac{1}{6}\epsilon_{ijk}\epsilon_{lmn}, \\
 \mathcal{F}_{ijk}^{\mathbf{10}\delta}\mathcal{F}_{lmn}^{\mathbf{10}\delta}  &= \frac{1}{6}\left(  \delta_{il}\delta_{jm}\delta_{kn} + \delta_{il}\delta_{jn}\delta_{km} \right. \\\nonumber
 &\hspace{30pt} \left. + \delta_{im}\delta_{jl}\delta_{kn} + \delta_{im}\delta_{jn}\delta_{kl} \right. \\\nonumber
 &\hspace{30pt} \left. + \delta_{in}\delta_{jm}\delta_{kl} + \delta_{in}\delta_{jl}\delta_{km} \right).
\end{align}
The color tensor relevant for the particular diagram shown in Fig.~\ref{fig:3bodypot} is
\begin{equation}
\begin{split}
     \mathcal{D}_{ijklmn}^{3\psi,3} =  \frac{1}{2} \left[ T^d_{im}T^a_{jl}T^b_{kr}T^e_{rn}f^{bdc}f^{aec} \right. \\
     \left. + T^d_{im}T^a_{jl}T^e_{kr}T^b_{rn}f^{bdc}f^{aec} \right],
     \end{split}
\end{equation}
and the tensors for its permutations can be obtained using $ \mathcal{D}_{ijklmn}^{3\psi,3} =  \mathcal{D}_{mnklij}^{3\psi,1}$ and $ \mathcal{D}_{ijklmn}^{3\psi,2} =  \mathcal{D}_{ijmnkl}^{3\psi,3}$.
Evaluating Eq.~\eqref{eq:colorfac} for these diagrams shows that the $\mathbf{1}$ and $\mathbf{10}$ color factors do not depend on the permutation label $q$ and are given by
 $\mathcal{C}_{\mathbf{1}}^{3\psi,q} = -\frac{1}{2}$ and $\mathcal{C}_{\mathbf{10}}^{3\psi,q} = -\frac{1}{4}$~\cite{Brambilla:2009cd}.
Evaluating Eq.~\eqref{eq:colorfac} for the adjoint operators leads to
\begin{align}
    \begin{pmatrix} \mathcal{C}_{\mathbf{8}\text{A}\text{A}}^{3\psi,1} & \mathcal{C}_{\mathbf{8}\text{A}\text{S}}^{3\psi,1} \\ \mathcal{C}_{\mathbf{8}\text{S}\text{A}}^{3\psi,1} &  \mathcal{C}_{\mathbf{8}\text{S}\text{S}}^{3\psi,1} \end{pmatrix} &= \begin{pmatrix} \frac{1}{16} & -\frac{\sqrt{3}}{8} \\ -\frac{\sqrt{3}}{8} & \frac{5}{16} \end{pmatrix} , \\
    \begin{pmatrix} \mathcal{C}_{\mathbf{8}\text{A}\text{A}}^{3\psi,2} & \mathcal{C}_{\mathbf{8}\text{A}\text{S}}^{3\psi,2} \\ \mathcal{C}_{\mathbf{8}\text{S}\text{A}}^{3\psi,2} &  \mathcal{C}_{\mathbf{8}\text{S}\text{S}}^{3\psi,2} \end{pmatrix} &= \begin{pmatrix} \frac{1}{16} & \frac{\sqrt{3}}{8} \\ \frac{\sqrt{3}}{8} & \frac{5}{16} \end{pmatrix}, \\
    \begin{pmatrix} \mathcal{C}_{MAA}^{3\psi,3} & \mathcal{C}_{\mathbf{8}\text{A}\text{S}}^{3\psi,3} \\ \mathcal{C}_{\mathbf{8}\text{S}\text{A}}^{3\psi,3} &  \mathcal{C}_{\mathbf{8}\text{S}\text{S}}^{3\psi,3} \end{pmatrix} &= \begin{pmatrix} \frac{7}{16} & 0 \\ 0 & -\frac{1}{16} \end{pmatrix},
\end{align}
which completes the construction of $L^{\text{pot},N_c=3}_{3\psi}$ to NNLO.
The potential for three-quark states in the adjoint representation is computed at LO in Ref.~\cite{Brambilla:2005yk}, and the presence of mixing between states created with $\bm{8}_{\text{A}}$ and $\bm{8}_{\text{S}}$ operators are discussed in Ref.~\cite{Brambilla:2009cd}. The NNLO $3\psi$ potentials for the adjoint representation are reported here for the first time. 
While the \textbf{1} and \textbf{10} three-quark potentials are always attractive, the adjoint three-quark potentials are repulsive for some configurations.

The action of this three-quark potential can be reproduced using the pNRQCD Lagrangian term
\begin{equation}
    \begin{split}
        L^{\text{pot},N_c=3}_{3\psi} =& -\int d^3\bs{r}_1d^3\bs{r}_2d^3\bs{r}_3\,  \psi^{\dagger}_i(t,\bs{r}_1)\psi^{\dagger}_j(t,\bs{r}_2)\psi^{\dagger}_k(t,\bs{r}_3)\\
        &\times \psi_l(t,\bs{r}_3)\psi_m(t,\bs{r}_2)\psi_n(t,\bs{r}_1) \\
    &\times \left[ \mathcal{F}_{ijk}^{\mathbf{1}}\mathcal{F}_{lmn}^{\mathbf{1}}  \frac{1}{6} V^{3\psi}_{\mathbf{1}}  + \mathcal{F}_{ijk}^{\mathbf{10}}\mathcal{F}_{lmn}^{\mathbf{10}} \frac{1}{6}   V^{3\psi}_{\mathbf{10}}   \right. \\
    &\hspace{15pt} \left. + \mathcal{F}_{ijk}^{\mathbf{8}\text{A}a} \mathcal{F}_{lmn}^{\mathbf{8}\text{A}a}W^{3\psi}_{\mathbf{8}\text{A}}  + \mathcal{F}_{ijk}^{\mathbf{8}\text{S}a} \mathcal{F}_{lmn}^{\mathbf{8}\text{S}a}W^{3\psi}_{\mathbf{8}\text{S}}  \right],
    \end{split} \label{eq:L3psiNc3}
\end{equation}
where the functions $W_{\mathbf{8}u}^{3\psi}$ defined below are related to but not identical to the adjoint potentials  $V_{\mathbf{8}uv}^{3\psi}$.
The action of either the symmetric or antisymmetric adjoint potential operator on the corresponding symmetric or antisymmetric adjoint state leads to a linear combination of symmetric and antisymmetric adjoint states arising from non-trivial Wick contractions.
Direct computation of the matrix elements of the adjoint operators in $L^{\text{pot},N_c=3}_{3\psi}$ between states creates by operators involving $\mathcal{F}_{ijk}^{\mathbf{8}\text{A}a}$ and $\mathcal{F}_{ijk}^{\mathbf{8}\text{S}a}$ shows that the desired potentials  $V_{\mathbf{8}u}^{3\psi}$ are reproduced using
\begin{equation}
\begin{split}
W_{\mathbf{8}\text{A}}^{3\psi} ={}&\alpha\left(\frac{\alpha}{4\pi}\right)^2 \left[ -\frac{1}{48} v_3(\bs{r}_{12},\bs{r}_{13})
    \right. \\ {}&\left.
    - \frac{1}{48} v_3(\bs{r}_{12},\bs{r}_{23})  +  \frac{11}{48} v_3(\bs{r}_{13},\bs{r}_{23}) \right],
    \end{split} \label{eq:WA3psidef}
\end{equation}
and
\begin{equation}
\begin{split}
W_{\mathbf{8}\text{S}}^{3\psi} ={}&\alpha\left(\frac{\alpha}{4\pi}\right)^2 \left[ \frac{7}{48} v_3(\bs{r}_{12},\bs{r}_{13})
    \right. \\ {}&\left.
    + \frac{7}{48} v_3(\bs{r}_{12},\bs{r}_{23})  -  \frac{5}{48} v_3(\bs{r}_{13},\bs{r}_{23}) \right].
    \end{split} \label{eq:WA3psidef}
\end{equation}

\begin{figure*}[t!]
    \includegraphics[width=0.8\textwidth]{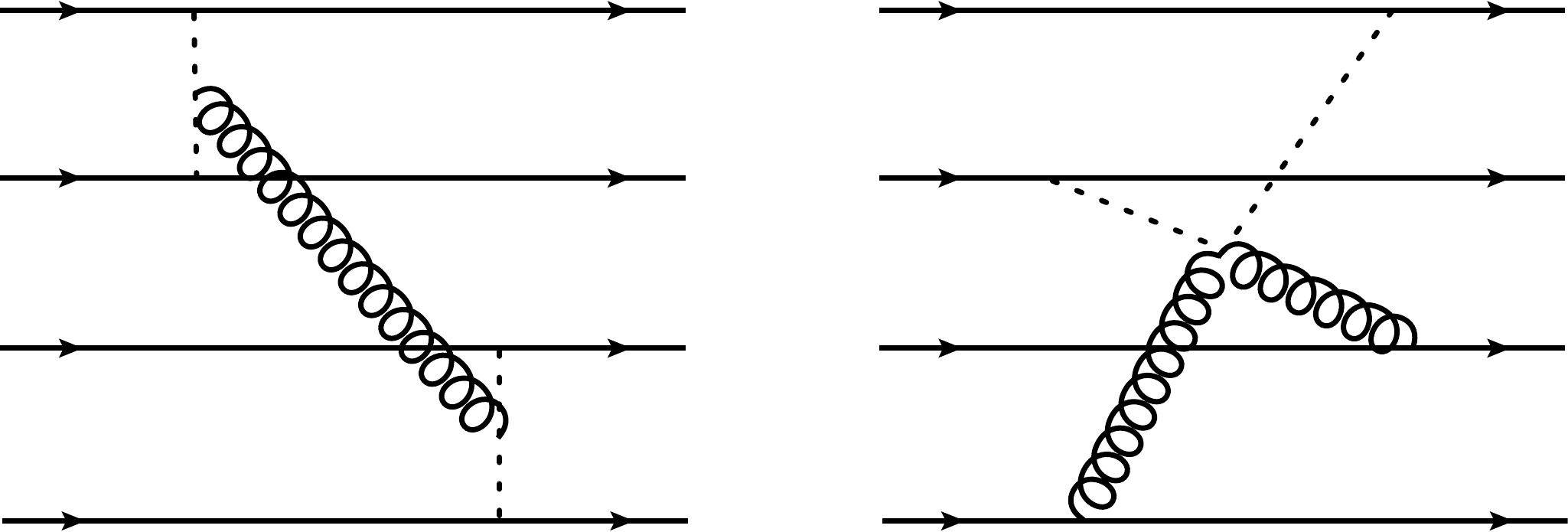}
    \caption{Example NNLO diagrams for $N=4$ which contribute to the $4$-body potential and demonstrate that $\alpha_s^{N}$ suppression at each higher body order is not necessarily respected beyond 3-body.}
    \label{fig:nbodypot}
\end{figure*}

This construction can be generalized\footnote{The case of $N_c = 2$ must be treated separately and is not considered here.} to $N_c \geq 3$.
Mixed-symmetry color adjoint tensors satisfying the same normalization condition as the $N_c=3$ case above can be defined in general by
\begin{align}
   \mathcal{F}_{ijkq_1\ldots q_{N_c-3}}^{\text{MA}a} &= \frac{1}{\sqrt{T_f (N_c-1)!}} \epsilon_{ijpq_1\ldots q_{N_c-3}}T^a_{pk},  \\
    \mathcal{F}_{ijkq_1\ldots q_{N_c-3}}^{\text{MS}a} &= \frac{1}{\sqrt{{2T_f N_c(N_c-2)!}}}\left( \epsilon_{ikpq_1\ldots q_{N_c-3}} T^a_{pj} \right. \\\nonumber
    &\hspace{30pt} \left. + \epsilon_{jkpq_1\ldots q_{N_c-3}} T^a_{pi} \right).
\label{eqn:mstensors}
\end{align}
The totally antisymmetric and totally symmetric tensors satisfy~\cite{Brambilla:2009cd}
\begin{align}
 \mathcal{F}_{ijk}^{\text{A}}\mathcal{F}_{lmn}^{\text{A}}  &= \frac{1}{N_c! }\epsilon_{ijko_1\ldots o_{N_c-3}}\epsilon_{lmno_1\ldots o_{N_c-3}}, \\
 \mathcal{F}_{ijk}^{\text{S}\delta}\mathcal{F}_{lmn}^{\text{S}\delta}  &= \mathcal{S}(N_c) \left(  \delta_{il}\delta_{jm}\delta_{kn} + \delta_{il}\delta_{jn}\delta_{km} \right. \\\nonumber
 &\hspace{30pt} \left. + \delta_{im}\delta_{jl}\delta_{kn} + \delta_{im}\delta_{jn}\delta_{kl} \right. \\\nonumber
 &\hspace{30pt} \left. + \delta_{in}\delta_{jm}\delta_{kl} + \delta_{in}\delta_{jl}\delta_{km} \right),
\end{align}
where
\begin{equation}
\begin{split}
    \mathcal{S}(N_c) &= \frac{1}{N_c^3 + 3N_c^2 + 2N_c} {2N_c - 1 \choose N_c} \\
    &= \frac{(2N_c-1)!}{(N_c!)^2 (N_c^2 + 3N_c + 2)}.
    \end{split}
\end{equation}
The structure of the potential in all cases is given by Eq.~\eqref{eq:V3psidef}.
Color factors can be obtained for general $N_c \geq 3$ using Eq.~\eqref{eq:colorfac} with the results
\begin{align}
\mathcal{C}_{\text{A}}^{3\psi,q} &= -\frac{N_c + 1}{8}, \\
\mathcal{C}_{\text{S}}^{3\psi,q} &=  -\frac{N_c - 1}{8},   \label{eq:colorfac3A}
\end{align}
which agree with the general $N_c$ results of Ref.~\cite{Brambilla:2009cd}, and
\begin{align}
    \begin{pmatrix} \mathcal{C}_{\text{MAA}}^{3\psi,1} & \mathcal{C}_{\text{MAS}}^{3\psi,1} \\ \mathcal{C}_{\text{MSA}}^{3\psi,1} &  \mathcal{C}_{\text{MSS}}^{3\psi,1} \end{pmatrix} &= \begin{pmatrix} \frac{1}{8(N_c-1)} & -\frac{\sqrt{N_c}}{4 \sqrt{2} \sqrt{N_c-1}} \\ -\frac{\sqrt{N_c}}{4 \sqrt{2} \sqrt{N_c-1}} & \frac{N_c+2}{16} \end{pmatrix} , \\
    \begin{pmatrix} \mathcal{C}_{\text{MAA}}^{3\psi,2} & \mathcal{C}_{\text{MAS}}^{3\psi,2} \\ \mathcal{C}_{\text{MSA}}^{3\psi,2} &  \mathcal{C}_{\text{MSS}}^{3\psi,2} \end{pmatrix} &= \begin{pmatrix} \frac{1}{8(N_c-1)} & \frac{\sqrt{N_c}}{4 \sqrt{2} \sqrt{N_c-1}} \\ \frac{\sqrt{N_c}}{4 \sqrt{2} \sqrt{N_c-1}} & \frac{N_c+2}{16} \end{pmatrix}, \\
    \begin{pmatrix} \mathcal{C}_{\text{MAA}}^{3\psi,3} & \mathcal{C}_{\text{MAS}}^{3\psi,3} \\ \mathcal{C}_{\text{MSA}}^{3\psi,3} &  \mathcal{C}_{\text{MSS}}^{3\psi,3} \end{pmatrix} &= \begin{pmatrix} \frac{2 N_c + 1}{8 (N_c - 1)} & 0 \\ 0 & \frac{N_c-4}{16} \end{pmatrix}.
\end{align}
These potentials are attainable with a pNRQCD Lagrangian term
\begin{equation}
    \begin{split}
        L^{\text{pot}}_{3\psi} =& -\int d^3\bs{r}_1d^3\bs{r}_2d^3\bs{r}_3\,  \psi^{\dagger}_i(t,\bs{r}_1)\psi^{\dagger}_j(t,\bs{r}_2)\psi^{\dagger}_k(t,\bs{r}_3)\\
        &\times \psi_l(t,\bs{r}_3)\psi_m(t,\bs{r}_2)\psi_n(t,\bs{r}_1) \\
    &\times \left[  \frac{N_c}{36}(N_c-1)(N_c-2) \mathcal{F}_{ijk}^{\text{A}}\mathcal{F}_{lmn}^{\text{A}}  V^{3\psi}_{\text{A}} \right. \\
     &\hspace{15pt} \left.  + \frac{1}{36 \mathcal{S}(N_c)}    \mathcal{F}_{ijk}^{\text{S}\delta}\mathcal{F}_{lmn}^{\text{S}\delta} V^{3\psi}_{\text{S}}   \right. \\
    &\hspace{15pt} \left. + \mathcal{F}_{ijk}^{\text{MA}a} \mathcal{F}_{lmn}^{\text{MA}a}W^{3\psi}_{\text{MA}}  + \mathcal{F}_{ijk}^{\text{MS}a} \mathcal{F}_{lmn}^{\text{MS}a}W^{3\psi}_{\text{MS}}  \right].
    \end{split} \label{eq:L3psiNc3}
\end{equation}
Direct computation of the matrix elements  involving $\mathcal{F}_{ijk}^{\text{MA}a}$ and $\mathcal{F}_{ijk}^{\text{MS}a}$ shows that the desired potentials  $V_{\text{MA}u}^{3\psi}$ and $V_{\text{MS}u}^{3\psi}$ are reproduced using
\begin{equation}
\begin{split}
W_{\text{MA}}^{3\psi} ={}&\alpha\left(\frac{\alpha}{4\pi}\right)^2 \left[ -\frac{N_c-1}{16 ((N_c-2) N_c+3)}v_3(\bs{r}_{12},\bs{r}_{13})
    \right. \\ {}&\left.
   -\frac{N_c-1}{16 ((N_c-2) N_c+3)} v_3(\bs{r}_{12},\bs{r}_{23}) \right. \\ {}&\left.
   +  \frac{(N_c-1) (2 (N_c-2) N_c+5)}{16 ((N_c-2)N_c+3)}v_3(\bs{r}_{13},\bs{r}_{23}) \right],
    \end{split} \label{eq:WA3psidef}
\end{equation}
and
\begin{equation}
\begin{split}
W_{\text{MS}}^{3\psi} ={}&\alpha\left(\frac{\alpha}{4\pi}\right)^2 \left[ \frac{N_c+4}{48} v_3(\bs{r}_{12},\bs{r}_{13})
    \right. \\ {}&\left.
    + \frac{N_c+4}{48} v_3(\bs{r}_{12},\bs{r}_{23})  +  \frac{N_c-8}{48} v_3(\bs{r}_{13},\bs{r}_{23}) \right].
    \end{split} \label{eq:WA3psidef}
\end{equation}
This completes the construction of the pNRQCD Lagrangian required to describe three-quark forces in generic hadron or multi-hadron states at NNLO.

Three-antiquark potentials are identical to three-quark potentials by symmetry.
However, it is noteworthy that additional $\psi\psi\psi^\dagger$ and $\psi\psi^\dagger\psi^\dagger$ potentials with distinct color factors are required to describe tetraquarks and other multi-hadron states containing both heavy quarks and heavy antiquarks at NNLO.
Even higher-body potentials involving combinations of four quark and antiquark fields are also relevant for such systems and are discussed next.

\subsection{Four- and more-quark potentials}\label{sec:4quark}

Four-quark and higher-body potentials that do not factorize into iterated insertions of two-quark and three-quark potentials must arise at some order in $\alpha_s$ during matching between pNRQCD and NRQCD and will need to be included in pNRQCD calculations of multi-hadron states at that order.
Perhaps surprisingly, the $\mathcal{O}(\alpha_s)$ suppression of three-quark potentials in comparison with quark-quark potentials does not extend to four-quark potentials: for generic multi-hadron systems, four-quark potentials arise at NNLO and therefore at the same order at three-quark potentials.
This can be seen by considering the diagrams in Fig.~\ref{fig:nbodypot}.
The transverse gluon propagator in the four-quark analog of the H-diagram shown leads to momentum dependence that does not factorize into products of fewer-body potentials, and for both diagrams shown, the color structures of all four quarks are correlated by the gluon interactions in a way that does not factorize.
Matching the contributions from these diagrams in pNRQCD therefore requires the introduction of four-quark potentials at NNLO.
Barring unexpected cancellations between diagrams, this four-quark potential -- and analogous four-body potentials involving one or more heavy antiquarks -- must be obtained and included in pNRQCD calculations of generic multi-hadron systems at NNLO.

Although a complete determination of the NNLO four-quark potential is beyond the scope of this work, it is straightforward to show that the potentials relevant for quarks in $SU(N_c)$ single-baryon systems greatly simplify and that for four-quark potentials vanish at NNLO for these special cases.
For $N_c \leq 3$, there are fewer than four quarks in a baryon, and it follows trivially that four-quark forces do not contribute to single-baryon observables.\footnote{Three- and four-body forces do not contribute to single-meson observables for any $N_c$ for the same reason.}
For $N_c \geq 4$, the absence of four-quark forces at NNLO for single-baryon systems is non-trivial, and we argue below that it follows from the color structure of single-baryon states.
These single-baryon states contain $N_c$ quarks in a color-singlet configuration and can therefore be constructed from linear combinations of states of the form
\begin{equation}
    \ket{B(\bs{r}_1,\ldots,\bs{r}_{N_c})} \equiv \frac{\epsilon_{i_1\ldots i_{N_c}}}{\sqrt{N_c!}}  \psi_{i_i}^\dagger(\bm{r}_1)  \cdots \psi_{i_{N_c}}^\dagger(\bm{r}_{N_c}) \ket{0}. \label{eq:Bdef}
\end{equation}
The antisymmetry of $\epsilon_{i_1\ldots i_{N_c}}$ implies that contributions from any potential operator to single-baryon observables will be totally antisymmetrized over the color indices of all $\psi_i$ and $\psi^\dagger_i$ fields arising in the operator. 
This means that only $V_{\text{A}}^{\psi\psi}$ and $V_{\text{A}}^{3\psi}$ contribute to the quark-quark and three-quark potentials for single-baryon states, respectively.
Further, the color structures of the four-quark potential diagrams shown in Fig.~\ref{fig:nbodypot} involve factors of
\begin{equation}
\begin{split}
T^a_{ik} T^b_{jl} f^{abc},
\end{split}
\end{equation}
where $i$ and $j$ ($k$ and $l$) label the color indices of any two of the incoming (outgoing) quark lines.
Contracting with the color tensors for single-baryon initial and final states leads to
\begin{equation}
\begin{split}
& T^a_{ik} T^b_{jl} f^{abc} \epsilon^{ik m_1\ldots m_{N_c-2}} \epsilon^{jl m_1\ldots m_{N_c-2}}   \\
& =-T^b_{ik} T^a_{jl} f^{abc} \epsilon^{ik m_1\ldots m_{N_c-2}} \epsilon^{jl m_1\ldots m_{N_c-2}}   \\
& = -T^b_{jl} T^a_{ik} f^{abc} \epsilon^{ik m_1\ldots m_{N_c-2}} \epsilon^{jl m_1\ldots m_{N_c-2}} \\
& = 0,
\end{split}
\end{equation}
where the antisymmetry of $f^{abc}$ has been used in going from the first to the second line, and the antisymmetry of $\epsilon_{i_1\ldots i_{N_c}}$ has been used in subsequently going to the third line.

For $N_c \geq 4$ single-baryon systems, diagrams with additional gluon propagators\footnote{An example of such a diagram can be obtained from Fig.~\ref{fig:3bodypot} by adding a fourth quark interacting with a potential gluon that is connected to the transverse gluon by a three-gluon interaction.}
lead to four-body forces at N$^3$LO that are not expected to vanish.
For multi-hadron systems, including tetraquarks and bound or scattering states of heavy baryons, total color antisymmetry of initial and final state quarks does not apply, and we emphasize that these four-quark potentials that have not yet been determined are required for complete NNLO calculations.

Five-quark (and higher-body) interactions require an additional gluon propagator compared to four-quark interactions and do not arise until N$^{3}$LO.

\subsection{pNRQCD Hamiltonian}\label{sec:hammy}

The Lagrangian formulation of pNRQCD described above can be readily converted to a nonrelativistic Hamiltonian form.
The generic kinetic and potential operators needed to construct the pNRQCD Hamiltonian are explicitly defined below. The action of the potential operator greatly simplifies when acting on quarkonium states and baryon states, and the particular structures of these states are also discussed in this section. For concreteness, unit-normalized quarkonium states are defined by
\begin{equation}
 \ket{Q\overline{Q}(\bm{r}_1,\bm{r}_2)} = \frac{1}{\sqrt{N_c}} \ket{\psi_m(\bm{r}_1),\chi^\dagger_{n}(\bm{r}_{2})} \delta_{mn}, \label{eq:QQbarState}
\end{equation}
while baryon states are defined by Eq.~\eqref{eq:Bdef}.
{ At LO in pNRQCD the ground-state of quarkonium must take the form of Eq.~\eqref{eq:QQbarState} because there are no other ways to construct a color-singlet from a product of $Q$ and $\overline{Q}$ fields, but at higher orders additional terms including ultrasoft fields are present. In general, Eq.~\eqref{eq:QQbarState} should be viewed as a ``trial wavefunction'' with the correct quantum numbers for describing quarkonium and both the spatial wavefunction and additional contribution to the ground-state can be obtained by solving for the lowest-energy state of the pNRQCD Hamiltonian with these quantum numbers.}
\begin{widetext}
The nonrelativistic potential operator $\hat{V}^{\psi\chi}$ appearing in the pNRQCD Hamiltonian is just $-L_{\psi\chi}^{\rm pot}$ with fermion fields replaced by Hilbert space operators. 
Its action on a quark-antiquark state is given by
\begin{equation}
\begin{split}
    &\hat{V}^{\psi\chi} \ket{\psi_m(\bm{r}_1),\chi^\dagger_{n}(\bm{r}_{2})}  \\
    &= \int d^3\bm{s}_1 d^3\bm{s}_2  \left[ \frac{1}{N_c}\delta_{ij}\delta_{kl} V^{\psi\chi}_{\mathbf{1}}(\bs{s}_{12})  + \frac{1}{T_F} T^a_{ij} T^a_{kl} V^{\psi\chi}_{\text{Ad}}(\bs{s}_{12}) \right] \\
    &\hspace{10pt}\times \psi^\dagger_i(t,\bs{s}_1)\chi_j(t,\bs{s}_2) \chi_k^\dagger(t,\bs{s}_2)\psi_l(t,\bs{s}_1)\ket{\psi_m(\bm{r}_1),\chi^\dagger_{n}(\bm{r}_{2})} \\
    &=   \left[ \frac{1}{N_c}\delta_{ij}\delta_{kl} V^{\psi\chi}_{\mathbf{1}}(\bs{r}_{12})  + \frac{1}{T_F}  T^a_{ij} T^a_{kl} V^{\psi\chi}_{\text{Ad}}(\bs{r}_{12}) \right]  \ket{\psi_i(\bm{r}_1),\chi^\dagger_j(\bm{r}_{2})} \delta_{km} \delta_{ln} 
    \end{split}
\end{equation}
The action of the potential operator on quarkonium states therefore simplifies to
\begin{equation}
\begin{split}
    &\hat{V}^{\psi\chi} \ket{Q\overline{Q}(\bm{r}_1,\bm{r}_2)} = \hat{V}^{\psi\chi} \ket{\psi_m(\bm{r}_1),\chi^\dagger_{n}(\bm{r}_{2})} \delta_{mn} \\
    &=  \left[ \frac{1}{N_c}\delta_{ij}\delta_{kl} V^{\psi\chi}_{\mathbf{1}}(\bs{r}_{12})  + \frac{1}{T_F}  T^a_{ij} T^a_{kl} V^{\psi\chi}_{\text{Ad}}(\bs{r}_{12}) \right]  \ket{\psi_i(\bm{r}_1),\chi^\dagger_j(\bm{r}_{2})} \delta_{km} \delta_{ln} \delta_{mn} \\
     &= .V^{\psi\chi}_{\mathbf{1}}(\bs{r}_{12}) \frac{1}{N_c}\delta_{ij}\delta_{kl} \delta_{kl} \ket{\psi_i(\bm{r}_1),\chi^\dagger_j(\bm{r}_{2})}  \\
     &= .V^{\psi\chi}_{\mathbf{1}}(\bs{r}_{12}) \delta_{ij}\ket{\psi_i(\bm{r}_1),\chi^\dagger_j(\bm{r}_{2})} \\
     &= V^{\psi\chi}_{\mathbf{1}}(\bs{r}_{12})\ket{Q\overline{Q}(\bm{r}_1,\bm{r}_2)},
    \end{split}\label{eq:V2psichiOp}
\end{equation}
where $T^a_{kl}\delta_{kl} = 0$ has been used to eliminate the color-adjoint term.
\end{widetext}
The action on a color-adjoint $Q\overline{Q}$ state $\ket{\psi_i(\bs{r}_1),\chi^\dagger_j(\bs{r}_2)} T^a_{ji} / \sqrt{T_F}$ analogously eliminates the color-singlet piece and is equivalent to multiplying the state by $V^{\psi\chi}_{\text{Ad}}(\bs{r}_{12})$ because $T^a_{kl} T^b_{lk} = T_F \delta^{ab}$.
This establishes that the terms in $-L_{\psi\chi}^{\rm pot}$ are correctly normalized to reproduce pNRQCD matching calculations for color-singlet and color-adjoint quark-antiquark states~\cite{Collet:2011kq,Kniehl:2004rk,Anzai:2013tja}.

The action of the quark-antiquark potential on more complicated multi-hadron states is given by applying the same operator $\hat{V}^{\psi\chi}$ to these states. For instance, the potential for a heavy tetraquark state is given by a color contraction of the action of the potential on a generic state with two heavy quarks and two heavy antiquarks,
\begin{widetext}
\begin{equation}
\begin{split}
    &\hat{V}^{\psi\chi} \ket{\psi_{n_1}(\bm{r}_1),\chi^\dagger_{n_2}(\bm{r}_{2}) \psi_{n_3}(\bm{r}_3),\chi^\dagger_{n_4}(\bm{r}_{4}) }  \\
    &= \int d^3\bm{s}_1 d^3\bm{s}_2  \left[ \frac{1}{N_c}\delta_{ij}\delta_{kl} V^{\psi\chi}_{\mathbf{1}}(\bs{s}_{12})  + \frac{1}{T_F} T^a_{ij} T^a_{kl} V^{\psi\chi}_{\text{Ad}}(\bs{s}_{12}) \right] \\
    &\hspace{10pt}\times \psi^\dagger_i(t,\bs{s}_1)\chi_j(t,\bs{s}_2) \chi_k^\dagger(t,\bs{s}_2)\psi_l(t,\bs{s}_1)\ket{\psi_{n_1}(\bm{r}_1),\chi^\dagger_{n_2}(\bm{r}_{2}) \psi_{n_3}(\bm{r}_3),\chi^\dagger_{n_4}(\bm{r}_{4}) } \\
    &=   \left[ \frac{1}{N_c}\delta_{ij}\delta_{n_1 n_2} V^{\psi\chi}_{\mathbf{1}}(\bs{r}_{12})  + \frac{1}{T_F}  T^a_{ij} T^a_{n_1 n_2} V^{\psi\chi}_{\text{Ad}}(\bs{r}_{12}) \right] \ket{\psi_{i}(\bm{r}_1),\chi^\dagger_{j}(\bm{r}_{2}) \psi_{n_3}(\bm{r}_3),\chi^\dagger_{n_4}(\bm{r}_{4}) } \\
    &\hspace{10pt} + \left[ \frac{1}{N_c}\delta_{ij}\delta_{n_1 n_4} V^{\psi\chi}_{\mathbf{1}}(\bs{r}_{14})  + \frac{1}{T_F}  T^a_{ij} T^a_{n_1 n_4} V^{\psi\chi}_{\text{Ad}}(\bs{r}_{14}) \right] \ket{\psi_{i}(\bm{r}_1),\chi^\dagger_{n_2}(\bm{r}_{2}) \psi_{n_3}(\bm{r}_3),\chi^\dagger_{j}(\bm{r}_{4}) } \\
    &\hspace{10pt} +  \left[ \frac{1}{N_c}\delta_{ij}\delta_{n_2 n_3} V^{\psi\chi}_{\mathbf{1}}(\bs{r}_{23})  + \frac{1}{T_F}  T^a_{ij} T^a_{n_2 n_3} V^{\psi\chi}_{\text{Ad}}(\bs{r}_{23}) \right] \ket{\psi_{n_1}(\bm{r}_1),\chi^\dagger_{j}(\bm{r}_{2}) \psi_{i}(\bm{r}_3),\chi^\dagger_{n_4}(\bm{r}_{4}) } \\
    &\hspace{10pt} + \left[ \frac{1}{N_c}\delta_{ij}\delta_{n_3 n_4} V^{\psi\chi}_{\mathbf{1}}(\bs{r}_{34})  + \frac{1}{T_F}  T^a_{ij} T^a_{n_3 n_4} V^{\psi\chi}_{\text{Ad}}(\bs{r}_{34}) \right] \ket{\psi_{n_1}(\bm{r}_1),\chi^\dagger_{n_2}(\bm{r}_{2}) \psi_{i}(\bm{r}_3),\chi^\dagger_{j}(\bm{r}_{4}) } .
\end{split} \label{eq:tet}
\end{equation}

The action of a generic $SU(N_c)$ quark-quark potential operator $\hat{V}^{\psi\psi}$ on an $N_Q$ quark state is analogously given by $-L_{\psi\psi}^{\rm pot}$ in Eq.~\eqref{eq:Lpsipsi} with fermion fields replaced by Hilbert-space operators and has the color decomposition
\begin{equation}
\begin{split}
    &\hat{V}^{\psi\psi} \ket{\psi_1(\bm{r}_1),\ldots,\psi_{N_q}(\bm{r}_{N_q})} \\
    &= \sum_{\bs{\rho}\in \{\text{A},\text{S}\}} \int d^3\bm{s}_1 d^3\bm{s}_2 V^{\psi\psi}_{\bs{\rho}}(\bm{s}_{12}) \left( \mathcal{F}^{\bs{\rho}}_{ij} \right)^* \mathcal{F}^{\bs{\rho}}_{kl} \psi^{\dagger}_i(t,\bs{s}_1)\psi^{\dagger}_j(t,\bs{s}_2) \psi_k(t,\bs{s}_2)\psi_l(t,\bs{s}_1) \ket{\psi_{n_1}(\bm{r}_1),\ldots,\psi_{n_{N_q}}(\bm{r}_{N_q})} \\
    &= \sum_{I \neq J} \sum_{\bs{\rho}\in \{\text{A},\text{S}\}} V^{\psi\psi}_{\bs{\rho}}(\bs{r}_{IJ})  \left( \mathcal{F}^{\bs{\rho}}_{m_I m_J} \right)^* \mathcal{F}^{\bs{\rho}}_{n_I n_J} \ket{\psi_{n_1}(\bm{r}_1),\ldots,\psi_{m_I}(\bm{r}_I),\ldots,\psi_{m_J}(\bm{r}_J),\ldots, \psi_{n_{N_q}}(\bm{r}_{N_q})},
    \end{split}\label{eq:V2psiOp}
\end{equation}
The action of a three-quark potential operator is analogous,
\begin{equation}
\begin{split}
     &\hat{V}^{3\psi} \ket{\psi_1(\bm{r}_1),\ldots,\psi_{N_q}(\bm{r}_{N_q})} \\
    &= \sum_{\bs{\rho} \in \{\text{A},\text{S},\text{MA},\text{MS}\}} \int d^3\bm{s}_1 d^3\bm{s}_2 d^3\bm{s}_3  V^{3\psi}_{\bs{\rho}}(\bm{s}_{12},\bm{s}_{13},\bm{s}_{23}) \left( \mathcal{F}^{\bs{\rho}}_{ijk} \right)^* \mathcal{F}^{\bs{\rho}}_{lmn} \psi^{\dagger}_i(t,\bs{s}_1)\psi^{\dagger}_j(t,\bs{s}_2)\psi^{\dagger}_k(t,\bs{s}_3) \\
    &\hspace{10pt} \times \psi_l(t,\bs{s}_3)\psi_m(t,\bs{s}_2)\psi_n(t,\bs{s}_1) \ket{\psi_{n_1}(\bm{r}_1),\ldots,\psi_{n_{N_q}}(\bm{r}_{N_q})} \\
    &=  \sum_{I \neq J \neq K} \sum_{\bs{\rho} \in \{\text{A},\text{S},\text{MA},\text{MS}\}} V^{3\psi}_{\bs{\rho}}(\bs{r}_{IJ},\bs{r}_{IK},\bs{r}_{JK})  \left( \mathcal{F^{\bs{\rho}}}_{m_I m_J m_K} \right)^* \mathcal{F}^{\bs{\rho}}_{n_I n_J n_K} \\
    &\hspace{10pt} \times \ket{\psi_{n_1}(\bm{r}_1),\ldots,\psi_{m_I}(\bm{r}_I),\ldots,\psi_{m_J}(\bm{r}_J),\ldots, \psi_{m_K}(\bm{r}_K),\ldots,  \psi_{n_{N_q}}(\bm{r}_{N_q})}.
    \end{split}\label{eq:V3psiOp}
\end{equation}
\end{widetext}
The four-quark potential operator $\hat{V}^{4\psi}$ can be defined analogously, although its explicit form at NNLO has not yet been computed.
These can be combined to define a total potential operator
\begin{equation}
\begin{split}
\hat{V} &= \hat{V}^{\psi\chi} + \hat{V}^{\psi\psi} + \hat{V}^{3\psi} + \hat{V}^{4\psi} \\
&\hspace{10pt} + V^{\psi\psi\chi} + V^{\psi\psi\psi\chi} + V^{\psi\psi\chi\chi} + \psi\leftrightarrow \chi  + \ldots,
\end{split}
\end{equation}
where 5-quark and higher-body potentials that do not contribute at NNLO are omitted, and $\psi\leftrightarrow \chi$ refers to antiquark-antiquark, 3-antiquark, and 4-antiquark potentials obtained by taking $\psi\leftrightarrow \chi$ in the quark-quark, 3-quark, and 4-quark potential operators.
Note that besides the 3-quark and 4-quark operators described above there are analogs of 3-quark and 4-quark potentials where only some of the quarks are replaced with antiquarks that enter the total potential at NNLO and arise for example in heavy tetraquark systems.
In conjunction with the usual nonrelativistic kinetic energy operator
\begin{equation}
\begin{split}
&\hat{T} \ket{\psi_{n_1}(\bm{r}_1),\ldots,\psi_{n_{N_q}}(\bm{r}_{N_q})} \\
&= \sum_{I} \frac{\bs{p}^2_I }{2m_Q}  \ket{\psi_{n_1}(\bm{r}_1),\ldots,\psi_{N_q}(\bm{r}_{n_{N_q}})} \\
&= -\sum_{I} \frac{\bs{\nabla}^2_I }{2m_Q}  \ket{\psi_{n_1}(\bm{r}_1),\ldots,\psi_{n_{N_q}}(\bm{r}_{N_q})},
\end{split}
\end{equation}
this potential operator can be used to construct the pNRQCD Hamiltonian operator
\begin{equation}
\hat{H} = \hat{T} + \hat{V},
\end{equation}
which is the basic ingredient used in the many-body calculations discussed below.

The eigenvalues of the nonrelativistic Hamiltonian $\hat{H}$ are equal to the total energies of the corresponding eigenstates minus the rest masses of any heavy quarks and antiquarks appearing in the state,
since the rest mass is removed from the Hamiltonian by the transformation in Eq.~\eqref{eq:NRpsi}.
The ground state of the sector of pNRQCD Hilbert space containing $N_Q$ heavy quarks, denoted $\ket{Q_1 \ldots Q_{N_Q},0}$, with mass or total energy $M_{Q_1\ldots Q_N}$ therefore has Hamiltonian matrix elements
\begin{equation}
  \begin{split}
    \Delta E_{Q_1\ldots Q_N} &\equiv \mbraket{Q_1 \ldots Q_{N_Q} ,0}{\hat{H}}{Q_1 \ldots Q_{N_Q} ,0} \\
    &= M_{Q_1\ldots Q_N} - N_Q m_Q.
  \end{split}
\end{equation}
The pNRQCD Hamiltonian and therefore $\Delta E_{Q_1\ldots Q_N}$ will depend on the definition of $m_Q$ above and, in particular, whether it is a bare or renormalized mass.
Although the unphysical nature of the pole mass $m_Q$ appearing in Eq.~\eqref{eq:NRpsi} and the pNRQCD Hamiltonian, therefore, leads to ambiguities in the definition of the nonrelativistic energy $\Delta E_{Q_1\ldots Q_N}$, the total energy $M_{Q_1\ldots Q_N}$ is independent of the prescription used to define $m_Q$ up to perturbative truncation effects.
Analogous considerations apply to pNRQCD states containing heavy quarks and antiquarks (assuming their separate number conservation), for example,
\begin{equation}
  \Delta E_{Q\overline{Q}} \equiv \mbraket{Q\overline{Q},0}{\hat{H}}{Q\overline{Q},0} = M_{Q\overline{Q}} - 2 m_Q.
\end{equation}
Once the value of $m_Q$ in a given scheme is determined, for example by matching $M_{Q\overline{Q}}$ or another hadron mass to experimental data or lattice QCD calculations, it can be used to predict other physical hadron masses from pNRQCD calculations of $\hat{H}$ eigenvalues and for example predict $M_{Q_1\ldots Q_N}$ from $\Delta E_{Q_1\ldots Q_N}$.

For baryon states, the quark-quark potential involves the color-tensor contraction
$\mathcal{F}^{\bs{\rho}}_{m_I m_J} \mathcal{F}^{\bs{\rho}}_{n_I n_J} \epsilon_{n_1\ldots n_{N_c}}$, which vanishes for the symmetric potential  involving $\mathcal{F}^{\text{S}}_{n_I n_J}$ and is equal to $\epsilon_{n_1\ldots m_I \ldots m_J \ldots n_{N_c}} /2$ for the antisymmetric potential using the color tensors defined in Eq.~\eqref{eq:colorfacdef}. Inserting this in into Eq.~\eqref{eq:V2psiOp} applied to the baryon state defined in Eq.~\eqref{eq:Bdef} gives
\begin{equation}
\begin{split}
  \hat{V}^{\psi\psi} \ket{B} &= \frac{1}{2}\sum_{I \neq J} V^{\psi\psi}_{\text{A}}(\bs{r}_{IJ}) \ket{B} \\
  &=\sum_{I<J} V^{\psi\psi}_{\text{A}}(\bs{r}_{IJ}) \ket{B}, \label{eq:Bpot}
  \end{split}
\end{equation}
where the coordinate dependence of $\ket{B(\bs{r}_1,\ldots,\bs{r}_{N_c})}$ has been suppressed for brevity and the $I \leftrightarrow J$ symmetry of the potential has been used in going from the first to the second line.
The analogous contraction for the three-quark potential $\mathcal{F}^{\bs{\rho}}_{m_I m_J m_K} \mathcal{F}^{\bs{\rho}}_{n_I n_J n_K} \epsilon_{n_1\ldots n_{N_c}}$ vanishes for all potentials except the totally antisymmetric case with $\bs{\rho}=\text{A}$. In this case the color-tensor contraction is equal to $ \epsilon_{n_1\ldots m_I \ldots m_J \ldots m_K \ldots n_{N_c}} / 3!$, which gives
\begin{equation}
\begin{split}
\hat{V}^{3\psi} \ket{B} &= \frac{1}{3!}\sum_{I \neq J \neq K} V^{3\psi}_{\text{A}}(\bs{r}_{IJ},\bs{r}_{IK},\bs{r}_{JK}) \ket{B} \\
&= \sum_{I<J<K} V^{3\psi}_{\text{A}}(\bs{r}_{IJ},\bs{r}_{IK},\bs{r}_{JK}) \ket{B}.
\end{split}
\end{equation}
Since the 4-quark interaction color tensors are orthogonal to $\epsilon_{ij\ldots}$ as discussed in Sec.~\ref{sec:4quark}
\begin{equation}
\hat{V}^{4\psi} \ket{B} = 0,
\end{equation}
at NNLO with non-zero contributions possible at N${}^3$LO.
These results establish that the color-antisymmetric two- and three-quark potential operators are correctly normalized to reproduce the pNRQCD matching calculations performed using baryon-level Lagrangian operators in Refs~\cite{Brambilla:2005yk,Brambilla:2009cd}. 
It can be shown similarly that the mixed-symmetry adjoint potential operators defined above are correctly normalized so that their action on an adjoint baryon state is equivalent to matrix multiplication by $V^{3\psi}_{\bs{\rho} uv}$.

We end this section with an interesting cross-check discussed for $N_c = 3$ in Ref.~\cite{Brambilla:2009cd}: the antisymmetric two-quark potential can be obtained to NNLO (including two-loop diagrams) using the NNLO three-body potential (which only includes one-loop diagrams) and setting $N_c-1$ quarks to be at the same position. These $N_c-1$ quarks then behave as an antiquark in color space, and thus a color-singlet quarkonium state arises.
Baryon states with $N_c-1$ co-located quarks can be defined by
\begin{equation}
\ket{M(\bs{r}_1,\bs{r}_2)} \equiv \ket{B(\bs{r}_1,\bs{r}_2=\ldots=\bs{r}_{N_c})}.
\end{equation}
The correspondence between an $N_c-1$ quark color source and an antiquark color source suggests that matrix elements can be equated between quarkonium states $\ket{Q\overline{Q}}$ and heavy baryon states with $N_c - 1$ quark positions identified,
\begin{equation}
\mbraket{Q\overline{Q}}{\hat{V}}{Q\overline{Q}}  = \mbraket{M}{ \hat{V} }{ M },
\end{equation}
at least to leading order in $1/m_Q$ where heavy quarks are equivalent to static color sources.
This provides a relation between the quark-antiquark and multi-quark potentials in each representation that make non-zero contributions in quarkonium and baryon states,
\begin{equation}
\begin{split}
    &\mbraket{Q\overline{Q}(\bs{r}_1,\bs{r}_2)}{\hat{V}^{\psi\chi}_{\bs{1}}}{Q\overline{Q}(\bs{r}_1,\bs{r}_2)} \\
    &= \mbraket{M(\bs{r}_1,\bs{r}_2)}{ \hat{V}_{A}^{\psi\psi}+\hat{V}_{A}^{3\psi} }{ M(\bs{r}_1,\bs{r}_2) } \\
    &= \sum_{I}V^{\psi\psi}_{\text{A}}(\bs{r}_{1 I}) +\sum_{I<J} V_{A}^{3\psi}(\bs{r}_{1 I},\bs{r}_{1 J},\bs{0}),
    \end{split}
\end{equation}
where potentials with all quark fields located at the same point have been removed since these correspond to local counterterms. There is only one four-quark separation $\bs{r} = \bs{r}_{12} = \bs{r}_{13} = \ldots$, and so the sums can be evaluated as
\begin{equation}
\begin{split}
  V^{\psi\chi}_{\bs{1}}(\bs{r}) &=(N_c - 1)V^{\psi\psi}_{\text{A}}(\bs{r})  \\
&\hspace{10pt} + 
\frac{1}{2}(N_c-1)(N_c-2) V_{A}^{3\psi}(\bs{r},\bs{r},\bs{0}),
\end{split}
\end{equation}
where the counting factor arises from the ${N_c-1 \choose 2} = (N_c-1)(N_c-2)/2$ three-body interactions between the $N_c-1$ identically located quarks and the quark at a specific position. 
Solving for the quark-quark antisymmetric potential, inserting the form of the three-quark potential in Eq.~\eqref{eq:V3psidef} with singular factors of $v_3(\bs{r},\bs{0})$ removed (by local counterterms), and
noting that the three-quark color factor given in Eq.~\eqref{eq:colorfac3A} and the quark-antiquark color factor $-C_F$ are related by
\begin{equation}
C_{\text{A}}^{3\psi,q} = -C_F \left[ \frac{N_c}{4(N_c-1)} \right],
\end{equation}
the quark-quark potential can be obtained in terms of the quark-antiquark potential and the three-quark potential function as
\begin{equation}
\begin{split}
   V^{\psi\psi}_{\text{A}}(\bs{r}) &= \frac{1}{N_c-1}\ \left[ V_{\bm{1}}^{\psi\chi}(\bs{r}) \right. \\
   & \hspace{10pt} \left. + C_F \alpha_s  \left( \frac{\alpha_s}{4\pi} \right)^2 \frac{N_c(N_c-2)}{8} v_3(\bs{r},\bs{r}) \right].
   \end{split}
\end{equation}
The three-quark potential function with equal arguments simplifies to
\begin{equation}
v_3(\bs{r},\bs{r}) = -\frac{4\pi^2(\pi^2-12)}{|\bs{r}|},
\end{equation}
which relates the quark-quark and quark-antiquark potentials at NNLO as
\begin{equation}
\begin{split}
   V^{\psi\psi}_{\text{A}}(\bs{r}) &= \frac{1}{N_c-1}\ \left[  V_{\bm{1}}^{\psi\chi}(\bs{r}) \right. \\
   & \hspace{10pt} \left. - \frac{\alpha_s C_F}{|\bs{r}|} \left( \frac{\alpha_s}{4\pi} \right)^2 \frac{N_c(N_c-2)}{2}\pi^2(\pi^2-12) \right].
   \end{split}
\end{equation}
This is consistent with the antisymmetric quark-quark potential attained previously in Eq.~\ref{eqn:delta2} and matches the result obtained for the case of $N_c=3$ in Ref.~\cite{Brambilla:2009cd}.
Note that for $N_c \geq 4$, this agreement is further consistent with the result above that four-quark potentials do not contribute to $N_c$-color baryon states at NNLO.

\section{Many-body methods}
\label{sec:manyb}

A wide range of techniques have been developed for solving nonrelativistic quantum many-body problems in nuclear and condensed matter physics. Quantum Monte Carlo methods provide stochastic estimates of energy spectra and other observables of quantum many-body states with systematic uncertainties that can be quantified and reduced with increased computational resources~\cite{Carlson:2014vla,Yan_2017,Gandolfi:2020pbj}.
In particular, the variational Monte Carlo approach allows upper bounds to be placed on many-body ground state energies that can be numerically optimized using a parameterized family of trial wavefunctions.
The Green's function Monte Carlo approach augments VMC by including imaginary time evolution that exponentially suppresses excited-state contributions and allows exact ground-state energy results to be obtained from generic trial wavefunctions (more precisely any trial wavefunction not orthogonal to the ground state) in the limit of large imaginary-time evolution.
The statistical precision of GFMC calculations is greatly improved by a good choice of the trial wavefunction that has a large overlap with the ground state, and often the optimized wavefunctions resulting from VMC calculations are used as the initial trial wavefunctions in subsequent GFMC calculations~\cite{Carlson:2014vla,Gandolfi:2020pbj}.
Ground-state energy results obtained using GFMC are themselves variational upper bounds on the true ground-state energy, as discussed further below. This combination of methods leverages the desirable features of VMC while using GFMC to remove hard-to-quantify systematic uncertainties associated with the Hilbert space truncation induced by a wavefunction parameterization with a finite number of parameters.

Previous works have used few-body methods, for example based on Fadeev equations, and variational methods to calculate quarkonium and baryon masses using potential models~\cite{Brambilla:1999ja,Brambilla:2022fqa,Martynenko:2007je,Roberts:2007ni,Silvestre-Brac:1996myf}.
Two previous works have applied variational methods to calculate baryon masses using pNRQCD potentials: Ref.~\cite{Jia:2006gw} uses the LO potential and a one-parameter family of analytically integrable variational wavefunctions, and Ref.~\cite{Llanes-Estrada:2011gwu} uses potentials up through NNLO with a two-parameter family of variational wavefunctions. 
Here, we extend these results by performing GFMC calculations with trial wavefunctions obtained using VMC in order to obtain reliable predictions for quarkonium and triply-heavy baryon masses across a wide range of $m_Q$ for QCD as well as $SU(N_c)$ gauge theories of dark mesons and baryons with $N_c \in \{2,\ldots,6\}$.
The methods used here are very computationally efficient -- by generating Monte Carlo ensembles for VMC and GFMC by applying the Metropolis algorithm with optimized trial wavefunctions used for importance sampling, we achieve more than an order of magnitude more precise results than previous calculations with modest computational resources.
The techniques developed here can further be applied straightforwardly to systems with more than three heavy quarks.
The remainder of this section discusses the formalism required to apply VMC and GFMC methods to pNRQCD for these systems and beyond.

\subsection{Variational Monte-Carlo}

The quantum mechanical state $\ket{\Psi}$ of a system containing $N_Q$ heavy quarks/antiquarks can be described by a coordinate space wavefunction $\Psi(\bs{R}) \equiv \braket{\bm{R}}{\Psi}$ where $\bs{R}~\equiv~(\bs{r}_1,\ldots,\bs{r}_{N_Q})$ is a vector of coordinates. The normalization condition
\begin{equation}
  \begin{split}
    1 &= \braket{\Psi}{\Psi} =  \int d\bm{R} \, \braket{\Psi}{\bm{R}}\braket{\bm{R}}{\Psi} \\
    &= \int d\bm{R} \, |\Psi(\bs{R})|^2,
  \end{split}
\end{equation}
will be used throughout this work.
The LO pNRQCD Hamiltonian is simply the Coulomb Hamiltonian, which is known to be bounded from below, and this boundedness will be assumed for the pNRQCD Hamiltonian at higher orders below and verified \emph{a posteriori}.
This implies that there is a set of unit-normalized energy eigenstates $\ket{n}$ with $H \ket{n} = \Delta E_n \ket{n} $ (note that we continue using $\Delta E$ to denote nonrelativistic energies here and below) that can be ordered such that $\Delta  E_0 \leq \Delta  E_1 \leq  \ldots $,
from which the well-known
Rayleigh-Ritz variational bound follows,
\begin{equation}
  \begin{split}
    \mbraket{\Psi}{H}{\Psi} = \sum_n |\braket{\Psi}{n}|^2 \Delta  E_n \geq \Delta  E_0.
  \end{split}
\end{equation}
This variational principle is the starting point for VMC methods.

Any trial wavefunction $\Psi(\bs{R};\bs{\omega})$ depending on a set of parameters $\omega = (\omega_1,\ldots)$ satisfies the variational principle,
\begin{equation}
\begin{split}
    \Delta E_0 &\leq \mbraket{\Psi_T(\bs{\omega})}{H}{\Psi_T(\bs{\omega})} \\
    &= \int d^3\bs{R}\, \Psi_T(\bs{R};\bs{\omega})^*H(\bs {R}) \Psi_T(\bs{R};\bs{\omega}), \label{eq:var}
    \end{split}
\end{equation}
where $\mbraket{\bs{R}}{H}{\bs{R}'} = H(\bs{R}) \delta(\bs{R}-\bs{R}')$.
By iteratively varying $\bs{\omega}$ using a numerical optimization procedure, the upper bound on $\Delta E_0$ provided by a parameterized family of trial wavefunctions can be successively improved.
If the trial wavefunction is sufficiently expressive as to describe the true ground-state wavefunction for some set of parameters, then the true ground-state energy and wavefunction can be determined using such an optimization procedure.
This is generally not the case for complicated many-body Hamiltonians and numerically tractable trial wavefunctions, and in this generic case, variational methods provide an upper bound on $\Delta E_0$ rather than a rigorous determination of the ground-state energy.

The integral in Eq.~\eqref{eq:var} is $3 N_Q$ dimensional and is challenging to compute exactly for many-body systems.
Instead, VMC methods apply Monte Carlo integration techniques to stochastically approximate the integral in Eq.~\eqref{eq:var}.
The magnitude of the trial wavefunction can be used to define a probability distribution,
\begin{equation}
\mathcal{P}(\bs{R};\bs{\omega}) = |\Psi_T(\bs{R};\bs{\omega})|^2,
\end{equation}
from which coordinates $\bs{R}$ can be sampled.
The standard Metropolis algorithm can then be used to approximate the integral in Eq.~\eqref{eq:var}: coordinates $\bs{R}_0$ are sampled from $\mathcal{P}(\bs{R};\bs{\omega})$, updated coordinates $\bs{R}_1 = \bs{R}_0 + \varepsilon \bs{x}$ are chosen using, for example, zero-mean and unit-variance Gaussian random variables $\bs{x}$ and a step size $\varepsilon$ discussed further below. The updated coordinates are accepted with probability 
$w_1 = \mathcal{P}(\bs{R}_1;\bm{\omega}) / \mathcal{P}(\bs{R}_0;\bm{\omega})$ or with probability 1 if $w_1 > 1$, and they are rejected otherwise.
If the coordinates are accepted, then $\bs{R}_1$ is added to an ensemble of coordinate values, while if they are rejected, then $\bs{R}_0$ is added.
This procedure is repeated with coordinates $\bm{R}_{i+1}$ updated analogously from the latest coordinates $\bm{R}_i$ in the ensemble. The new coordinates are accepted with probability $w_{i+1} = \mathcal{P}(\bs{R}_{i+1};\bm{\omega}) / \mathcal{P}(\bs{R}_i;\bm{\omega})$ (or probability 1 if $w_{i+1} > 1$).
The resulting ensemble is approximately a set of random variables drawn from $\mathcal{P}(\bm{R};\bm{\omega})$ if the coordinates from an initial thermalization period of $N_{\rm therm}$ updates are omitted, and they are approximately statistically independent if $N_{\rm skip}$ update steps are skipped between successive members of the final coordinate ensemble, where $N_{\rm skip}$ is chosen to be longer than the autocorrelation times of observables of interest.\footnote{
  Below, we find $N_{\rm skip} \gtrsim 100$ to be sufficient to achieve negligible autocorrelations in $\mbraket{\Psi_T(\bs{\omega})}{H}{\Psi_T(\bs{\omega})}$ using $\epsilon$ on the order of the Bohr radius of the Coulombic trial wavefunctions discussed in Sec.~\ref{sec:trial}.} 

An ensemble of $N_{\rm var}$ such coordinates can then be used to approximate the integral in Eq.~\eqref{eq:var} as
\begin{equation}
\mbraket{\Psi_T(\bs{\omega})}{H}{\Psi_T(\bs{\omega})} 
    \approx \frac{1}{N_{\rm var}} \sum_{i=1}^{N_{\rm var}} H(\bm{R}_i). \label{eq:stochastic}
\end{equation}
In VMC methods, this approximation of $\mbraket{\Psi_T(\bs{\omega})}{H}{\Psi_T(\bs{\omega})}$ is used as a loss function to be minimized using numerical optimization techniques. 
For a complete review of VMC and its implementation, see Ref.~\cite{Carlson:2014vla}.

In the VMC calculation below, we use the Adam optimizer~\cite{kingma2017adam} to update our trial wavefunction parameters iteratively.
Default Adam hyperparameters are used with a step size initially chosen to be $10^{-2}$.
After the change in loss function fails to improve for 10 updates, the step size is reduced by a factor of 10.
After two such reductions of the step size, optimization is restarted using the best trial wavefunction parameters from the previous optimization round and step sizes of $10^{-3}$ and subsequently $10^{-4}$ in order to refresh the Adam momenta and improve convergence to optimal parameters without overshooting.
Gradients of the loss function are stochastically estimated in analogy to Eq.~\eqref{eq:stochastic} using auto-differentiation techniques implemented in the Python package \texttt{PyTorch}~\cite{paszke2019pytorch}.

\subsection{Green's Function Monte Carlo} \label{sec:gfmc}

The optimal trial wavefunction $\Psi_T(\bs{R},\bs{\omega})$ obtained using VMC methods still may not provide an accurate determination of $\Delta E_0$ because of the limited expressiveness of a finite-parameter function suitable for numerical optimization.
To overcome this limitation, we use the standard QMC strategy of taking the optimal trial wavefunction obtained from VMC as the starting point for subsequent GFMC calculation~\cite{Carlson:2014vla,Gandolfi:2020pbj}.
GFMC calculations use evolution\footnote{This evolution is often described as diffusion because the free particle nonrelativistic imaginary-time Schr{\"o}dinger equation is the diffusion equation.} in imaginary time $\tau$ to exponentially suppress excited-state components of $\ket{\Psi_T}$, which is analogous to the imaginary-time evolution used in lattice QCD calculations.
In the limit of infinite imaginary-time evolution, the ground state with a given set of quantum numbers can be obtained from any trial wavefunction with the same quantum numbers,
\begin{equation}
    \ket{0}=\lim_{\tau \to \infty} e^{-H \tau}\ket{\Psi_T}.
    \label{eqn:imtproj}
\end{equation}
In our case, imaginary-time evolution can be used to determine the ground-state energy and wavefunction of a system with $N_Q$ heavy quarks/antiquarks using a pNRQCD Hamiltonian with conserved heavy quark/antiquark numbers.

In general, directly computing the propagator in~\eqref{eqn:imtproj}
is not feasible for arbitrary $\tau$. However, taking small imaginary time, $\delta\tau=\tau/N$ for $N\gg 1$ and recovering the full projection in large time can be achieved by a Lie-Trotter product~\cite{Trotter:1959},
\begin{align}
    \Psi(\tau,\bs{R}_N)={}&\int\prod_{i=0}^{N-1}d\bs{R}_i\bra{\bs{R}_N}e^{-H\delta\tau}\ket{\bs{R}_{N-1}}\nonumber\\{}&\times \cdots \times \bra{\bs{R}_1}e^{-H\delta\tau}\ket{\bs{R}_{0}}\braket{\bs{R}_0}{\Psi_T},
\end{align}
making the computation feasible. We can then define the GFMC wavefunction in integral form at an imaginary time $\tau+\delta{\tau}$ 
\begin{equation} 
    \Psi(\tau+d\tau,\bs{R})=\int d\bs{R}' \, G_{\delta\tau}(\bs{R},\bs{R}')\Psi(\tau,\bs{R}'),
    \label{eqn:gfmcwvfn}
\end{equation}
in terms of a Green's function,
\begin{equation}
    G_{\delta\tau}(\bs{R},\bs{R}' )=\bra{\bs{R}}e^{-H\delta\tau}\ket{\bs{R}'}.
\end{equation}
Practically, one approximates short-time propagation with the Trotter-Suzuki expansion,
\begin{align}
    G_{\delta\tau}(\bs{R},\bs{R}')={}&e^{-V(\bs{R})\delta\tau/2}\bra{\bs{R}}e^{-T\delta\tau}\ket{\bs{R}'}\times\nonumber\\
    {}&e^{-V(\bs{R}')\delta\tau/2}+\mathcal{O}(\delta\tau^2), \label{eq:GTrott}
\end{align}
where $V$ is the potential in configuration space and $T$ is the kinetic energy which defines the free-particle propagator, which for nonrelativistic systems is expressible as a Gaussian distribution in configuration space,
\begin{equation} \label{eq:Gauss}
    \bra{\bs{R}}e^{-T\delta\tau}\ket{\bs{R}'} = \left(\frac{1}{\lambda^3\pi^{3/2}}\right)^{N_Q} e^{-(\bs{R}-\bs{R}')^2/\lambda^2},
\end{equation}
where $\lambda^2 = 2 \delta \tau / m_Q$~\cite{Carlson:2014vla,Gandolfi:2020pbj}. 
The integral in Eq.~\eqref{eqn:gfmcwvfn} describing the action of a single Trotter step to the wavefunction is therefore computed by sampling $\bs{R}-\bs{R}'$ from Eq.~\eqref{eq:Gauss} and then explicitly multiplying by the potential factors appearing in Eq.~\eqref{eq:GTrott}.
In order to reduce the variance of GFMC results, a further resampling step is applied in which $\bs{R}-\bs{R}'$ and $-(\bs{R}-\bs{R}')$ are both proposed as possible updates, and a Metropolis sampling step is used to select one proposed update as described in more detail in Ref.~\cite{Gandolfi:2020pbj}.

If the action of the pNRQCD potential on a given state can be described by a spin- and color-independent potential depending only on $\bs{R}$, then it is straightforward to exponentiate the potential as indicated in Eq.~\eqref{eq:GTrott}.
Conveniently, precisely this situation arises for meson and baryon states at $\mathcal{O}(m_Q^0)$ as shown in Sec.~\ref{sec:hammy}.
In applications of pNRQCD to multi-hadron systems, this will not usually be the case because generic states are not eigenstates of a single color tensor operator but instead will include contributions from multiple color tensor operators in the potential.
Calculations including $\mathcal{O}(1/m_Q^2)$ effects will also have spin-dependent potentials even in the single-meson and single-baryon cases. 
In generic applications including color- and spin-dependent potentials it will be necessary to expand the exponential, for instance as a Taylor series $e^{-V \delta \tau/2} \approx 1 - V \delta \tau / 2 + V^2 (\delta \tau)^2 / 8 + \ldots$.
Since the potential appearing in these expressions is a $2 N_c N_Q \times 2 N_c N_Q$ matrix, the accuracy of this expansion will have to be balanced against the computational cost of its evaluation when deciding how many terms to include.
More details on including matrix-valued potentials in GFMC calculations can be found in Refs.~\cite{Carlson:2014vla,Gandolfi:2020pbj}.
Different treatment will be required for momentum-dependent potentials at $\mathcal{O}(1/m_Q^2)$.

Applying an operator $\mathcal{O}$ to the imaginary-time-evolved wavefunction $\Psi_T(\bm{R},\tau)$ leads to the
mixed expectation values
\begin{equation}
  \mbraket{\Psi_T}{\mathcal{O}}{\Psi_T(\tau)} = \mbraket{\Psi_T}{\mathcal{O} e^{-H \tau}}{\Psi_T}.
\end{equation}
Expectation values involving symmetric insertions of imaginary-time evolution operators can also be computed from the mixed expectation values $\mbraket{\Psi_T}{\mathcal{O}}{\Psi_T(\tau)}$ and $\mbraket{\Psi_T(\tau)}{\mathcal{O}}{\Psi_T}$~\cite{Pervin:2007sc,Carlson:2014vla}.
Since $H$ commutes with $e^{-H\tau}$, Hamiltonian matrix elements are automatically symmetric,
\begin{equation}
  \begin{split}
    \mbraket{\Psi_T}{H}{\Psi_T(\tau)} &= \mbraket{\Psi_T}{e^{-H\tau/2} H e^{-H\tau/2}}{\Psi_T} \\
    &= \mbraket{\Psi_T(\tau/2)}{H}{\Psi_T(\tau/2)}.
  \end{split}
\end{equation}
By Eq.~\eqref{eq:var}, this implies that GFMC binding-energy determinations provide variational upper bounds on the energy of the ground state $E_0$ with quantum numbers of $\Psi_T$.
It further implies that GFMC Hamiltonian matrix elements have the spectral representation
\begin{equation}
\mbraket{\Psi_T}{H}{\Psi_T(\tau)} = \sum_n \Delta E_n |Z_n|^2 e^{-\Delta E_n \tau},
\end{equation}
where $Z_n = \braket{n}{\Psi_T}$. In the large-$\tau$ limit, dependence on $Z_0$ can be removed by dividing by $\braket{\Psi_T}{\Psi_T(\tau)}$ since
\begin{equation}
  \braket{\Psi_T}{\Psi_T(\tau)} = \sum_n |Z_n|^2 e^{-\Delta E_n \tau}.
\end{equation}
Defining the GFMC approximation to the Hamiltonian matrix element as
\begin{equation}
  \left< H(\tau) \right> \equiv \frac{ \mbraket{\Psi_T}{H}{\Psi_T(\tau)} }{ \braket{\Psi_T}{\Psi_T(\tau)} },
\end{equation}
and the excitation gap
\begin{equation}
    \delta \equiv \Delta E_1 - \Delta E_0,
\end{equation}
this shows that in the large $\tau$ limit
\begin{equation}
  \begin{split}
    \left< H(\tau) \right> &=  \frac{ \sum_n 
    \Delta E_n |Z_n|^2 e^{-\Delta E_n \tau} }{ \sum_n |Z_n|^2 e^{-
    \Delta E_n \tau} } \\
    &= 
    \Delta E_0 + \left| \frac{Z_1}{Z_0} \right|^2 \delta \ e^{- \delta  \tau} + \ldots ,
  \end{split}
\end{equation}
where $\ldots$ denotes terms exponentially suppressed by $e^{-\delta \tau}$ for $n > 1$.
Corrections to $\left< H(\tau) \right> \approx \Delta E_0$ are therefore exponentially suppressed by $\delta \tau$, and GFMC calculations can achieve accurate ground-state energy estimates even if $Z_1 / Z_0$ is not small as long as $\tau \gg 1/\delta$.

The computational simplicity of pNRQCD, particularly for mesons and baryons at $\mathcal{O}(m_Q^0)$, makes it straightforward to achieve $\tau \gg 1/\delta$ in the numerical calculations below.
Constant fits to $\left< H(\tau) \right>$ using correlated $\chi^2$-minimization are therefore used below to fit ground-state energies from GFMC results.
To avoid contamination from $\mathcal{O}(e^{-\delta \tau})$ excited-state effects, the minimum imaginary time used for fitting $\tau_{\rm min}$ was varied, and in particular 30 different $\tau_{\rm min}$ were chosen from $[0, L_\tau-1]$.
The covariance matrices for these fits are ill-conditioned due to the large number of imaginary time steps used, and results are therefore averaged over windows of consecutive $\tau$ before performing fits.
Linear shrinkage~\cite{stein1956,Ledoit:2004} is used with the diagonal of the coviance matrix as the shrinkage target in order to further improve the numerical stability of covariance matrix estimation.
The results $\Delta E^f$ obtained from $\chi^2$-minimization for each choice of fit range $[\tau_{\rm min}^f,L_\tau - 1]$ enumerated by $f = 1,\ldots,30$ with corresponding $\chi^2$ minima $\chi^2_f$ are then averaged in order to penalize fits with poor goodness-of-fit (arising from non-negligible excited-state effects) using the Bayesian model averaging method of Ref.~\cite{Jay:2020jkz} with flat priors.
This corresponds to
\begin{equation}
\begin{split}
\Delta E &= \sum_f w_f \Delta E^f, \\ 
\end{split} \label{eq:averaging}
\end{equation}
where the normalized weights $w_f$ are defined by
\begin{equation}
\begin{split}
\tilde{w}_f &= \exp\left[ -\frac{1}{2}\left( \chi^2_f + 2\tau_{\rm min}^f \right) \right], \\
w_f &= \frac{\tilde{w}_f}{\sum_g \tilde{w}_g},
\end{split} \label{eq:weights}
\end{equation}
where a constant factor of two times the number of parameters that cancels from the weighted average defined in Eq.~\eqref{eq:averaging} below has been omitted.
The model averaged fit uncertainties $\delta \Delta E$ are then given in terms of the individual fit uncertainties $\delta \Delta E^f$ by~\cite{Jay:2020jkz}
\begin{equation}
\begin{split}
\delta \Delta E &= \sum_f w_f \delta \Delta E^f \\
&\hspace{10pt} + \sum_f w_f (\Delta E_f)^2 - \sum_f (w_f \Delta E_f)^2, \\ 
\end{split} 
\end{equation}
where the terms on the second line provide a measure of systematic uncertainty arising from the variance of the ensemble of fit results.
Finally, the size of the $\tau$ averaging window is varied in order to test the stability of covariance matrix determination, and stability of the final fit results after model averaging is tested for different choices of $\tau$ averaging window size starting with 2 and 4 and then continuing by doubling the window size until $1\sigma$ consistency between consecutive window-size choices is achieved.
In this manner, the model-averaged $\Delta E$ from the first $\tau$ window size consistent with the previous $\tau$ window size is taken as the final GFMC result quoted for all parameter choices below.

\section{Coulombic trial wavefunctions}
\label{sec:trial}

The QMC methods above require a parameterized family of trial wavefunctions $\psi_T(\bs{R},\bs{C})$ as the starting point for VMC.
At LO, the color-singlet quark-antiquark potential is identical to a rescaled Coulomb potential, and the ground-state wavefunction is known analytically.
Beyond LO, there are logarithmic corrections to the Coulombic shape of the potential.
To assess how accurately a given variational family of trial wavefunctions has described the ground state of these higher-order potentials, GFMC calculations are performed using these variationally optimized trial wavefunctions.
The amount of imaginary-time evolution required to converge toward the true ground-state energy, as well as the statistical precision of the GFMC calculations with a given trial wavefunction, provide quantitative measures of how close a given trial wavefunction is to the true ground state.
Several families of trial wavefunctions are considered for these systems below. A simple variational ansatz corresponding to Coulomb ground-state wavefunctions with appropriately tuned Bohr radii provides relatively stringent variational bounds on NLO and NNLO quarkonium energies while also leading to computationally efficient GFMC calculations.
Analogous variational and GFMC calculations for baryons show that products of Coulomb ground-state wavefunctions with appropriately tuned Bohr radii provide simple but remarkably effective trial wavefunctions for heavy baryons.

\subsection{Quarkonium}

The pNRQCD potential for quarkonium states is given at $\mathcal{O}(m_Q^{0})$ from Eq.~\eqref{eq:V2psichiOp} and Eq.~\eqref{eqn:singpotsOLO} by
\begin{equation}
  \begin{split}
    \hat{V}\ket{Q\overline{Q}(\bs{r}_1,\bs{r}_2)} &= V^{\psi\chi,(0)}_{\mathbf{1}}(\bs{r}_{12}) \ket{Q\overline{Q}(\bs{r}_1,\bs{r}_2)} \\
    &= - \frac{C_F \alpha_V(|\bs{r}_{12}|,\mu)}{|\bs{r}_{12}|}  \ket{Q\overline{Q}(\bs{r}_1,\bs{r}_2)}. \label{eq:QQbarpot}
  \end{split}
\end{equation}
At LO, $\alpha_V(|\bs{r}_{12}|,\mu) = \alpha_s(\mu)$ and Eq.~\eqref{eq:QQbarpot} takes the Coulombic form
\begin{equation}
  \begin{split}
    \hat{V}^{(\text{LO})}\ket{Q\overline{Q}(\bs{r}_1,\bs{r}_2)} &=  - \frac{C_F \alpha_s}{|\bs{r}_{12}|}  \ket{Q\overline{Q}(\bs{r}_1,\bs{r}_2)}.
  \end{split}
\end{equation}
Therefore, the pNRQCD Hamiltonian for quarkonium at LO is identical to a rescaled version of the Hamiltonian for positronium~\cite{Hylleraas:1947zza}.
The energy eigenstate wavefunctions $\psi_{nlm}(\bs{r}_{12})$ can therefore be classified by the same quantum numbers as the Hydrogen atom, $n\in\mathbb{N}$, $l=0,\ldots,n-1$ and $m=-l,\ldots,l$. They further share the same functional form as the Hydrogen atom wavefunctions with
\begin{equation}
    \psi_{100}(\bs{r};a)=\frac{1}{\sqrt{\pi}a^{3/2}}e^{-|\bs{r}|/a},
\end{equation}
where $a$ is a constant analogous to the Hydrogen atom Bohr radius that for quarkonium at LO is equal to
\begin{equation}
  a^{(\text{LO})} = \frac{2}{\alpha_s C_F m_Q}.
\end{equation}
The corresponding quarkonium ground-state energy is equal to 
\begin{equation}
  \Delta E_{Q\overline{Q}}^{(\text{LO})} = -\frac{\alpha_s^2 C_F^2 m_Q}{4}. \label{eq:EQQbarLO}
\end{equation}
Knowledge of the exact ground-state wavefunction for this case provides a powerful test of numerical QMC methods because
\begin{equation}
\begin{split}
  &\hat{H}^{(\text{LO})} \ket{\psi_i(\bs{r}_1)\chi_{ i}(\bs{r}_2)} \psi_{100}\left(\bs{r}_{12};a=\frac{2}{\alpha_s C_F m_Q}\right) \\
  &= \Delta E_{Q\overline{Q}}^{(\text{LO})}  \ket{\psi_i(\bs{r}_1)\chi_{ i}(\bs{r}_2)} \psi_{100}\left( \bs{r}_{12};a=\frac{2}{\alpha_s C_F m_Q}\right),
\end{split}
\end{equation}
for any $\bs{r}_1$, and $\bs{r}_2$.
Therefore QMC results must reproduce $\Delta E_{Q\overline{Q}}^{(\text{LO})}$ with zero variance when using $\psi_{100}$ with $a=2/(\alpha_s C_F m_Q)$ as a trial wavefunction.

A generic quarkonium wavefunction can be expanded in a basis of hydrogen wavefunctions as {
\begin{equation}
\Psi_T(\bs{r}_1,\bs{r}_2;\bs{C},a) = \sum_{n=1}^\Lambda \sum_{l=0}^{n-1} \sum_{m=-l}^l C_{nlm} \psi_{nlm}(\bs{r}_{12},a),\label{eq:psiTQQbar}
\end{equation} }
where $\Lambda$ provides a truncation of the complete infinite family of wavefunctions, leading to a finite-dimensional family of trial wavefunctions suitable for VMC calculations.
We have verified that variational calculations using the LO potential and { $\Lambda \in \{1,2,3\}$ } reproduce the exact LO ground-state energy within uncertainties and are consistent with {$C_{nlm} \propto \delta_{n1}\delta_{l0}\delta_{m0}$ } and $a=2/(\alpha_s C_F m_Q)$.
Beyond LO, we find that over a wide range of $\alpha_s \in [0.05,0.5]$ the best variational bounds obtained using generic wavefunctions with $\Lambda \in \{1,2,3\}$ are consistent with those where { $C_{nlm} \propto \delta_{n1}\delta_{l0}\delta_{m0}$ } .
Since the $\mathcal{O}(m_Q^0)$ potential is a central potential only depending on $|\bs{r}_{12}|$, orbital angular momentum is a conserved quantum number, and it is not surprising that the ground state is $S$-wave with only $l=0$ wavefunctions present.
Contributions to the ground-state from wavefunctions with { $n > 1$ } should arise in principle beyond LO; however, we find that including { $n > 1$ }  wavefunctions in our variational calculations leaves variational bound on $\Delta E_{Q\overline{Q}}$ unchanged with few percent precision over a wide range of $\alpha_s$.
Similarly, we find that trial wavefunctions described by sums of 2-3 exponentials or Gaussians do not achieve lower variational bounds than those with a single { $n = 1$ }  Coulomb wavefunction at the level of a few percent precision.

\begin{figure}[t!]
  \centering
  \subfigure{\includegraphics[width=\linewidth]{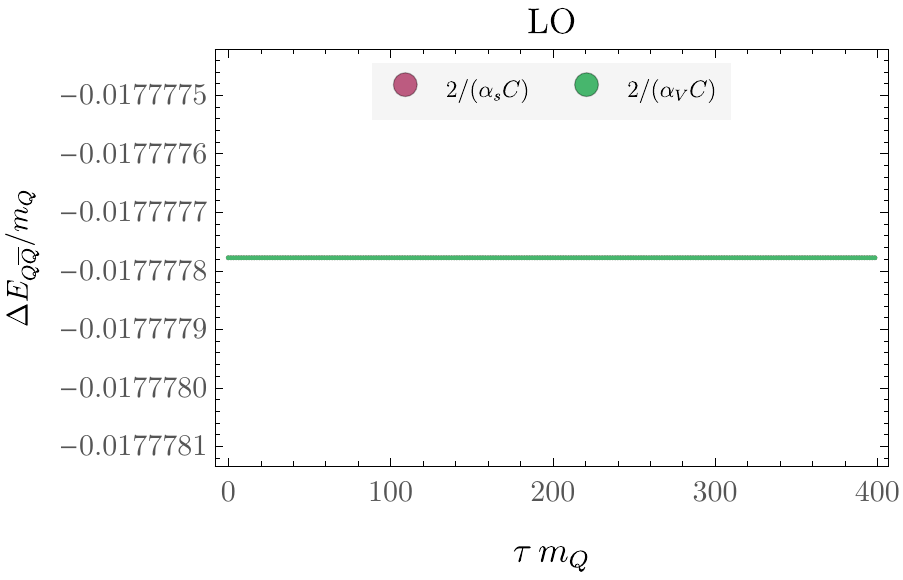}}
  \subfigure{\includegraphics[width=\linewidth]{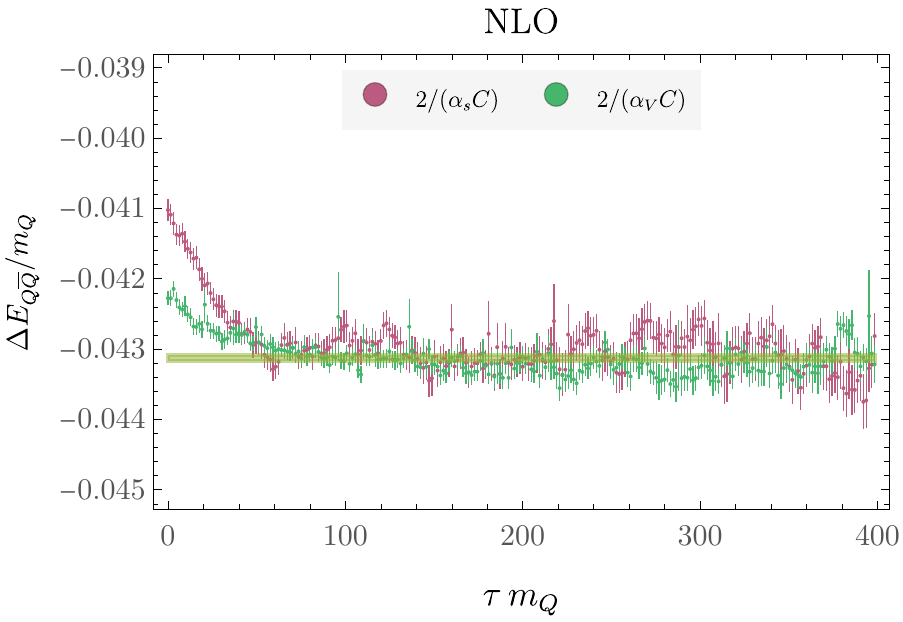}}
  \subfigure{\includegraphics[width=\linewidth]{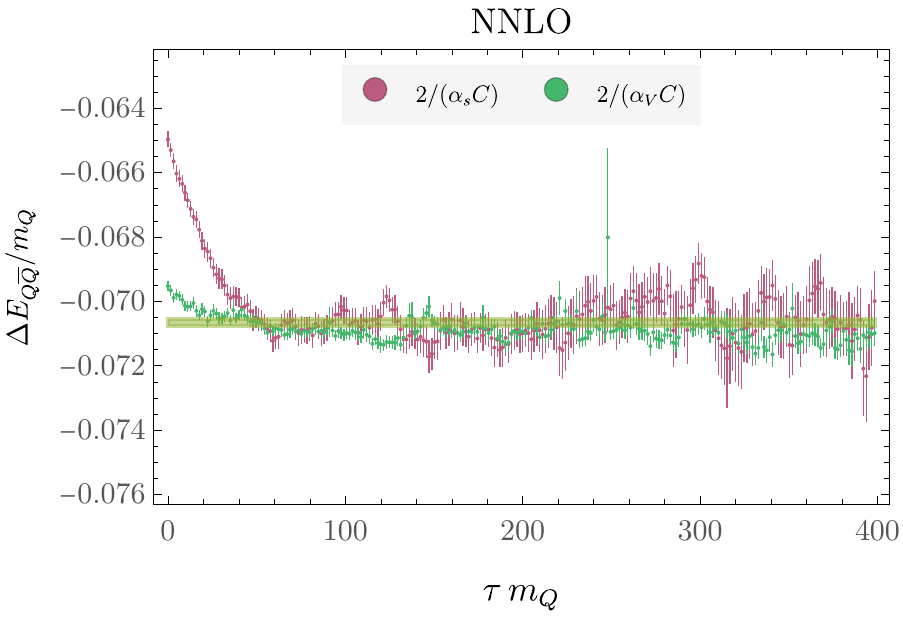}}
\caption{Heavy quarkonium binding energy GFMC results for $\left<H(\tau)\right>$ with $\alpha_s = 0.2$ as functions of $\tau m_Q$ using LO trial wavefunctions (green) and the trial wavefunctions obtained using VMC calculations (purple). The Hamiltonian includes the $\mathcal{O}(m_Q^0)$ pNRQCD potential with the different perturbative orders in $\alpha_s$ indicated.}
\label{fig:qqbar_EMP_QCD_0p2}
\end{figure}

\begin{figure}[t!]
  \centering
  \subfigure{\includegraphics[width=\linewidth]{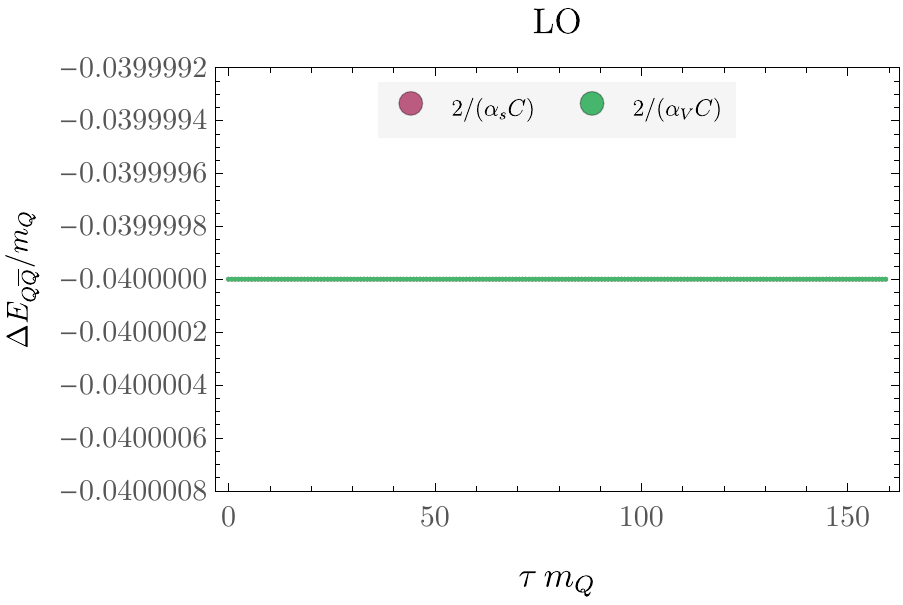}}
  \subfigure{\includegraphics[width=\linewidth]{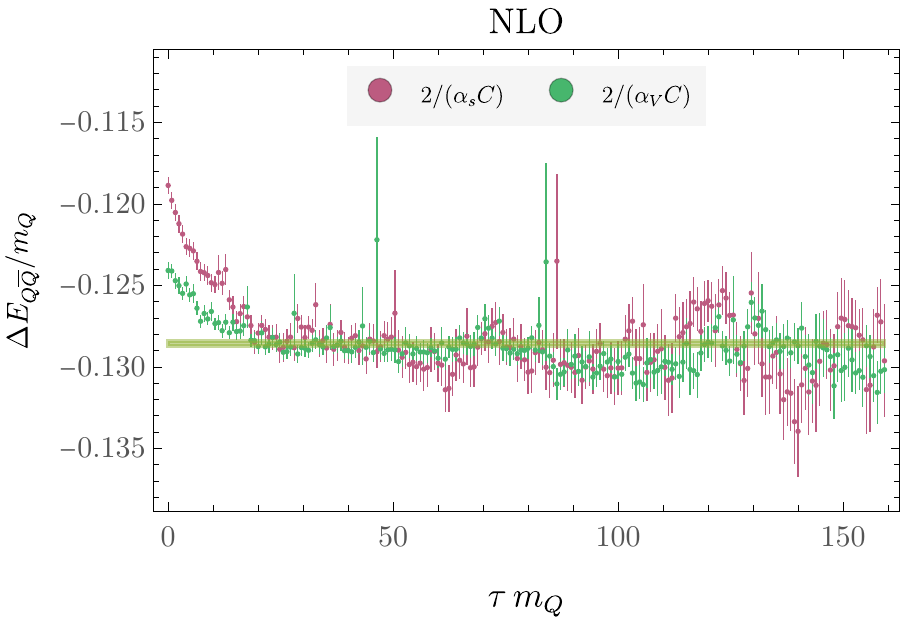}}
  \subfigure{\includegraphics[width=\linewidth]{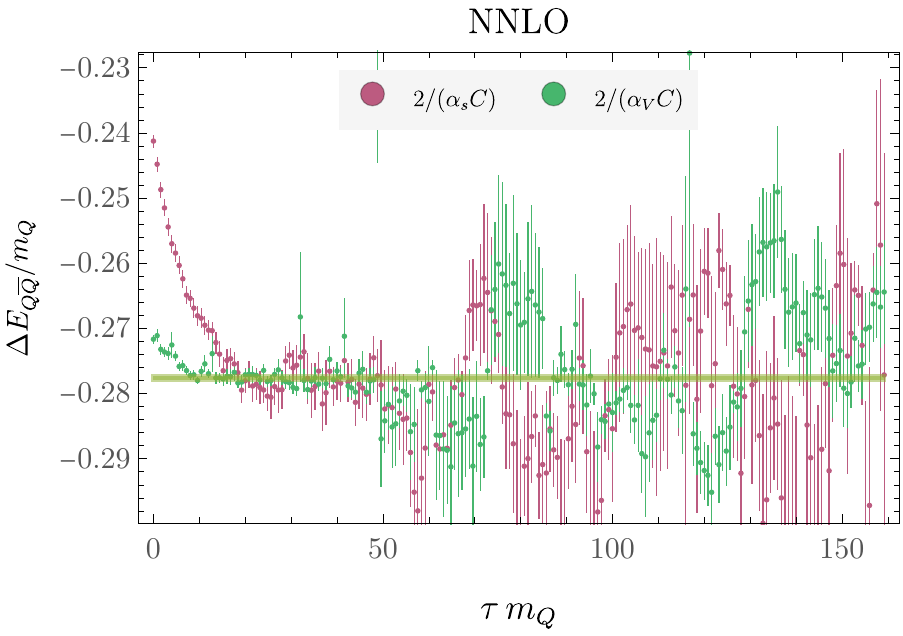}}
\caption{Heavy quarkonium binding energy GFMC results for $\left<H(\tau)\right>$ with $\alpha_s = 0.3$ analogous to those in Fig.~\ref{fig:qqbar_EMP_QCD_0p2}.}
\label{fig:qqbar_EMP_QCD_0p3}
\end{figure}

These results motivate the simple one-parameter wavefunction ansatz
{
\begin{equation}
\Psi_T(\bs{r}_1,\bs{r}_2;a) = \psi_{100}(\bs{r}_{12},a).
\end{equation}
}
Using VMC to determine the optimal $a$ for NLO and NNLO leads to significantly lower ground-state energies than those obtained with $a^{(\text{LO})}$.
The optimal $a$ are smaller than $a^{(\text{LO})}$, which is to be expected if the NLO potential is approximately Coulombic because $\alpha_V(|\bs{r}_{12}|,\mu) > \alpha_s(\mu)$ at NLO and beyond.
Assuming that $\mu$ is chosen to be on the order of $1/|\bs{r}_{12}|$ for distances where the wavefunction is peaked, contributions to the NLO potential proportional to $\ln(\mu |\bs{r}_{12}| e^{\gamma_E})$ can be approximated as a constant denoted $L_{\mu}$.
This corresponds to an approximation of the NLO potential as a Coulomb potential with $\alpha_s(\mu)$ replaced by the $|\bs{r}_{12}|$-independent constant $\alpha_V(|\bs{r}_{12}|, \mu = e^{L_{\mu}-\gamma_E}/|\bs{r}_{12})$.
The ground-state wavefunction under the approximation is { $\psi_{100}(|\bs{r}_{12};a(L_{\mu}))$ } with
\begin{equation}
  a(L_{\mu}) = \frac{2}{\alpha_V(|\bs{r}_{12}|,\mu = e^{L_{\mu}-\gamma_E}/|\bs{r}_{12}|) C_F m_Q}. 
  \label{eq:aL}
\end{equation}
Without assuming any approximation for the potential, {$\psi_{100}(|\bs{r}_{12};a(L_{\mu}))$ } can be viewed as a variational ansatz that is equivalent to { $\psi_{100}(|\bs{r}_{12};a)$ } with the only difference being that $L_{\mu}$ is the variational parameter to be explicitly optimized instead of $a$.
The advantage of the $a(L_{\mu})$ parameterization is that the dependence of the ground-state Bohr radius on $\alpha_s$ is approximately incorporated into $a(L_\mu)$ with constant $L_\mu$.
Empirically, { $\psi_{100}(|\bs{r}_{12};a(L_{\mu}))$ } with $L_\mu = 0$ is found to give ground-state energy results that are consistent at the few-percent level with optimal VMC results over a range of $\alpha_s \in [0.1,0.3]$ (somewhat larger $L_\mu \sim 0.5$ are weakly preferred for small $\alpha_s$).
We are therefore led to the simple trial wavefunction ansatz {
\begin{equation}
  \Psi_T(\bs{r}_1,\bs{r}_2) = \psi_{100}(\bs{r}_{12},a(L_\mu=0)).\label{eq:psiTQQbarSimp}
\end{equation}
}

GFMC results using the VMC trial wavefunctions $\Psi_T(\bs{r}_1,\bs{r}_2)$ are shown in Figs.~\ref{fig:qqbar_EMP_QCD_0p2}-\ref{fig:qqbar_EMP_QCD_0p3} for quark masses corresponding to $\alpha_s(\mu_p) = 0.2$ and $\alpha_s(\mu_p) = 0.3$ respectively, using the renormalization scale choice $\mu_p = 4 \alpha_s(\mu_p) m_Q$ discussed further below.
Results using the exact LO wavefunction with $a=2/(\alpha_s C_F m_Q)$ as GFMC trial wavefunctions are also shown for comparison.
Both results are identical at LO and reproduce the exact result, Eq.~\eqref{eq:EQQbarLO}, with zero variance at machine precision.

At NLO, the VMC wavefunctions give 3\% and 4\% lower variational bounds than LO wavefunctions for $\alpha_s = 0.2$ and $\alpha_s = 0.3$, respectively.
After GFMC evolution, both results approach energies 2\% lower than the VMC variational bounds for both $\alpha_s$.
Slightly less imaginary-time evolution is required to achieve ground-state saturation at a given level of precision for VMC wavefunctions than LO wavefunctions.
At NNLO, the VMC wavefunctions achieve more significant 7\% and 11\% lower variational bounds than LO wavefunctions for $\alpha_s = 0.2$ and $\alpha_s = 0.3$, respectively.
GFMC evolution again leads to 2\% lower energies than optimized variational wavefunctions for both $\alpha_s$.
Significantly less imaginary-time evolution is required to achieve ground-state saturation using optimized variational wavefunctions at NNLO.
For NLO potentials, the variance of $\left< H(\tau) \right>$ computed using VMC trial wavefunction is similar to that obtained using LO trial wavefunctions.
For NNLO potentials, the corresponding variance is 50\% smaller using VMC trial wavefunctions than using LO trial wavefunctions.

Notably, significantly more imaginary-time evolution is required to achieve ground-state saturation with $\alpha_s = 0.2$ than with $\alpha_s = 0.3$.
At both NLO and NNLO, $1\sigma$ agreement between model-averaged fit results and Hamiltonian matrix elements at particular $\tau$ is seen for $\tau \gtrsim 25 / m_Q$ with $\alpha_s = 0.3$ and is only seen for $\tau \gtrsim 50 / m_Q$ with $\alpha_s = 0.3$.
This scaling is consistent with theoretical expectations for a Coulombic system: the energy gap between the ground- and the first-excited state at LO is 
\begin{equation}
  \delta^{(\text{LO})} = \frac{3\alpha_s^2 C_F^2 m_Q}{16},
\end{equation}
and excited-state contributions to GFMC results are suppressed by $e^{-\delta \tau}$.
The observed scaling of $\delta$ in our GFMC results is consistent with $\delta \sim \alpha_s^2 m_Q$ holding approximately at higher orders.

\begin{figure}
\centering
  \subfigure{\includegraphics[width=\linewidth]{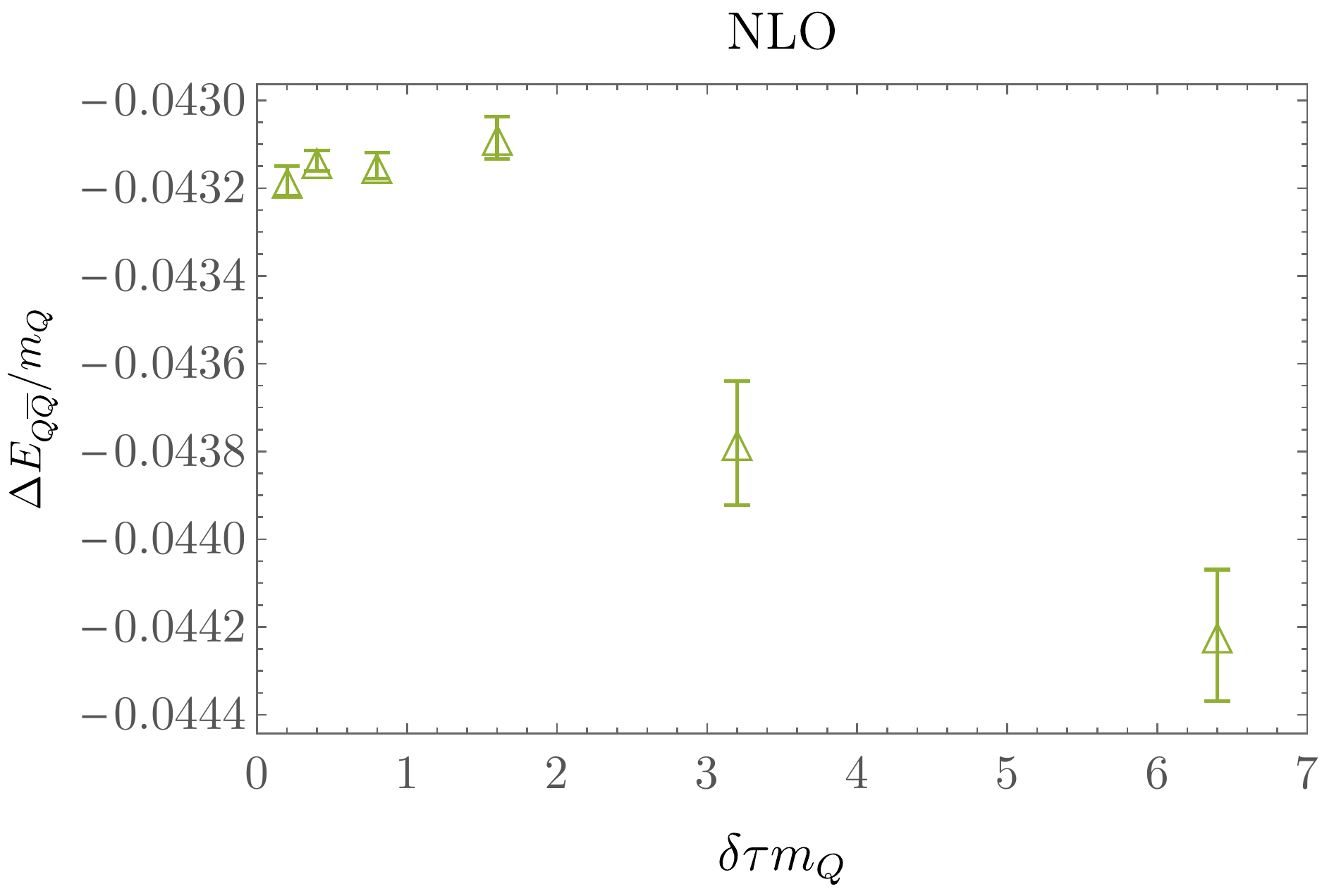}}
  \subfigure{\includegraphics[width=\linewidth]{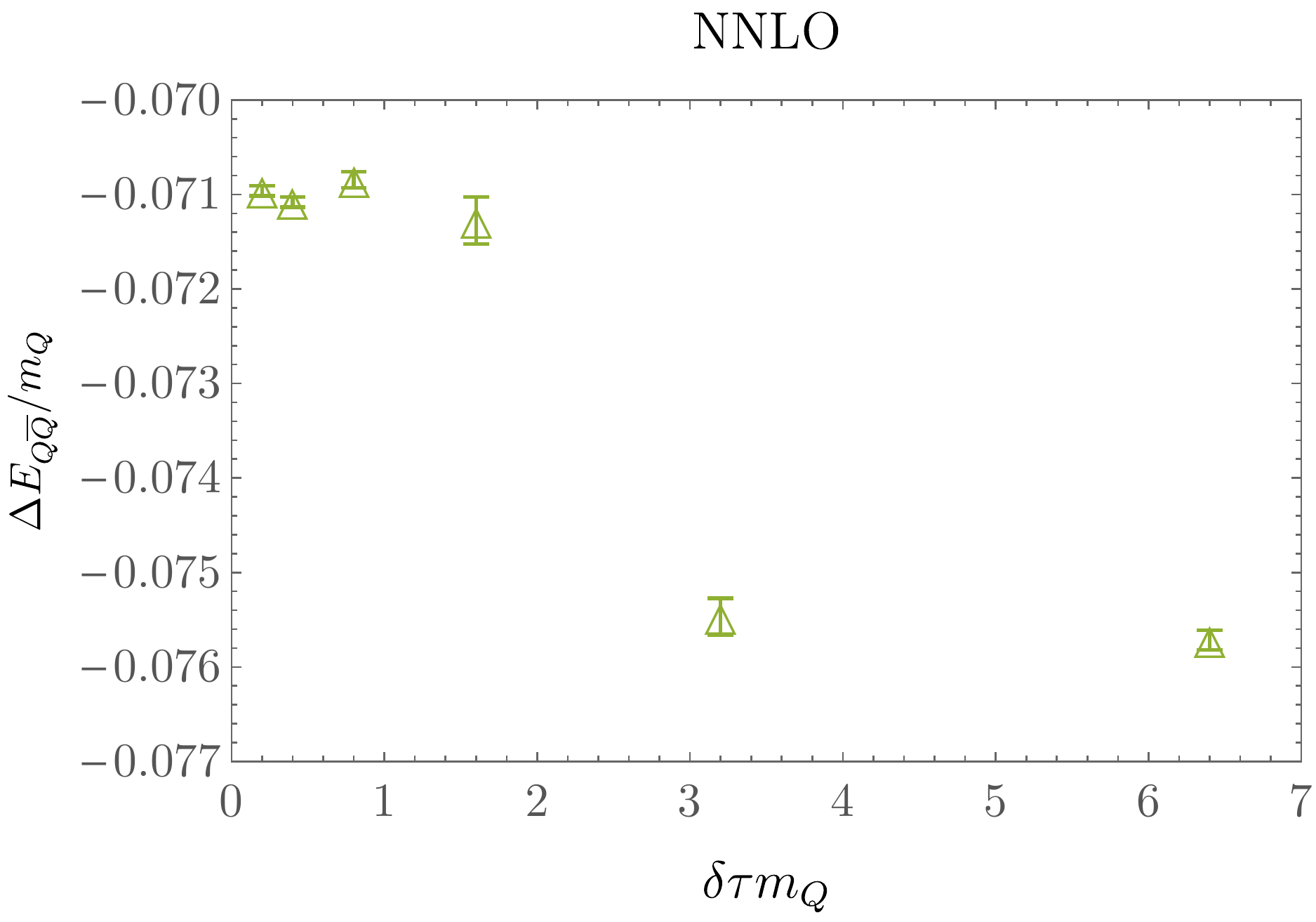}}
\caption{Heavy quarkonium binding energy results obtained from fits to the $\left<H(\tau)\right>$ results in Fig.~\ref{fig:qqbar_EMP_QCD_0p2} are shown for GFMC calculating with several different choices of Trotterization scale $\delta \tau$ as functions of $\delta \tau m_Q$ for NLO and NNLO pNRQCD potentials.
The LO results are not shown since the results are exact and therefore $\delta \tau$ independent.}
\label{fig:dtauscale}
\end{figure}

These GFMC results include discretization effects arising from the Trotterization of the imaginary-time evolution operator $e^{-\hat{H} \tau}$ discussed in Sec.~\ref{sec:gfmc} and were performed using $\delta \tau = 0.4 / m_Q$.
We repeated GFMC calculations using a wide range of $\delta \tau \in [0.2/m_Q, 6.4/m_Q]$ in order to study the size of these discretization effects; results for $\alpha_s = 0.2$ are shown in Fig.~\ref{fig:dtauscale}.
Discretization effects are found to be sub-percent level and smaller than our GFMC statistical uncertainties for $\tau m_Q \lesssim 2$ with evidence for few-percent discretization effects at larger $\delta \tau$.
Similar results are found for other $\alpha_s$ with the smallest $\delta \tau$ where discretization effects are visibly found to increase with decreasing $\alpha_s$ roughly as $1/\alpha_s$.
To validate this determination, we computed the expectation value of $[\hat{V}, \hat{T}]$ and found that the $\delta \tau$ scales where discretization effects become visible are roughly consistent with  $\Delta E_{Q\overline{Q}} / \mbraket{Q\overline{Q}}{[\hat{V}, \hat{T}]}{Q\overline{Q}}$ as expected from the Baker-Campbell-Hausdorff commutator corrections arising from approximating $e^{(\hat{T} + \hat{V})\delta \tau}$ as $e^{-\hat{T}\delta \tau} e^{ - \hat{V} \delta \tau}$~\cite{PhysRevX.11.011020}.

\subsection{Baryons}

The pNRQCD quark-quark potential acting on baryon states is given at $\mathcal{O}(m_Q^0)$ by Eq.~\eqref{eq:Bpot} and Eq.~\eqref{eq:Vpsipsi} by
\begin{equation}
  \begin{split}
    \hat{V}^{\psi\psi} \ket{B} &=  \sum_{I<J} V^{\psi\psi,(0)}_{\text{A}}(\bs{r}_{IJ})  \ket{B} \\
    &= - \sum_{I<K} \frac{C_B \alpha_V(|\bs{r}_{IJ}|,\mu)}{|\bs{r}_{IJ}|}  \ket{B}, \label{eq:QQQpot}
  \end{split}
\end{equation}
where $C_B = C_F / (N_c-1)$. As discussed above, three-quark potentials arise for baryons at NNLO; however the quark-quark potential arises at LO and can therefore be expected to play a dominant role.

The baryon quark-quark potential has a similar Coulombic form to the quarkonium potential, except that for the baryon case, there is a sum over Coulomb potentials for all relative coordinate differences.
A similar (though not identical) summation arises in the kinetic term if the baryon wavefunction is taken to be a linear combination of products of Coulomb wavefunctions,
{
\begin{equation}
  \Psi_T(\bs{R};\bs{C},a) = \prod_{I=1}^{N_c} \sum_{J<I} \sum_{n=1}^\Lambda \sum_{l=0}^{n-1} \sum_{m=-l}^l C_{nlm} \psi_{nlm}(\bs{r}_{IJ},a),
\label{eq:psiTQQQ}
\end{equation}
}
where $\bs{R} = (\bs{r}_1,\ldots,\bs{r}_{N_c})$.
Although VMC calculations are performed using { $\Lambda \in \{1,2,3\}$ }, the variational energy bounds obtained for $N_c = 3$ baryons are consistent with those obtained using ground-state wavefunctions where { $C_{nlm} \propto \delta_{n1} \delta_{l0} \delta_{m0}$ }.
Similarly, results using sums of one or two exponential or Gaussian corrections to a product of { $n=1$ } Coulomb wavefunctions are found to give consistent variational bounds at the one percent level across a wide range of $\alpha_s$.
This motivates the simple one-parameter family of trial wavefunctions 
{
\begin{equation}
  \Psi_T(\bs{R};a) = \prod_{I=1}^{N_c} \sum_{J<I} \psi_{100}(\bs{r}_{IJ},a).
\end{equation} }
Analogous results are found for (less systematic) VMC studies with $N_c \in \{4,5,6\}$.
This VMC ansatz is similar to the exponential wavefunction ansatz used in variational calculations of pNRQCD baryons at LO in Ref.~\cite{Jia:2006gw}. However, it differs significantly from the ansatz used in analogous NNLO calculations in Ref.~\cite{Llanes-Estrada:2011gwu}, which used a product of momentum-space exponentials that therefore have power-law decays at large separations to describe $N_c = 3$ baryons.
It is perhaps surprising that baryon ground-state energies are accurately described using a product of Coulomb ground-state wavefunctions even at NNLO with three-quark potentials present; however, as discussed in Sec.~\ref{sec:qcd_baryons} below the three-quark potentials lead to sub-percent corrections to results using just quark-quark potentials for $\alpha_s \lesssim 0.3$.

\begin{figure}[t!]
  \centering
  \subfigure{\includegraphics[width=\linewidth]{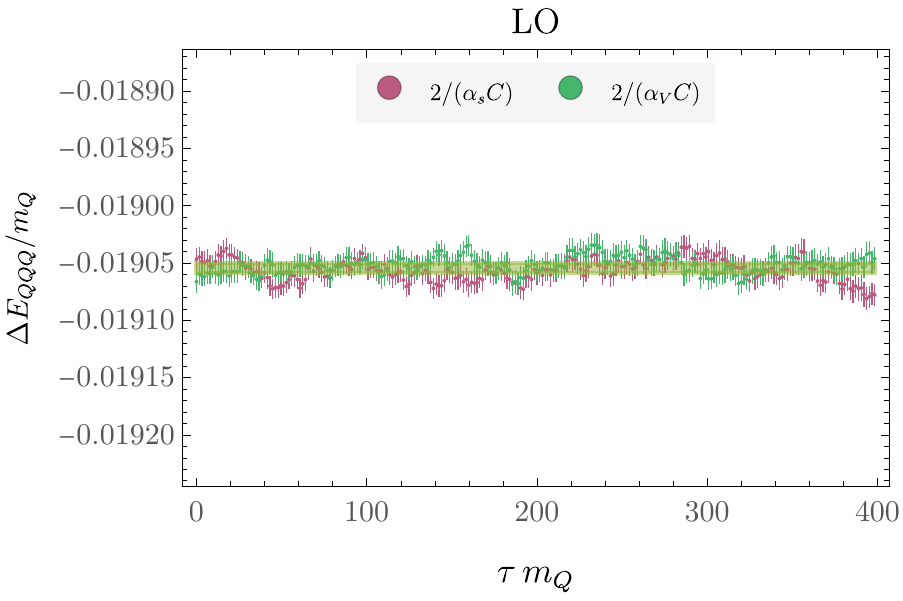}}
  \subfigure{\includegraphics[width=\linewidth]{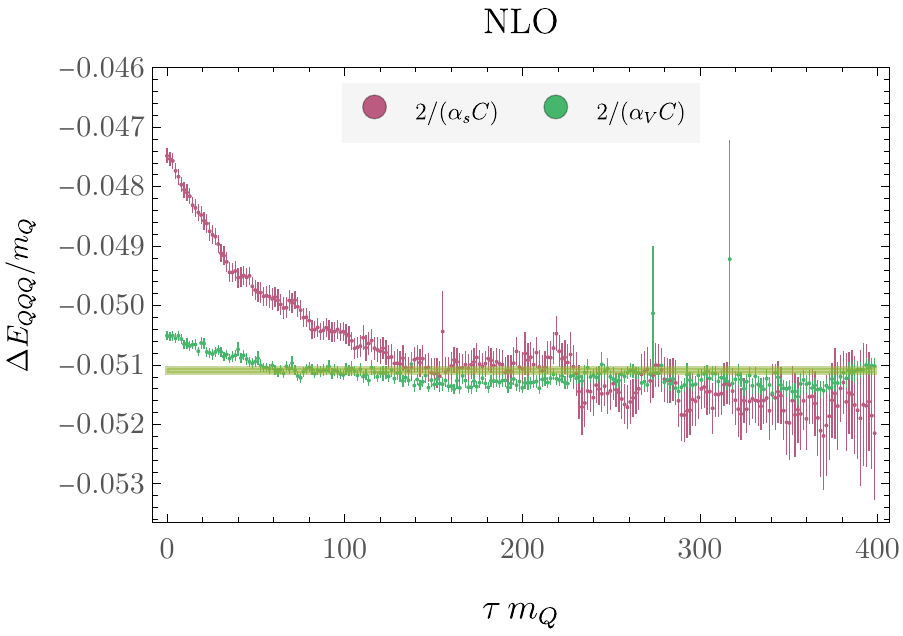}}
  \subfigure{\includegraphics[width=\linewidth]{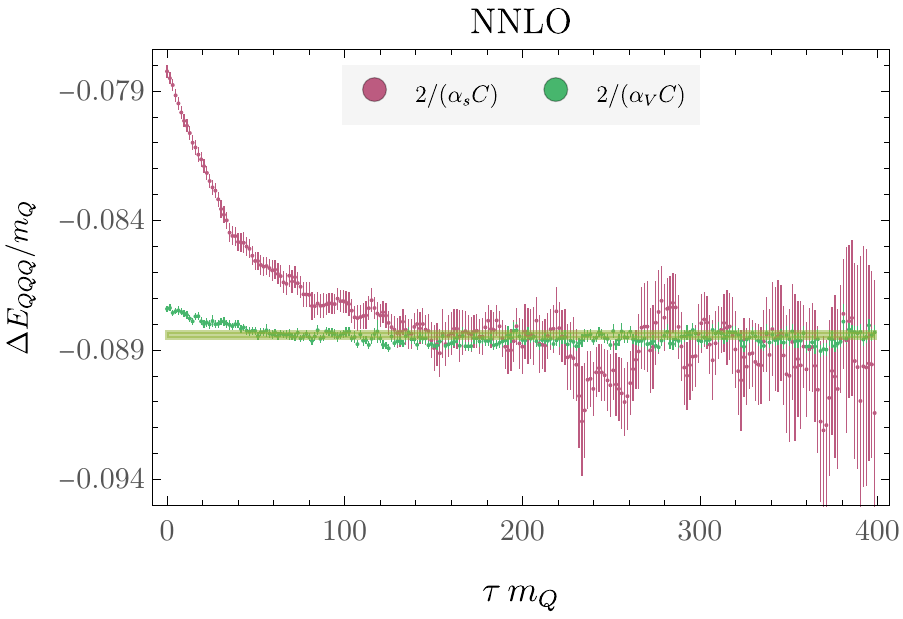}}
\caption{Triply-heavy baryon binding energy results for $\left<H(\tau)\right>$ with $\alpha_s = 0.2$ analogous to those in Fig.~\ref{fig:qqbar_EMP_QCD_0p2}.}
\label{fig:qqq_EMP_QCD_0p2}
\end{figure}

\begin{figure}[t!]
  \centering
  \subfigure{\includegraphics[width=\linewidth]{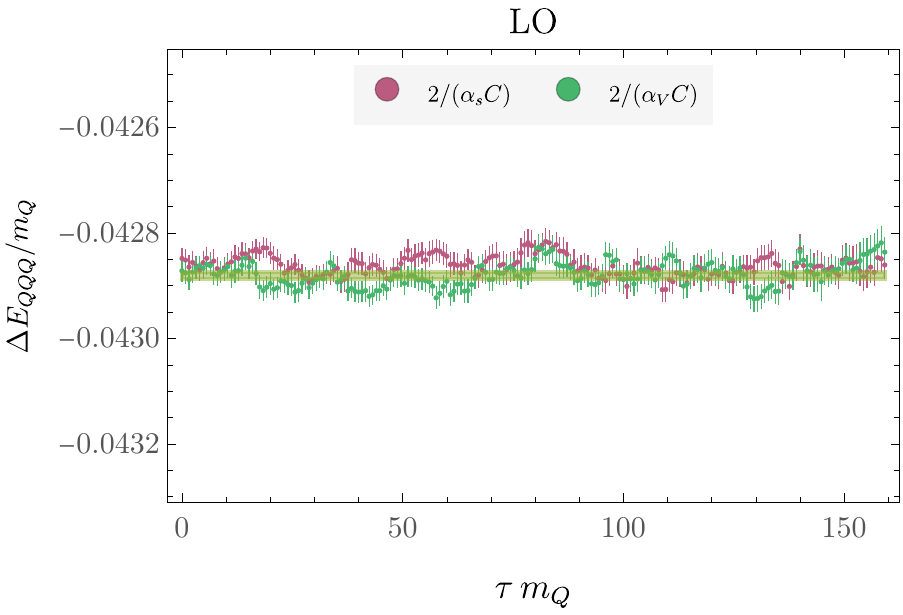}}
  \subfigure{\includegraphics[width=\linewidth]{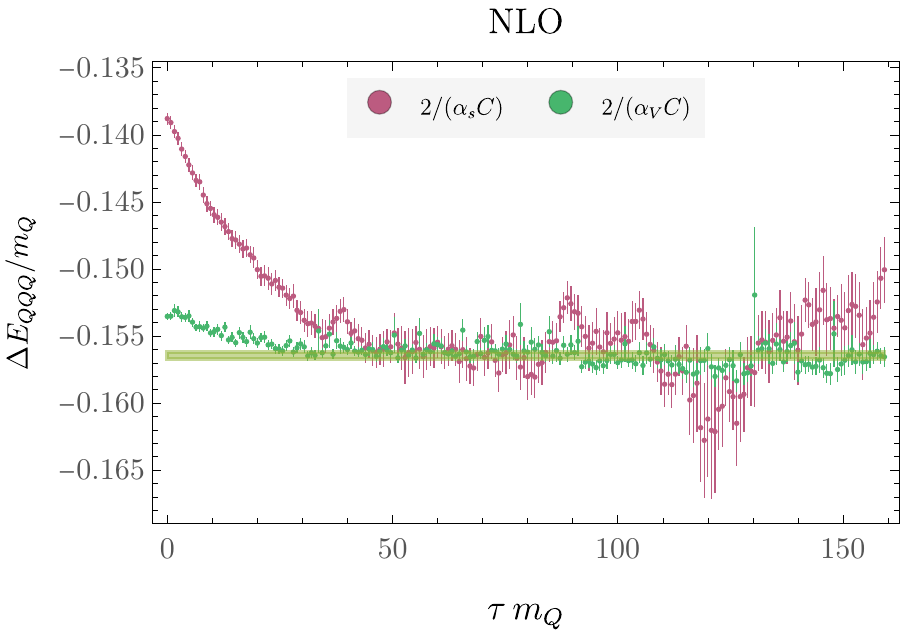}}
  \subfigure{\includegraphics[width=\linewidth]{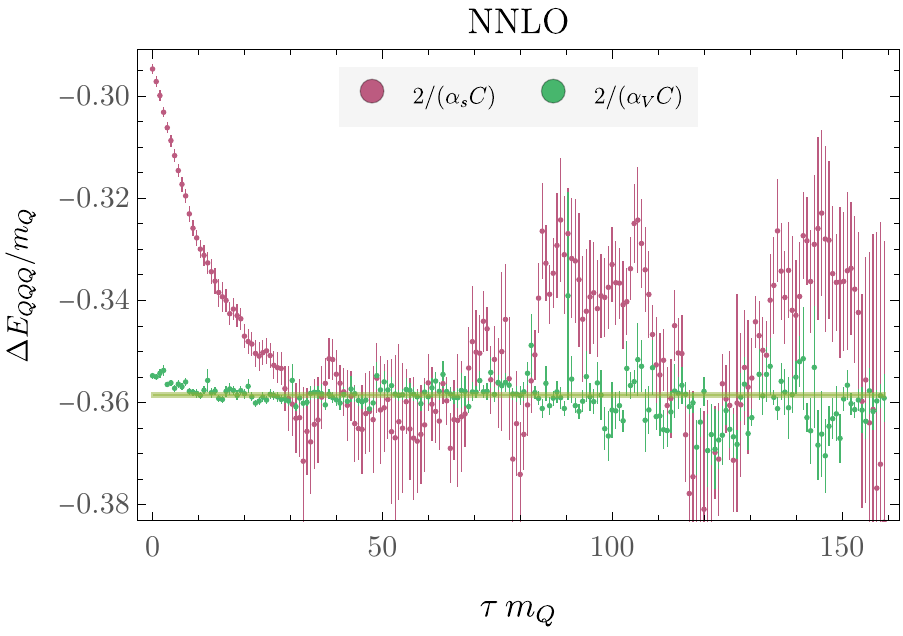}}
\caption{Triply-heavy baryon binding energy results for $\left<H(\tau)\right>$ with $\alpha_s = 0.3$ analogous to those in Fig.~\ref{fig:qqbar_EMP_QCD_0p2}.}
\label{fig:qqq_EMP_QCD_0p3}
\end{figure}

At LO, the optimal variational bounds obtained from VMC with this one-parameter trial wavefunction family are consistent with
\begin{equation}
  a^{(\text{LO})} = \frac{2}{\alpha_s C_B m_Q},
\end{equation}
which is the same Bohr radius appearing in the exact LO quarkonium result rescaled by the color factor applying in the baryon potential.
Beyond LO, we again parameterize the Bohr radius by $a(L_\mu)$ defined in Eq.~\eqref{eq:aL} where $L_\mu$ corresponds to the value of $\ln(\mu r e^{\mu_E})$ if logarithmic $r$ dependence is approximated as constant.
The optimal value of $L_\mu$ increases mildly with increasing $m_Q$, but across the range, $0.1 \leq \alpha_s \leq 0.3$ ground-state energy results with a constant value of $L_\mu = 0.5$ are within a few percent of optimal VMC ground-state energies (somewhat larger $L_\mu \sim 1$ are weakly preferred for small $\alpha_s$).
The GFMC calculations of QCD and $SU(N_c)$ baryons below therefore use the simple trial wavefunction ansatz {
\begin{equation}
  \Psi_T(\bs{R}) = \prod_{I=1}^{N_c} \sum_{J<I} \psi_{100}(\bs{r}_{IJ},a(L_\mu = 0.5)). \label{eq:psiTQQQSimp}
\end{equation}
}

GFMC results using the VMC trial wavefunctions $\Psi_T(\bs{r}_1,\bs{r}_2)$ are shown in Figs.~\ref{fig:qqq_EMP_QCD_0p2}-\ref{fig:qqq_EMP_QCD_0p3} for the same quark masses and renormalization scales as for quarkonium above.
Although the LO baryon wavefunction is not an eigenstate of $\hat{H}^{(\text{LO})}$, it provides remarkably precise and approximately $\tau$-independent Hamiltonian matrix elements with excited-state contamination not visible within $0.1\%$ statistical uncertainties.
Similar results are found with $N_c \in \{4,5,6\}$.
This suggests that the product form of the baryon trial wavefunction used here is suitable for describing multi-quark states with identical attractive Coulomb interactions between all quarks.

Beyond LO, similar patterns arise as in the quarkonium case above, but excited-state effects are more pronounced for baryons before VMC optimization.
VMC wavefunctions give 6\%  and 10\% lower variational bounds than LO wavefunctions for NLO potentials with $\alpha_s = 0.2$ and $\alpha_s = 0.3$, respectively.
Excited-state contamination is still visible in GFMC results using VMC wavefunctions for $\tau \lesssim 50 / m_Q$ with $\alpha_s = 0.2$ and $\tau \lesssim 25 / m_Q$ with $\alpha_s = 0.3$, which is similar to the corresponding $\tau$ required for similar suppression of quarkonium excited-states and shares the same $1/(\alpha_s^2 m_Q)$ scaling expected for Coulombic excited-state effects.
At least a factor of two larger $\tau$ is required to achieve the same level of excited-state suppression using LO baryon wavefunctions.
The fitted GFMC ground-state energy is $1\%$ and $2\%$ lower than the VMC wavefunction results for $\alpha_s=0.2$ and $\alpha_s = 0.3$, respectively.

At NNLO, VMC wavefunctions give 10\%  and 17\% lower variational bounds than LO wavefunctions with $\alpha_s = 0.2$ and $\alpha_s = 0.3$, respectively.
Excited-state effects are mild and similar to NLO using VMC wavefunctions with 1\% differences between VMC and fitted GFMC ground-state energy results, but very large excited-state effects and large variance increase with $\tau$ are both visible using LO baryon wavefunctions with NNLO potentials.
The reduction in variance between VMC and LO baryon wavefunctions is more than an order of magnitude for some $\tau$, and for large $\tau$, the signal using LO wavefunctions is lost while VMC wavefunctions have relatively mild variance increases.
It is perhaps not surprising that LO baryon wavefunctions do not provide a suitable trial wavefunction for GFMC calculations at NNLO, where in particular three-quark potentials enter. However, it is remarkable that simple VMC optimization of the Bohr radius of a product of Coulomb wavefunctions is sufficient to provide a trial wavefunction leading to high-precision GFMC results with few-percent excited-state effects only for $\tau \lesssim  2/(\alpha_s^2 m_Q)$.

\begin{figure}[t!]
\centering
  \subfigure{\includegraphics[width=\linewidth]{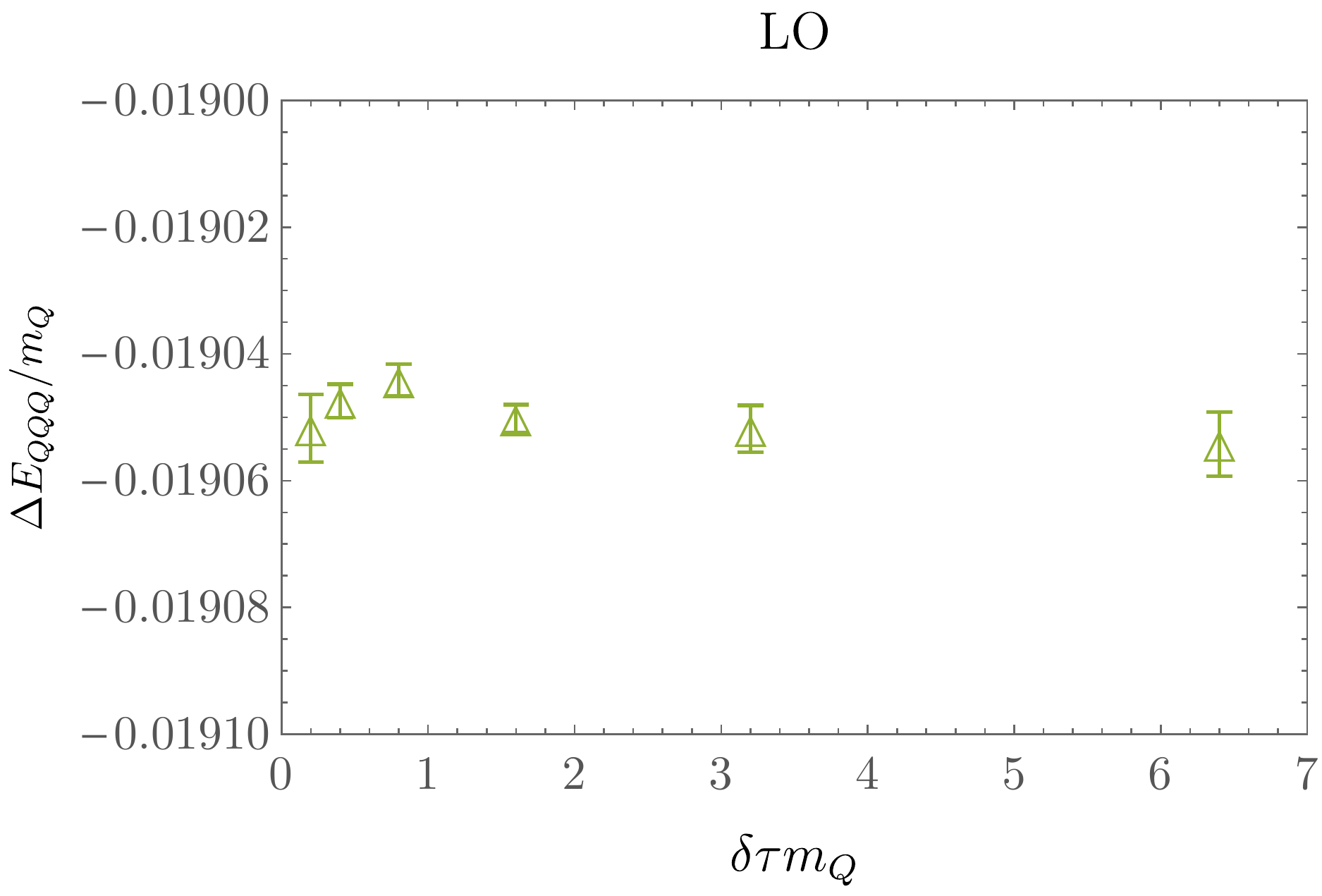}}
  \subfigure{\includegraphics[width=\linewidth]{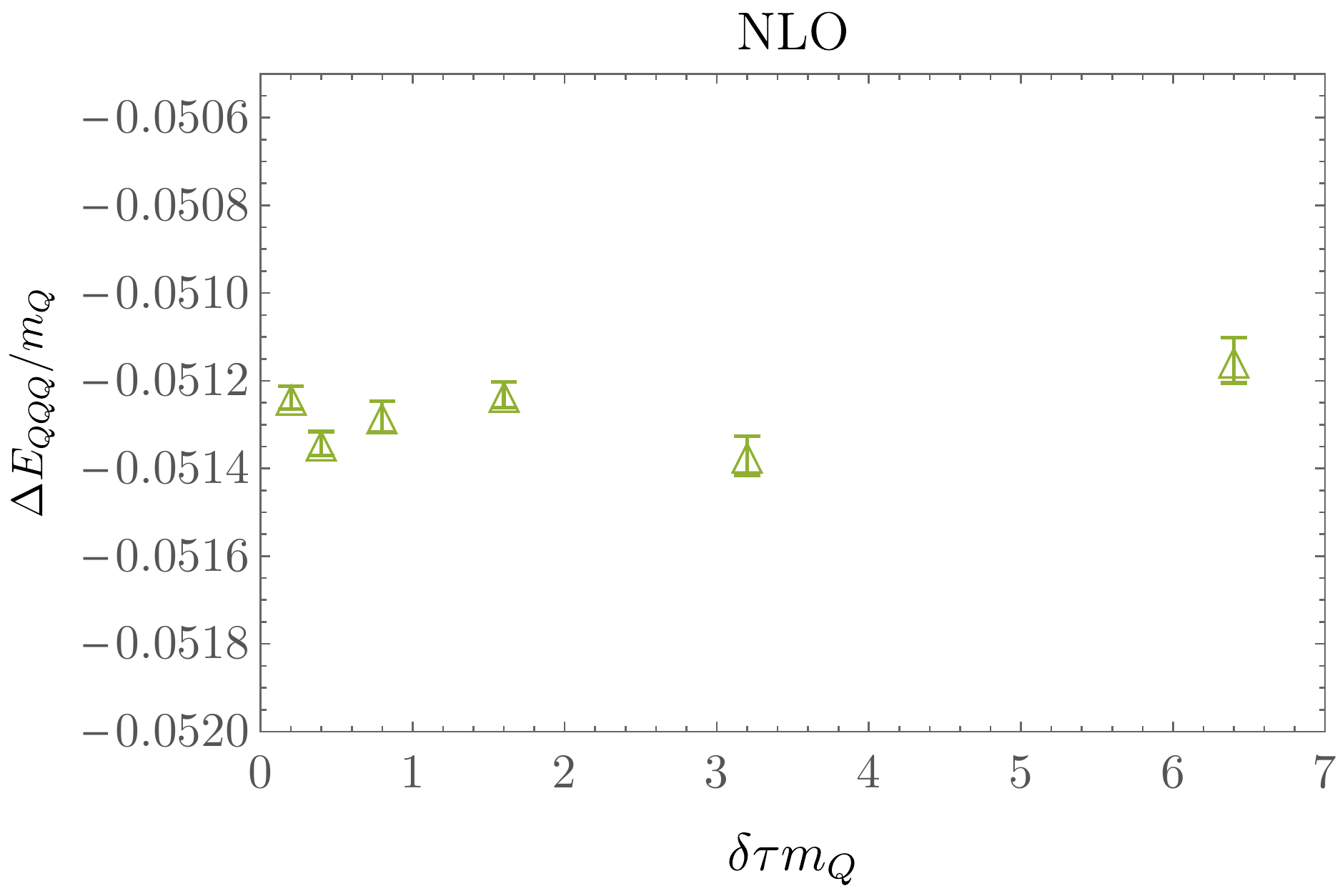}}
  \subfigure{\includegraphics[width=\linewidth]{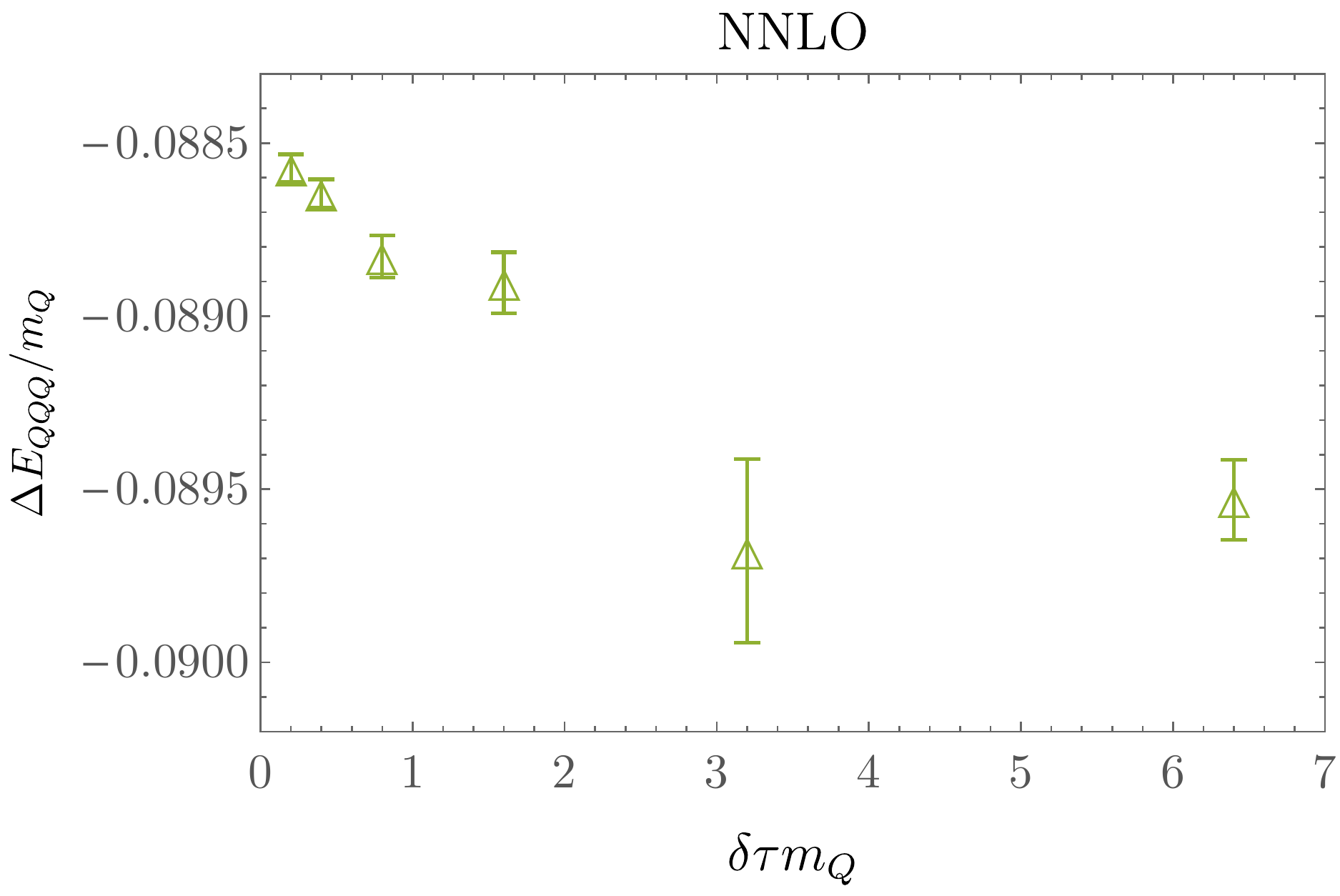}}
\caption{Triply-heavy baryon binding energy results obtained from fits to the $\left<H(\tau)\right>$ results in Fig.~\ref{fig:qqq_EMP_QCD_0p2} are shown for GFMC calculating with several different choices of Trotterization scale $\delta \tau$ as functions of $\delta \tau m_Q$ for each perturbative order studied.}
\label{fig:dtauscaleB}
\end{figure}

The dependence of fitted GFMC results on $\delta \tau$ is shown in Fig.~\ref{fig:dtauscaleB} for the example of $N_c = 3$ baryons with $\alpha_s = 0.2$.
Interestingly, LO baryon ground-state energy results are observed to be independent of $\delta \tau$ to percent-level precision for $\delta \tau \lesssim 100 / m_Q$ even though the LO baryon wavefunction is not exactly a LO energy eigenstate.
Discretization effects are also not clearly resolved at NLO for $\delta \tau \lesssim 6/m_Q$, although more significant effects appear for larger $\delta \tau$.
At NNLO, there are clear signals of percent-level discretization effects of $\delta \tau \gtrsim 1/m_Q$, but negligible sub-percent discretization effects are seen for smaller $\delta \tau$.
The calculations below target percent-level determinations of ground-state (nonrelativistic) energies and therefore use $\delta \tau = 0.4 / m_Q$ for QCD and $\delta \tau \in [0.4/m_Q, 0.8/m_Q]$ for exploring strongly coupled dark sectors for which these discretization effects are expected to be negligible.


\section{QCD binding energy results}
\label{sec:qcd}

The heavy quarkonium mass $M_{Q\overline{Q}} = 2m_Q + \Delta E_{Q\overline{Q}}$ is one of the simplest pNRQCD observables, and matching its calculated value to experimental results provides a way to fix the pNRQCD parameter $m_Q$.
The heavy quarkonium spectrum has been previously computed in pNRQCD for $b$ and $c$ mesons to $\rm{N^3LO}$~\cite{Pineda:1997hz,Kniehl:2002br} using perturbative quark mass definitions { such as the 1S mass}.
Here, we use an alternative quark-mass definition, analogous to definitions used in lattice QCD, in which we tune the pole mass $m_Q$ to reproduce experimental quarkonium masses. 
Once $m_Q$ is determined using this tuning procedure, pNRQCD can be used to make predictions for other hadron masses and matrix elements.
Below, the masses of triply-heavy baryons containing $b$ and $c$ quarks are computed and compared with lattice QCD results~\cite{Meinel:2010pw,Brown:2014ena} in order to validate the methods discussed above.
Further, it is straightforward and relatively computationally inexpensive to extend pNRQCD calculations over a wide range of $m_Q$, which allows the dependence of meson and baryon masses on $m_Q$ to be studied for a wide range of $m_Q \gg \Lambda_{QCD}$.

For each choice of $m_Q$, the renormalization scale $\mu$ is chosen to be in the range $\alpha m_Q < \mu < m_Q$ so that neither the logs of $\mu / m_Q$ arising in NRQCD matching or the logs of $\mu r$ explicitly appearing in the potential are too large~\cite{Brambilla:2004jw,Pineda:2011dg} since on average $r \sim 1/(v m_Q) \sim 1/(\alpha m_Q)$ as supported by the success of Hydrogen wavefunction with this value of the Bohr radius discussed above. In particular, the GFMC results below use a central value of the renormalization scale
\begin{equation}
\mu_p = 4 \alpha_s(\mu_p) m_Q, \label{eq:mup_def}
\end{equation}
which can be solved using iterative numerical methods to determine $\mu_p$ for a given value of $m_Q$.
In order to study the dependence on this choice of scale, GFMC calculations are performed with $\mu = 2\mu_p$ and $\mu = \mu_p / 2$ as well as with $\mu = \mu_p$.
The RG evolution of $\alpha_s(\mu)$ is solved using the $\beta$-function calculated at one order higher in perturbation theory than the pNRQCD potential, and in particular, the one-, two-, and three-loop $\beta$ functions are used along with the LO, NLO, and NNLO potentials.
The $\beta$-function coefficients, the values of the Landau pole scale $\Lambda_{QCD}$ required to reproduce the experimentally precisely constrained value $\alpha_s(M_Z) = 0.1184(7)$ for the three-loop $\alpha_s$, and the quark threshold matching factors related theories with $N_f$ and $N_f-1$ flavors are reviewed in Ref.~\cite{Bethke:2009jm}; the same initial condition is used to determine the values of $\Lambda_{QCD}$ used for one- and two-loop $\alpha_s$ in the LO and NLO results of this work.

Numerical results in this section use GFMC calculations with the trial wavefunction discussed in Sec.~\ref{sec:trial}.
Calculations use 8 equally spaced values of $m_Q \in [m_c,m_b]$ (using the $\overline{\text{MS}}$ masses~\cite{ParticleDataGroup:2022pth}) for which the $N_f =4$ potential is used (the renormalization scale satisfies $\mu_p > m_c$ for this range) and another 8 equally spaced values of $m_Q \in [m_b,m_t]$ for which the $N_f=5$ potential is used.
The Trotterization scale $\delta \tau = 0.4 / m_Q$ is chosen, which is expected to lead to sub-percent discretization effects on binding energies according to the results of Sec.~\ref{sec:trial}.
The total imaginary-time length of GFMC evolution is chosen to be $N_\tau \delta \tau = 8 / (\alpha_s^2 m_Q)$ in order to ensure that imaginary times much larger than the expected inverse excitation gap $\delta \sim 1/(\alpha_s^2 m_Q)$ are achieved, which the results of Sec.~\ref{sec:trial} indicate are sufficient to reduce excited-state contamination to the sub-percent level.
This corresponds to $N_\tau \in [200,1400]$ for $m_Q \in [m_c,m_t]$.
Relatively modest GFMC ensembles with $N_{\rm walkers} = 5,000$ are found to be sufficient to achieve sub-percent precision on binding energy determinations.

\begin{figure}[t!]
  \centering
  \subfigure{\includegraphics[width=\linewidth]{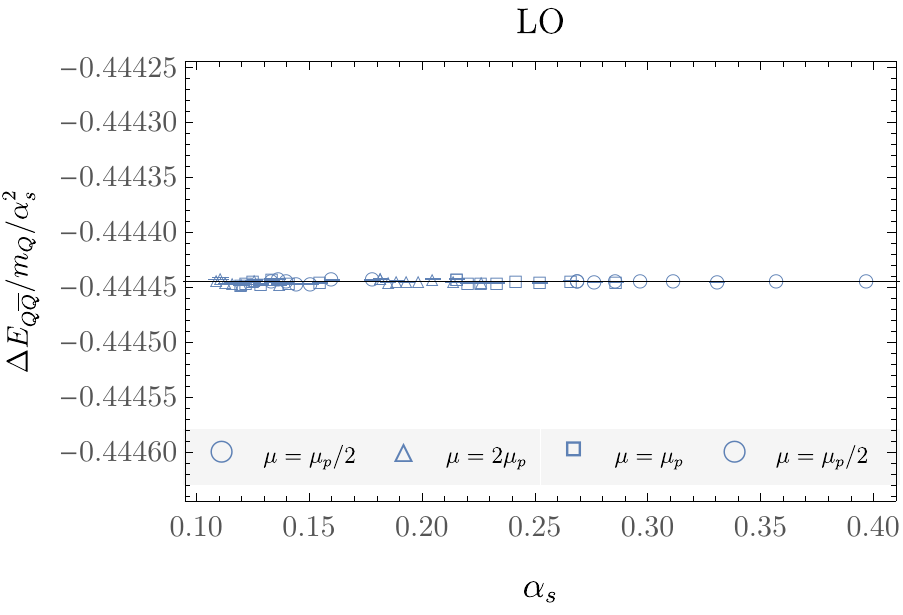}}
  \subfigure{\includegraphics[width=\linewidth]{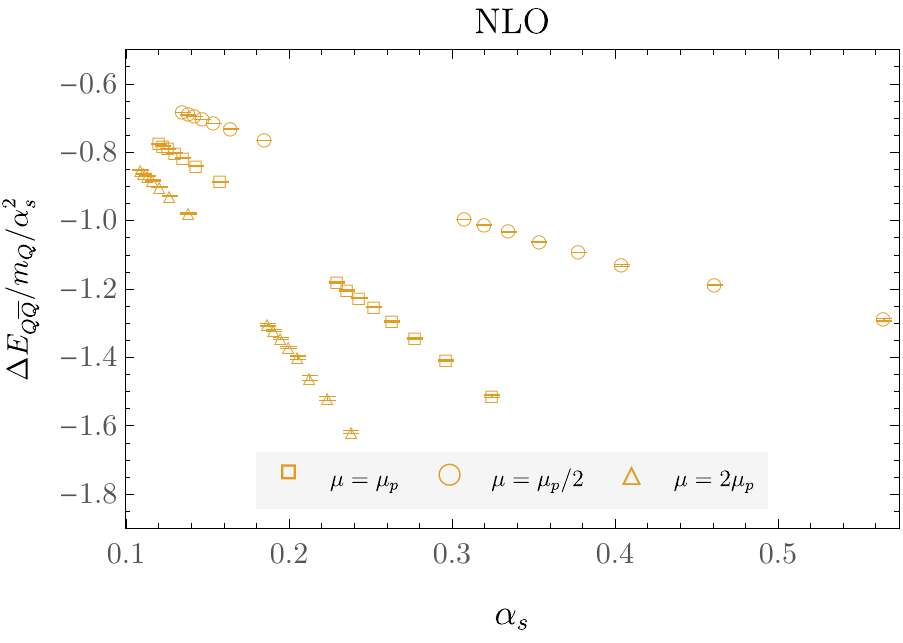}}
  \subfigure{\includegraphics[width=\linewidth]{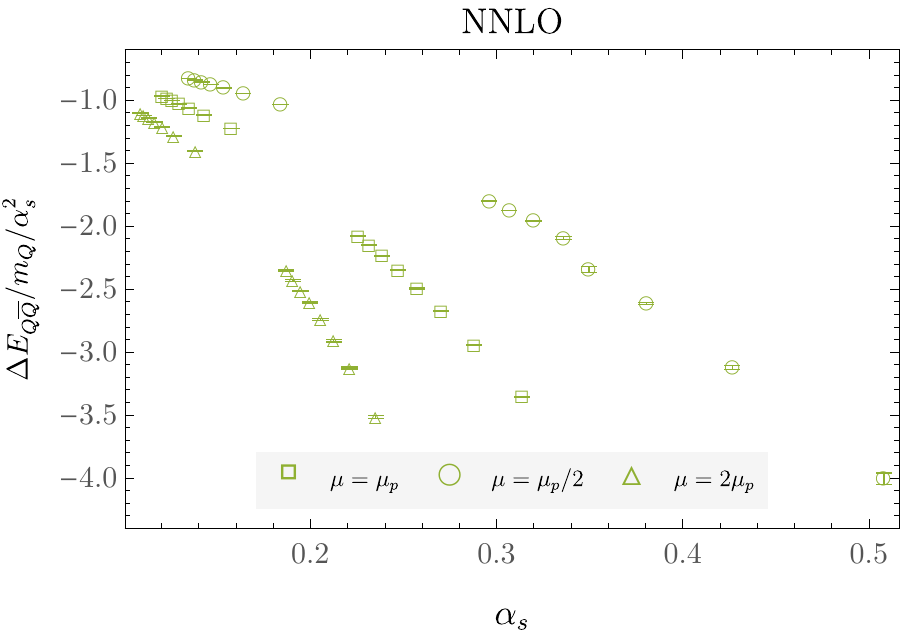}}
\caption{Heavy quarkonium binding energy results as functions of $\alpha_s$ (excluding points with $N_f=5$ and $m_q = m_c$ for clarity).}
\label{fig:qqbar_vs_alpha_QCD}
\end{figure}

\subsection{Heavy quarkonium}

\begin{figure}[t!]
  \includegraphics[width=\linewidth]{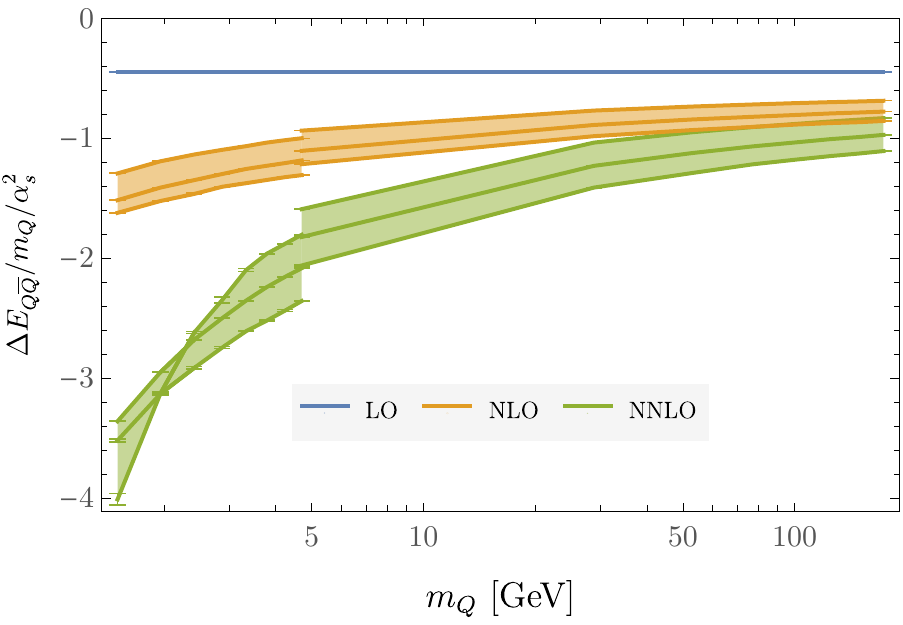}
\caption{Heavy quarkonium binding energy results as functions of $m_Q$. Fitted GFMC results are shown as points with error bars showing the statistical plus fitting systematic uncertainties discussed in the main text. Shaded bands connect results with renormalization scale choices $\mu \in \{\mu_p, 2\mu_p, \mu_p/2$\}. }
\label{fig:qqbar_vs_mQ_QCD}
\end{figure}

Results for the heavy quarkonium binding energy $\Delta E_{Q\overline{Q}}$ for the ranges of $\alpha_s$ above with $N_f = 4$ and $N_f = 5$ at LO, NLO, and NNLO in pNRQCD are obtained from fits to GFMC results as described above and shown as functions of $\alpha_s$ in Fig.~\ref{fig:qqbar_vs_alpha_QCD}.
At LO, the exact result $\Delta E_{Q\overline{Q}}^{(\text{LO})} / m_Q / \alpha_s^2 = -C_F^2 / 4$ is reproduced as discussed above.
At NLO and NNLO clear dependence on $\alpha_s$ can be seen in $\Delta E_{Q\overline{Q}} / m_Q / \alpha_s^2$.
For a Coulombic system, NLO corrections of $\mathcal{O}(\alpha_s)$ would lead to $\mathcal{O}(\alpha_s)$ and $\mathcal{O}(\alpha_s^2)$ corrections to the quarkonium binding energy.
Further corrections arise from the logarithmic differences between pNRQCD and Coulomb potentials, but as discussed in Sec.~\ref{sec:trial}, these differences are relatively mild for $\alpha_s \lesssim 0.3$ and the renormalization scale $\mu_p$ discussed above.
Quadratic fits to the NLO results in Fig.~\ref{fig:qqbar_vs_alpha_QCD} with constant terms fixed to $-C_F^2/4$ achieve $\chi^2/\text{dof} \sim 0.7$ for $N_f = 5$ results and $\chi^2/\text{dof} \sim 2.1$ for $N_f = 4$ results with $\mu = \mu_p$, indicating that logarithmic effects are not well-resolved for couplings in the $N_f = 5$ range but may be apparent for couplings in the $N_f = 4$ range.
Similarly, NNLO corrections to the potential should be approximately described by an $\mathcal{O}(\alpha_s^4)$ polynomial with constant term $-C_F^2/4$ and the same linear term as arises at NLO.
Fits of this form to the NNLO results in Fig.~\ref{fig:qqbar_vs_alpha_QCD} achieve $\chi^2/\text{dof} \sim 0.8$ for $N_f = 5$ results and $\chi^2/\text{dof} \sim 2.0$ for $N_f = 4$ results with $\mu = \mu_p$.
Performing analogous fits to results with $\mu = 2\mu_p$ leads to slightly better goodness-of-fit for $N_f = 4$ results with $\chi^2 / \text{dof} \sim 1$ and slightly worse goodness-of-fit with $\chi^2 / \text{dof} \sim 2$ for $N_f = 5$ results.
On the other hand, identical fits to results with $\mu = \mu_p/2$ achieve similar goodness of fit for $N_f = 5$ results, and unacceptably bad $\chi^2 / \text{dof}$ for NNLO results at $N_f = 4$.
These results suggest that the choice $\mu = \mu_p$ is effective at minimizing the size of logarithmic effects over the range of $m_Q \in [m_c,m_t]$ and in particular that large logarithmic effects arise for $m_Q \sim m_c$ and $\mu = \mu_p / 2$.

\begin{center}
\begin{table*}[tb]
\centering
\begin{tabular}{| c | c | c | c | c | c | c |} 
 \hline
 $1S$ mesons & Order & $\alpha_s(\mu)$  & $m_Q$ & $\chi^2/{\rm dof}$  &   $M_{Q\bar{Q}}$ & Measured $M_{Q\bar{Q}}$~\cite{ParticleDataGroup:2022pth} \\ [0.5ex]
 \hline\hline
  $(J/\psi,\eta_c)$ & LO (exact)  & 0.282678 & 1.56206 & {} & 3.06865 & 3.06865(10)   \\
$(J/\psi,\eta_c)$ & NLO & 0.313613 & 1.65413 & 1.1 & 3.0684(3) & 3.06865(10) \\
$(J/\psi,\eta_c)$ & NNLO & 0.297100 & 1.77159 & 0.8 & 3.0690(4) & 3.06865(10) \\
  $(\Upsilon,\eta_b)$ & LO (exact) & 0.214850 & { 4.77041} & {} & 9.44295 & 9.44295(90)   \\
 $(\Upsilon,\eta_b)$ & NLO & 0.227325 & 4.86831 & 1.1 & 9.4430(5) & 9.44295(90) \\
 $(\Upsilon,\eta_b)$ & NNLO  & 0.222492 & 4.96974 & 1.2 & 9.4422(5) & 9.44295(90) \\
 \hline
 \hline
\end{tabular}
\caption{Spin-averaged ${}^{1}S$ heavy quarkonium masses computed in this work for $c\overline{c}$ and $b\overline{b}$ systems are compared with experimental results. The errors quoted in the $M_{Q\overline{Q}}$ column show combined statistical and fitting systematic uncertainties (LO results are exact). The quark masses shown in the $m_Q$ column are tuned in order to achieve agreement between calculated and measured masses. The quoted $\chi^2/\text{dof}$ is a weighted average (using the weights in Eq.~\eqref{eq:weights} of the individual $\chi^2/\text{dof}$ from each fit to GFMC results performed as described in the main text.}
    \label{tab:mesonmass}
\end{table*}
\end{center}

The same results for $\Delta E_{Q\overline{Q}}/m_Q /\alpha_s^2$ at each order of pNRQCD and with $\mu \in \{\mu_p, 2\mu_p, \mu_p/2\}$ are shown as functions of $m_Q$ in Fig.~\ref{fig:qqbar_vs_mQ_QCD}.
Large differences are visible between LO and NLO results, with smaller but still significant differences between NLO and NNLO results.
The (exact) LO result is independent of the renormalization scale, $\Delta E_{Q\overline{Q}}^{(\text{LO})} /m_Q / \alpha_s^2 = -C_F^2/4$.
Non-trivial dependence on the renormalization scale enters at NLO.
The dependence on the renormalization scale is somewhat more significant at NNLO, 
with a sharp increase in $\Delta E_{Q\overline{Q}}^{(\text{NNLO})}/m_Q / \alpha_s^2$ at small $m_Q$ arising with $\mu = \mu_p / 2$.

\begin{figure}
\centering
  \subfigure{\includegraphics[width=\linewidth]{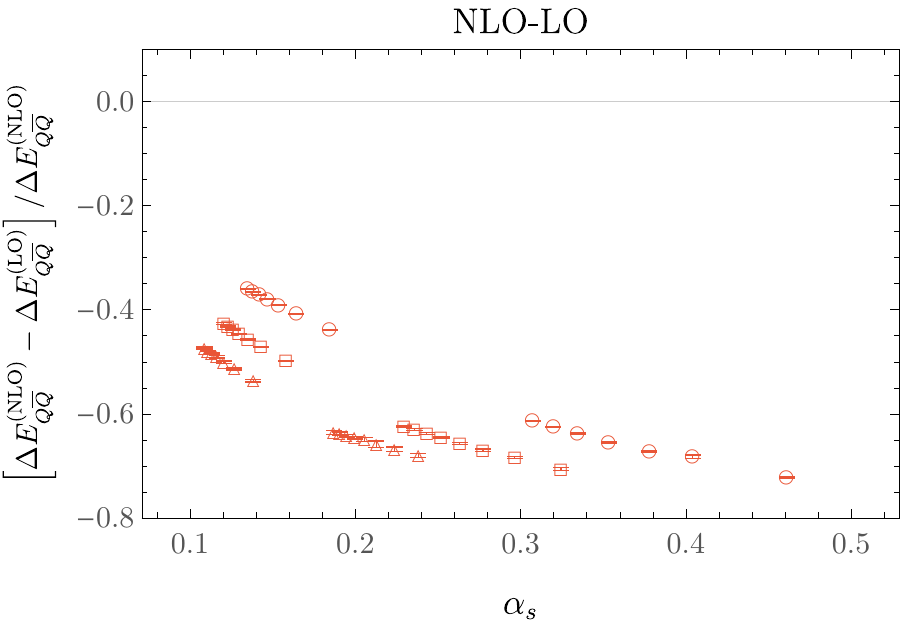}}
  \subfigure{\includegraphics[width=\linewidth]{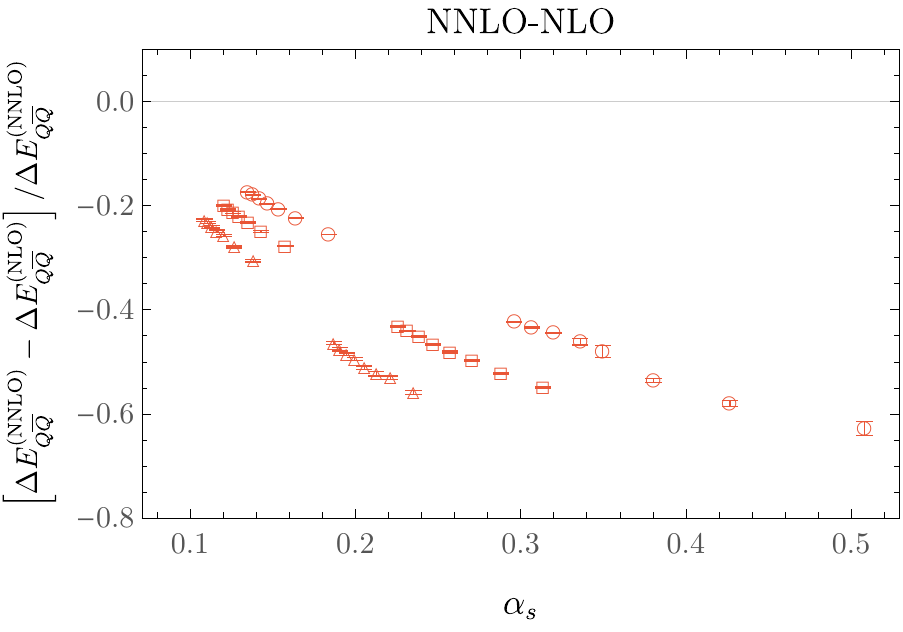}}
\caption{Relative differences between heavy quarkonium binding energies calculated at different orders of pNRQCD (excluding points with $N_f=5$ and $m_q = m_c$ for clarity).}
\label{fig:qqbar_pert_diff}
\end{figure}

The relative sizes of differences in quarkonium binding energies computed at different orders of pNRQCD are shown in Fig.~\ref{fig:qqbar_pert_diff}.
Large differences of 40-70\% are seen between LO and NLO over the range of $\alpha_s$ studied here.
Smaller but still significant differences of 20-50\% are seen between NLO and NNLO results.
This suggests that the perturbative expansion in $\alpha_s(\mu_p)$ does not converge rapidly over the range of $m_Q$ studied here, and even for $m_Q \sim m_t$, NLO and NNLO effects on the relation between $\Delta E_{Q\overline{Q}}$ and $m_Q$ are still 40\% and 20\% of LO results respectively.

These results for $\Delta E_{Q\overline{Q}}$ do not provide physical predictions until $m_Q$ has been specified. The parameter $m_Q$ appearing in the pNRQCD Lagrangian is a pole mass that can be fixed once it is related to a known observable.
Perturbation theory generally leads to slowly converging relations between pole mass definitions and physical observables due to infrared renormalon ambiguities~\cite{Beneke:1994sw,Beneke:1998ui}.
Better convergence can be expected for predictions of relationships between physical observables where renormalon effects cancel.
We, therefore, use the nonperturbative (in terms of treatment of the potential) results for $\Delta E_{Q\overline{Q}}$ provided by the GFMC calculations above to relate $M_{Q\overline{Q}}$ and $m_Q$ at each order of pNRQCD.
In particular, we define $m_c$ and $m_b$ by the values of $m_Q$ for which $M_{Q\overline{Q}}$ agrees with experimental determinations of the spin-averaged quarkonium mass combinations $M_{Q\overline{Q}} = 3/4 M_{Q\overline{Q}}^{{}^3S_1} + M_{Q\overline{Q}}^{{}^1S_0}$.
An iterative tuning procedure is used to determine $m_b$, and $m_c$ in which fits to the GFMC results above are used to provide initial guesses for the masses that are then refined by performing additional GFMC calculations with the current best-fit $m_b$ and $m_c$ and then re-fitting including these results.
This is repeated until the procedure has converged within our GFMC statistical uncertainties, which leads to the values of $m_b$ and $m_c$ at each order of pNRQCD shown in Table~\ref{tab:mesonmass}.
Large order-by-order shifts in the values of $m_b$ and $m_c$ needed to reproduce experimental quarkonium results are seen, as expected from the poor perturbative convergence of relations between quark pole masses and quarkonium masses.
Analogous effects arise in relations between quark pole masses and other hadron masses. With $m_b$ and $m_c$ fixed to reproduce quarkonium masses, further pNRQCD hadron mass predictions are effectively relations between hadron masses that should have better convergence than the relations between the individual hadron masses and the quark pole masses.

\subsection{Triply-heavy baryons}\label{sec:qcd_baryons}

\begin{figure}[t!]
  \centering
  \subfigure{\includegraphics[width=\linewidth]{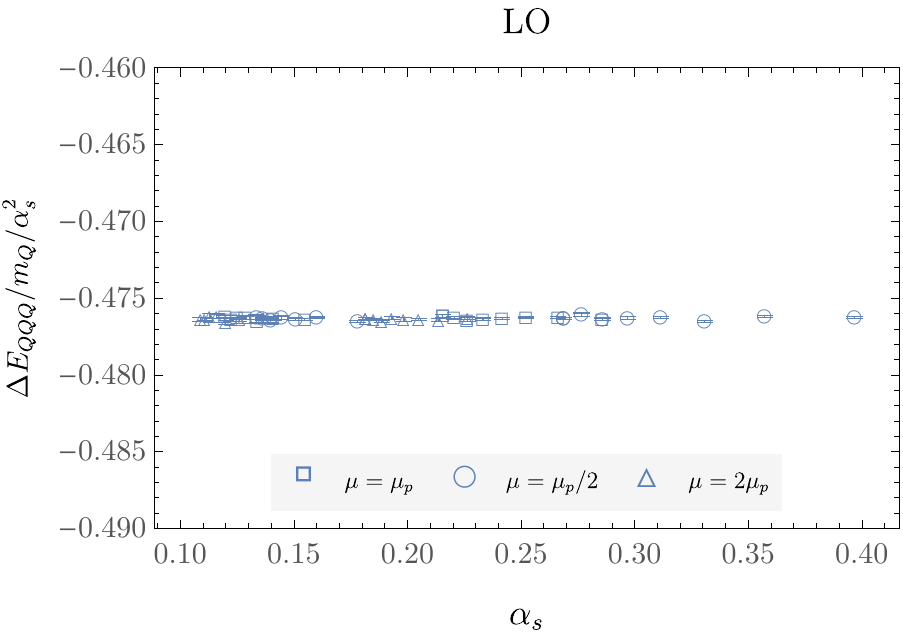}}
  \subfigure{\includegraphics[width=\linewidth]{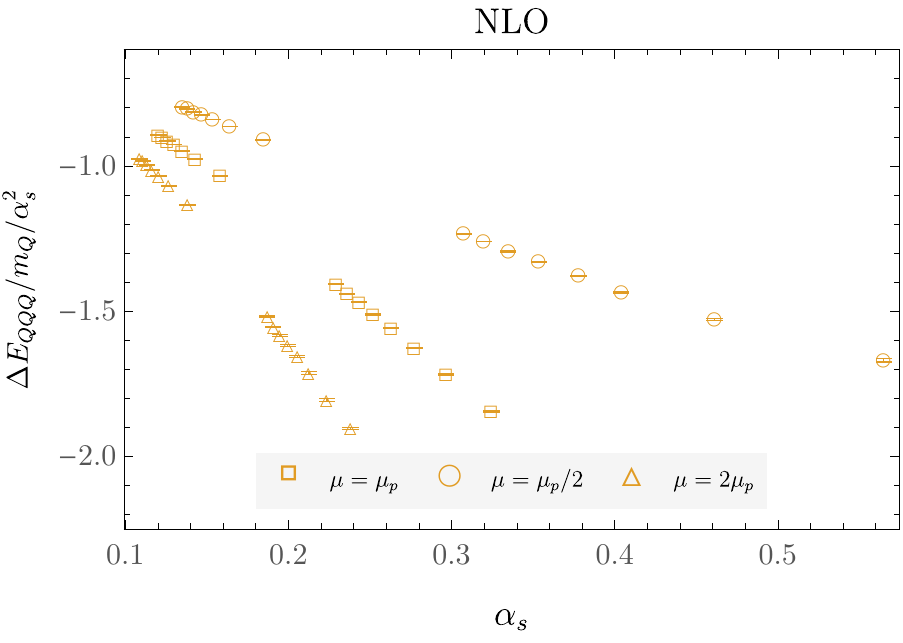}}
  \subfigure{\includegraphics[width=\linewidth]{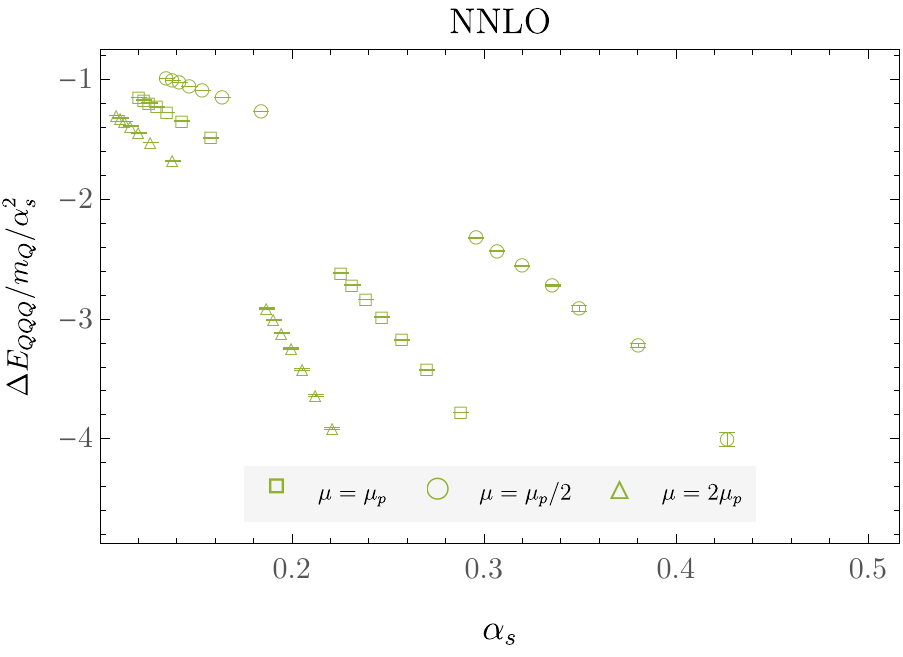}}
\caption{Triply-heavy baryon binding energy results as functions of $\alpha_s$ (excluding points with $N_f=5$ and $m_q = m_c$ for clarity).}
\label{fig:qqq_vs_alpha_QCD}
\end{figure}

\begin{figure}[t!]
  \includegraphics[width=\linewidth]{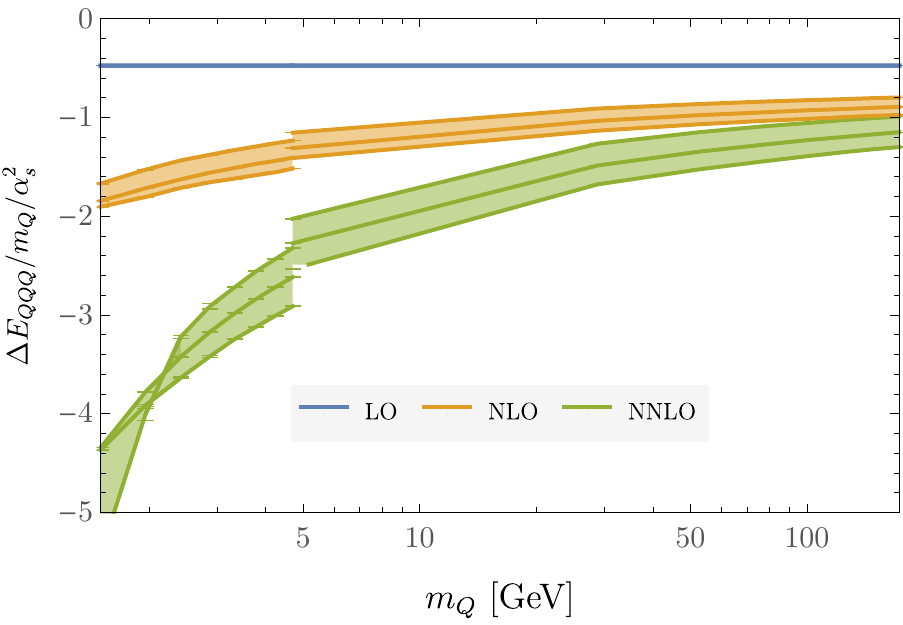}
\caption{Triply-heavy baryon binding energy results as functions of $m_Q$ with shaded bands connecting results with renormalization scale choices $\mu \in \{\mu_p, 2\mu_p, \mu_p/2$\}. }
\label{fig:qqq_vs_mQ_QCD}
\end{figure}

Results for triply-heavy baryon binding energies $\Delta E_{QQQ}$ over the same ranges of $\alpha_s$ with $N_f = 4$ and $N_f=5$ are shown in 
Fig~\ref{fig:qqq_vs_alpha_QCD}.
The same results for $\Delta E_{QQQ}/m_Q /\alpha_s^2$ at each order of pNRQCD and with $\mu \in \{\mu_p, 2\mu_p, \mu_p/2\}$ are shown as functions of $m_Q$ in Fig.~\ref{fig:qqq_vs_mQ_QCD}.
The order-by-order differences in the relation between $\Delta E_{QQQ}/m_Q /\alpha_s^2$ and $m_Q$ are similar to the case of heavy quarkonium discussed above.
Although LO results are not exactly renormalization scale independent for baryons, numerical results are found to be scale independent to better than 0.1\% precision.
Visible scale dependence appears at NLO, with slightly large scale dependence appearing at NNLO.

\begin{figure}
\centering
  \subfigure{\includegraphics[width=\linewidth]{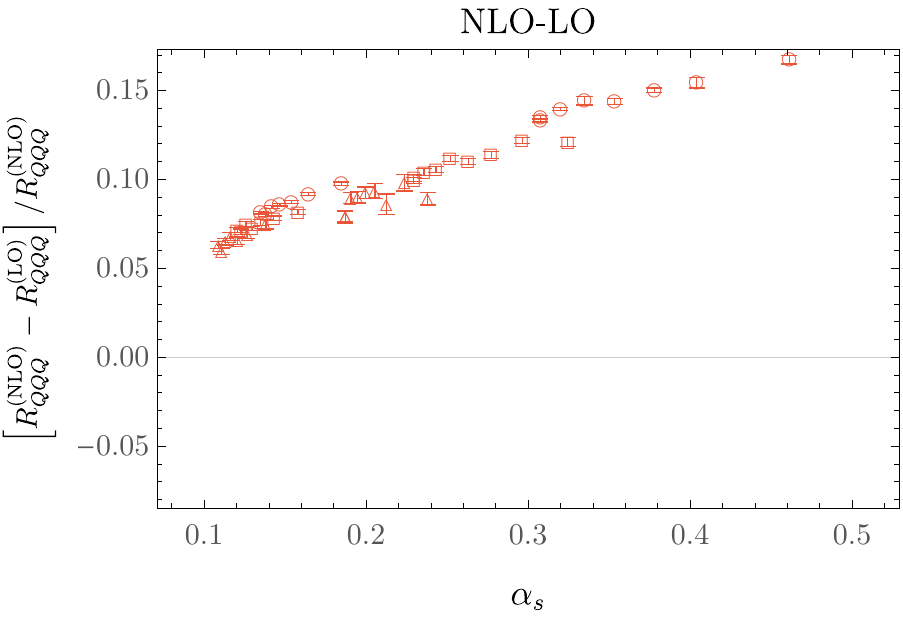}}
  \subfigure{\includegraphics[width=\linewidth]{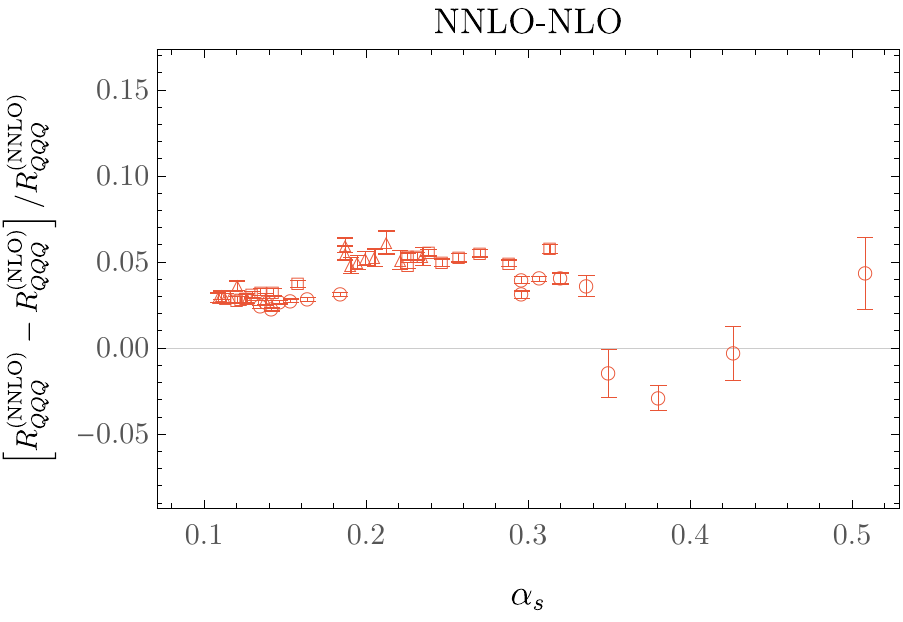}}
\caption{Relative differences between triply-heavy baryon binding energies calculated at different orders of pNRQCD (excluding points with $N_f=5$ and $m_q = m_c$ for clarity).}
\label{fig:qqq_pert_diff}
\end{figure}

The similarities between Fig.~\ref{fig:qqbar_vs_mQ_QCD} and Fig.~\ref{fig:qqq_vs_mQ_QCD} suggest that the large order-by-order shifts in the relations between the pole mass $m_Q$ and both the quarkonium and baryon masses are highly correlated and that predictions of the ratio of the baryon and quarkonium binding energies as a function of $m_Q$ have much better perturbative convergence than either binding energy individually.
This is confirmed by directly calculating the perturbative differences of these ratios shown in Fig.~\ref{fig:qqq_pert_diff}.
Although both quarkonium and baryon binding energies individually have 40-70\% differences between LO and NLO over the range of $\alpha_s$ studied here, the corresponding change in the ratio of baryon and meson binding energies,
\begin{equation}
  R_{QQQ} \equiv \frac{\Delta E_{QQQ} }{ \Delta E_{Q\overline{Q}}},
\end{equation}
is 5-10\%.
Similarly, both quarkonium and baryon binding energies have 20-50\% differences between NNLO and NLO, but $R_{QQQ}$ differences by only 3-8\%.

\begin{figure}
\centering
  \includegraphics[width=\linewidth]{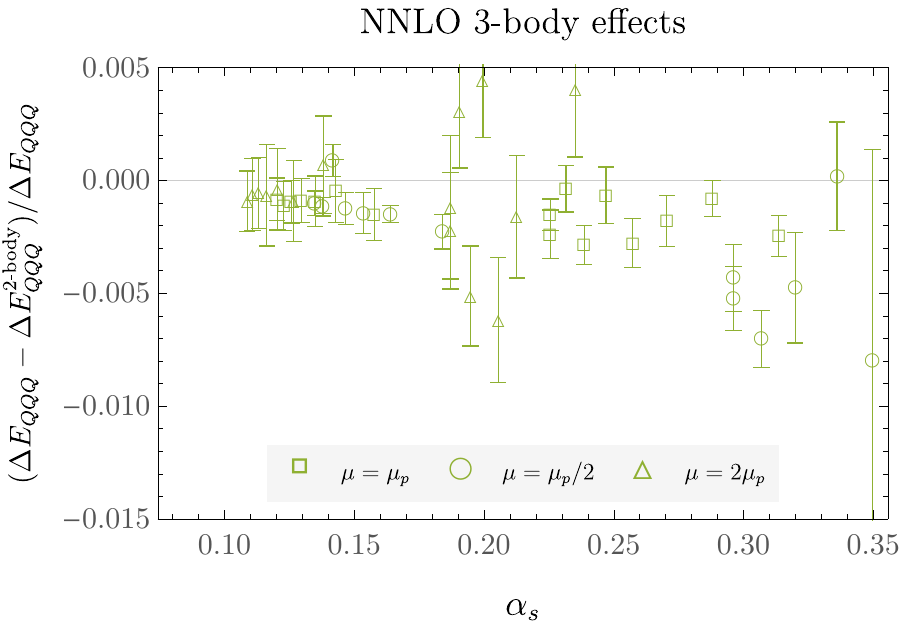}
\caption{Relative differences between triply-heavy baryon binding energies calculated using NNLO two-quark potentials only and full NNLO results including both two- and three-quark potentials.}
\label{fig:qqq_3body}
\end{figure}

It is further possible to separate the contributions to $\Delta E_{QQQ}$ arising from three-quark potentials from those arising from quark-quark potentials only.
The effects of three-body potentials, which first arise at NNLO, are isolated by performing GFMC calculations using only the NNLO quark-quark potentials and taking the difference with results obtained with three-quark potentials included.
The relative size of this difference is shown as a function of $\alpha_s$ in Fig.~\ref{fig:qqq_3body}.
Interestingly, including three-body potentials leads to sub-percent changes to NNLO heavy baryon binding energies for $\alpha_s \lesssim 0.3$, which is much smaller than the overall difference between NLO and NNLO binding energies.
Still, three-body potential effects of around 0.25\% - 1\% of NNLO binding energy results are well-resolved from zero and seen to lower baryon masses in comparison with results obtained using only quark-quark potentials, as expected since the color-antisymmetric three-quark potential is attractive.

\begin{figure}[t!]
  \centering
  \includegraphics[width=\linewidth]{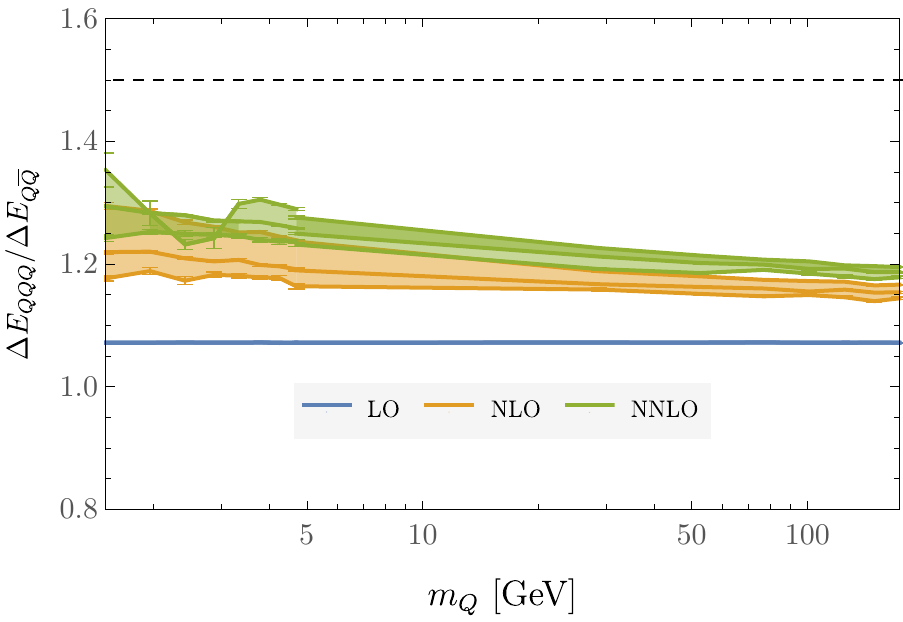}
\caption{Ratios of triply-heavy baryon and heavy quarkonium binding energy results as functions of $m_Q$ with shaded bands connecting results with renormalization scale choices $\mu \in \{\mu_p, 2\mu_p, \mu_p/2$\}. }
\label{fig:ratios_qcd}
\end{figure}

The binding-energy ratio $R_{QQQ}$ results are shown in Fig.~\ref{fig:ratios_qcd}. It is clear that $R_{QQQ}$ is approximately independent of $m_Q$ over the entire range of quark masses studied here.
At LO, constant fits to GFMC results with $N_f = 5$ and $\mu = \mu_p$ give
\begin{equation}
  R_{QQQ}^{(\text{LO})}  \approx 1.0717(1),
  \label{eq:LOratio}
\end{equation}
with $\chi^2 / \text{dof} = 1.6$ with consistent results obtained for other choices of $\mu$ and for $N_f=4$.
Beyond LO, mild $m_Q$ dependence can be resolved in $R_{QQQ}$ that can be described by an $\mathcal{O}(\alpha_s)$ linear correction.
At NLO, a linear fit to GMFC results with $N_f = 5$ and $\mu = \mu_p$ gives
\begin{equation}
  R_{QQQ}^{(\text{NLO})}  \approx 1.114(3) + 0.33(2) \alpha_s, \label{eq:NLOratio}
\end{equation}
with $\chi^2 / \text{dof} = 1.0$.
Results with other choices of $\mu$ lead to consistent constant terms with $\mathcal{O}(\alpha_s)$ terms ranging from 0.31 - 0.4.
Fits to $N_f = 4$ results are consistent with $N_f = 5$ results but have larger uncertainties and somewhat worse $\chi^2 / \text{dof} \sim 2$.
At NNLO, an analogous linear fit to $N_f =5$ results with $\mu = \mu_p$ gives
\begin{equation}
  R_{QQQ}^{(\text{NNLO})}  \approx 1.116(2) + 0.60(2) \alpha_s, \label{eq:NNLOratio}
\end{equation}
with $\chi^2 / \text{dof} = 1.4$. Other NNLO results are generally described similarly or slightly worse by linear fits.
However, NNLO results with $N_f = 4$ and $\mu = \mu_p/2$ show nonlinear features for small $m_Q$ in Fig.~\ref{fig:ratios_qcd} that are not accurately described by an $\mathcal{O}(\alpha_s)$ polynomial, which is not surprising because the corresponding results for $\Delta E_{Q\overline{Q}}$ and $\Delta E_{QQQ}$ show evidence for significant non-Coulombic effects at small $m_Q$.

These pNRQCD results can be compared with general constraints from QCD inequalities.
The Weingarten inequality, $M_N \geq m_\pi$~\cite{Weingarten:1983uj}, was extended by Detmold to $M_N \geq 3/2 m_\pi$~\cite{Detmold:2014iha} by showing that all maximal isospin multi-meson interactions are repulsive or vanishing at threshold and do not lead to bound states. 
The same arguments apply for $Q\overline{Q}$ multi-meson states if quark-antiquark annihilation is neglected because identical patterns of quark contractions arise in this case as for $u\overline{d}$.
Since neglecting $Q\overline{Q}$ annihilation is a valid approximation for heavy quarks up to $\mathcal{O}(1/m_Q^{2})$~\cite{Pineda:1998kj}, the corresponding heavy-quark meson and baryon mass inequality is
\begin{equation}
  M_{QQQ} \geq \frac{3}{2} M_{Q\overline{Q}} + \mathcal{O}(1/m_Q^{2}). 
\end{equation}
These bounds can be directly compared with the pNRQCD results obtained here.
Comparisons can also be made at the level of the meson and baryon binding energies, since 
\begin{equation}
\frac{M_{QQQ}}{M_{Q\overline{Q}}} = \frac{3 m_Q + \Delta E_{QQQ}}{2 m_Q + \Delta E_{Q\overline{Q}}} \geq \frac{3}{2}, \label{eq:Detmold}
\end{equation}
leads after multiplying by $M_{Q\overline{Q}}$ to
\begin{equation}
\frac{\Delta E_{QQQ}}{\Delta E_{Q\overline{Q}}} \leq \frac{3}{2}, \label{eq:DeltaDetmold}
\end{equation}
where $M_{Q\overline{Q}} > 0$ and  $\Delta E_{Q\overline{Q}} < 0$ have been assumed when forming ratios and $\mathcal{O}(1/m_Q^2)$ effects have been neglected.
Note that Eq.~\eqref{eq:Detmold} is necessarily saturated as $m_Q \rightarrow \infty$, where $\alpha_s$ at scales proportional to $m_Q$ vanishes and therefore $M_{QQQ} \rightarrow 3 m_Q$, $M_{Q\overline{Q}} \rightarrow 2 m_Q$, and $M_{QQQ}/M_{Q\overline{Q}} \rightarrow 3/2$.
However, the lack of saturation of Eq.~\eqref{eq:DeltaDetmold} for arbitrary $m_Q$ implies that the saturation of Eq.~\eqref{eq:Detmold} is only logarithmic as $m_Q \rightarrow \infty$.
Eqs.~\eqref{eq:LOratio}-\eqref{eq:NNLOratio} show that $\Delta E_{QQQ}/\Delta E_{Q\overline{Q}}$ is predicted to be 72-74\% of the way to saturating the Detmold inequality at LO-NNLO in pNRQCD, demonstrating that in the $m_Q \rightarrow \infty$ limit baryons in QCD are almost but not entirely as bound as is allowed by the positivity of the QCD path integral measure.

\begin{center}
\begin{table*}[tb]
\centering
\begin{tabular}{|c | c | c | c | c |} 
 \hline
Baryon & This work: $M_{QQQ}$ & This work: $\chi^2/{\text{dof}}$  & Variational Methods ($M_{QQQ}$)  & Lattice QCD ($M_{QQQ}$) \\ [0.5ex]
 \hline\hline
 $\Omega_{ccc}$ & \begin{tabular}{@{}c@{}} LO: 
4.62670(2)\\ NLO: 4.6718(7)\\ NNLO: 
 4.7070(9)\end{tabular} & \begin{tabular}{@{}c@{}} LO: 
 1.0\\ NLO: 1.4\\ NNLO: 
0.9\end{tabular} & \begin{tabular}{@{}c@{}} LO: 4.76(6)~\cite{Jia:2006gw}  \\ NNLO+mNLO: 4.97(20)~\cite{Llanes-Estrada:2011gwu} \end{tabular}  &  4.796(8)(18) ~\cite{Brown:2014ena}  \\\hline
 $\Omega_{ccb}$ & \begin{tabular}{@{}c@{}} LO: 
 7.81522(2)\\ NLO: 7.8667(6) \\ NNLO: 
 7.919(1) \end{tabular} & \begin{tabular}{@{}c@{}} LO: 
 1.5\\ NLO: 1.1 \\ NNLO: 
 1.0 \end{tabular} & \begin{tabular}{@{}c@{}} LO: 7.98(7)~\cite{Jia:2006gw}  \\ NNLO+mNLO: 8.20(15)~\cite{Llanes-Estrada:2011gwu} \end{tabular}  &
 \begin{tabular}{@{}c@{}} 8.007(9)(20)~\cite{Brown:2014ena}   \\ 8.005(6)(11)~\cite{Mathur:2018epb}  \end{tabular}
 \\\hline
 $\Omega_{cbb}$ & \begin{tabular}{@{}c@{}} LO: 
 11.03593(2) \\ NLO: 11.0957(8)\\ NNLO: 
 11.116(1) \end{tabular} & \begin{tabular}{@{}c@{}} LO: 
 1.3 \\ NLO: 1.1 \\ NNLO: 
 1.0 \end{tabular} & \begin{tabular}{@{}c@{}} LO: 11.48(12)~\cite{Jia:2006gw}  \\ NNLO+mNLO: 11.34(26)~\cite{Llanes-Estrada:2011gwu} \end{tabular}  & 
 \begin{tabular}{@{}c@{}} 11.195(8)(20)~\cite{Brown:2014ena}    \\ 11.194(5)(12)~\cite{Mathur:2018epb}
 \end{tabular}
 \\\hline
 $\Omega_{bbb}$ & \begin{tabular}{@{}c@{}} LO: 
 14.20641(3)\\ NLO: 
 14.2573(7)\\ NNLO:
 14.287(1)\end{tabular} & \begin{tabular}{@{}c@{}} LO: 
1.2\\ NLO: 
 1.4\\ NNLO: 
 1.4\end{tabular} & \begin{tabular}{@{}c@{}} LO: 14.76(18)~\cite{Jia:2006gw}  \\ NNLO+mNLO: 14.57(25)~\cite{Llanes-Estrada:2011gwu} \end{tabular} & \begin{tabular}{@{}c@{}} 
 14.371(4)(12)~\cite{Meinel:2010pw}  \\ 14.366(9)(20)~\cite{Brown:2014ena} \\ 14.366(7)(9)~\cite{Mathur:2022nez}
 \end{tabular}  \\
 \hline
 \hline
\end{tabular}
\caption{Comparison of the triply-heavy baryon mass results obtained here with results from other pNRQCD and LQCD calculations. All masses are given in GeV and obtained using $\alpha_s$ and $m_Q$ from Table~\ref{tab:mesonmass} and the $\chi^2 / \text{dof}$ correspond to weighted averages analogous to the quarkonium results. }
    \label{tab:baryonmass}
\end{table*}
\end{center}

\begin{figure}[t!]
  \centering
  \subfigure{\includegraphics[width=\linewidth]{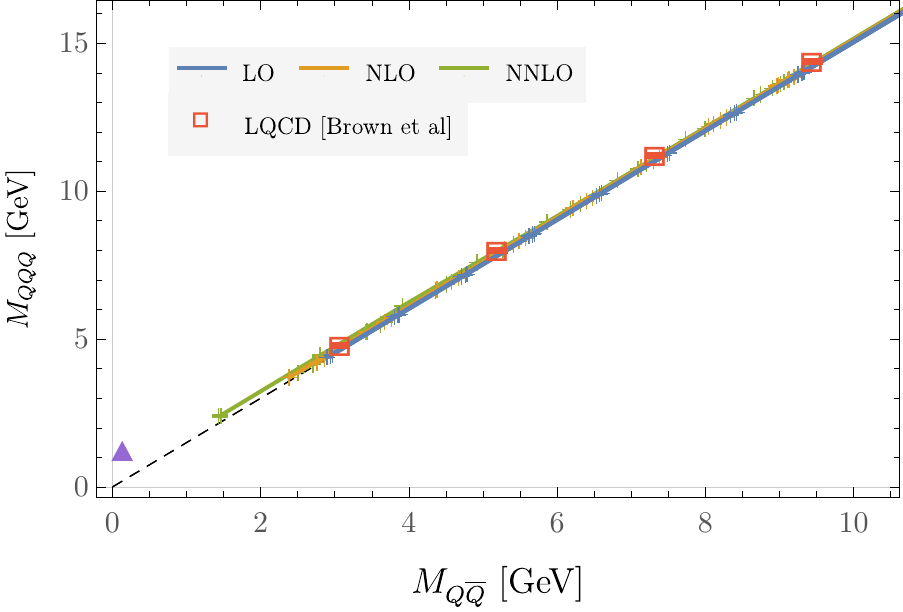}}
  \subfigure{\includegraphics[width=\linewidth]{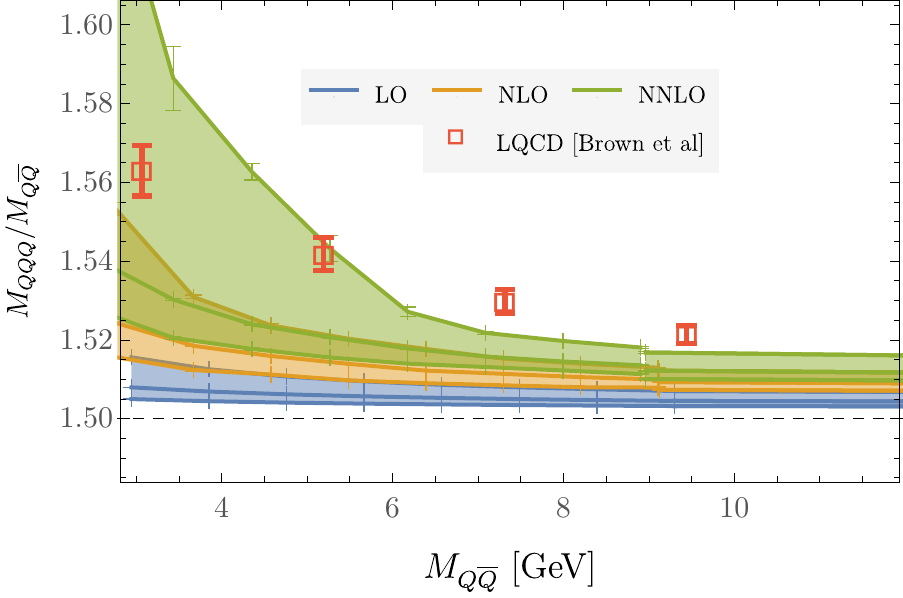}}
\caption{The top panel shows triply-heavy baryon masses $M_{QQQ}$ as functions of $M_{Q\overline{Q}}$ with renormalization scale choices $\mu \in \{\mu_p, 2\mu_p, \mu_p/2$\}. The bottom panel shows the ratio $M_{QQQ} / M_{Q\overline{Q}}$ analogously. The LQCD results of Ref.~\cite{Brown:2014ena} calculated using NRQCD including $\mathcal{O}(1/m_Q^2)$ effects are shown for comparison as red points with error bands showing total statistical plus systematic uncertainties. Experimental results for $M_N / m_\pi$ are also shown for reference on the top panel as a purple triangle.}
\label{fig:boundQCD}
\end{figure}

Precise pNRQCD predictions for $ccc$, $ccb$, $bbc$, and $bbb$ baryon masses can be made using the values of $m_c$ and $m_b$ tuned to reproduce $M_{c\overline{c}}$ and $M_{b\overline{b}}$ and given in Table~\ref{tab:mesonmass}.
Due to the exchange symmetry of the Coulomb trial wavefunctions used here, $\nabla^2_I \Psi_T(\bm{r}_1,\bm{r}_2,\bm{r}_3)$ is independent of $a$. The correct kinetic-energy operator for $bbc$ baryons is therefore obtained by considering three equal-mass quarks with mass equal to
\begin{equation}
  m_{bbc} = \frac{3}{2}\left( \frac{1}{2m_b} + \frac{1}{m_c} \right)^{-1} = \frac{3 m_b m_c}{2m_b + m_c}. \label{eq:mbbc}
\end{equation}
An analogous $ccb$ reduced mass $m_{ccb}$ is obtained by taking $b \leftrightarrow c$ in Eq.~\eqref{eq:mbbc}.
Corresponding renormalization scales are defined as usual and for example $\mu_{bbc} = 4 \alpha_s(\mu_{bbc}) m_{bbc}$.
GFMC results for triply-heavy baryon masses using $m_Q \in \{m_c, m_{ccb}, m_{bbc}, m_{bbb}\}$ therefore lead to pNRQCD predictions for $\Omega_{ccc}$, $\Omega_{ccb}$, $\Omega_{bbc}$, and $\Omega_{bbb}$ baryon masses shown in Table~\ref{tab:baryonmass}.
These pNRQCD predictions are compared with LQCD results~\cite{Brown:2014ena, Meinel:2010pw,Mathur:2018epb,Mathur:2022nez} for these baryon masses and found to underpredict LQCD by about 200 MeV at LO and 100 MeV at NNLO for all baryon masses considered.
The differences between NNLO and NLO results are significantly smaller than those between NLO and LO results, suggesting good convergence for the $\alpha_s$ expansion of the pNRQCD potential.
The remaining differences between NNLO and LQCD results likely arise primarily from the $1/m_Q$ effects neglected in this work.
In particular, the calculations of $\Omega_{bbb}$ in Refs.~\cite{Brown:2014ena, Meinel:2010pw,Mathur:2022nez} employ lattice NRQCD actions with $\mathcal{O}(1/m_Q^2)$ terms included, and therefore the differences in $\Omega_{bbb}$ mass predictions must arise from $\mathcal{O}(1/m_Q)$, $\mathcal{O}(1/m_Q^2)$, and higher-order $\alpha_s$ corrections.
Relative differences { between pNRQCD and LQCD} baryon mass predictions decrease with increasing quark mass as roughly $1/m_Q$ and at NNLO ranges from 2 \% for the $\Omega_{ccc}$ to 0.7 \% for the $\Omega_{bbb}$.
It is noteworthy that our GFMC pNRQCD predictions have 10-100 times smaller statistical uncertainties than LQCD results with both relativistic and NR quark actions; however, it is clear that systematic uncertainties from neglected effects in the pNRQCD potential are much larger than statistical uncertainties in either case and require the inclusion of $1/m_Q$ effects to be reduced.

A further measure of the size of systematic uncertainties arising from perturbative truncation effects is provided by comparing predictions for heavy baryon and meson masses with different choices of $\mu \in \{\mu_p, 2\mu_p, \mu_p/2\}$.
As seen in Fig.~\ref{fig:boundQCD}, the perturbative convergence of $M_{QQQ}/M_{Q\overline{Q}}$ as a function of $M_{Q\overline{Q}}$ is better than the convergence of either mass individually, and differences between different scales are reduced.
However, significant $\mu$ dependence arises at NNLO for $m_Q \sim m_c$ due to the nonlinear dependence of both $M_{Q\overline{Q}}$ and $M_{QQQ}$ on $\alpha_s$ for $\mu = \mu_p / 2$ with relatively small $m_Q$.
Since $m_Q$ does not enter this comparison, it is straightforward to compare to LQCD results, and the differences between NNLO pNRQCD results and LQCD results are seen to be comparable to the differences between pNRQCD results with different $\mu$ choices. 
Both pNRQCD and LQCD results obey Eq.~\eqref{eq:Detmold}.

\section{Dark hadrons}
\label{sec:bsm}

Inspired by the stability of the proton, a dark sector with non-Abelian gauge interactions can give rise to a stable, neutral dark matter candidate -- the dark baryon -- as reviewed in Refs.~\cite{Kribs:2016cew,Cline:2021itd,Cline:2016nab,DeGrand:2019vbx}.
A simple UV-complete model of dark baryons is a hidden $SU(N_c)$ dark sector with $N_c$ dark colors. If one includes dark quarks, then a dark QCD sector, charged under $SU(N_c)$ or $G_{{\text{SM}}}\times SU(N_c)$ with $n_d$ dark flavors arises. The pure hidden sector Lagrangian is then given by,
\begin{equation}
  \mathcal{L}_{D}=-\frac{1}{2} {\text{Tr}}G_{\mu\nu}^2+\sum_{i=1}^{n_d} \overline{Q_d^i} \left[ i{\slashed{D}}+m_{d}^i \right] Q_d^i
\end{equation}
with masses $m_d^i$ and coupling $\alpha_d=g_d^2/(4\pi)$, dark gauge fields $A_d$ and dark fermions, $Q_d^i$, and $D^{\mu}=\partial^{\mu}-ig_dA_d^{\mu,a}T^a$. 
A global $U(1)$ symmetry leads to a conserved dark baryon number and, therefore, the stability of dark baryons, denoted $B_d$ below.
A dark composite sector also arises naturally for BSM extensions in which the Higgs boson is composite~\cite{Andersen:2011yj,Bellazzini:2014yua}.

As in QCD, at renormalization scales $\mu$ well above the dark confinement scale, $\mu\gg\Lambda_d$, the perturbative relation, 
\begin{equation}
  \Lambda_d^{(\text{LO})} =\mu\exp{\left(-\frac{2\pi}{\beta_d\alpha_d(\mu)}\right)},
    \label{eqn:lamdark}
\end{equation}
defines the relationship between $\alpha_d$ and $\Lambda_d$ at the lowest order, where $\beta_d$ is the one-loop dQCD beta function, with analogous expressions arising at higher order~\cite{Bethke:2009jm}. For $n_d \ll 4N_c$ the theory is confining.
Below we consider $n_d = 1$ for simplicity and denote the dark quark mass as $m_d$.
In the regime $m_d\gg\Lambda_d$,
the pNRQCD formalism and numerical methods discussed above can be used to make reliable perturbative predictions for the masses, lifetimes, and other properties of hidden-sector composite particles referred to as dark hadrons below.

One can further weakly couple the dark sector to the visible sector in various ways, leading to direct detection signatures~\cite{Lee:2015gsa,Hambye:2009fg,Bai:2010qg}. 
If dark sector quarks are changed under parts of the SM, production, and decay of dark quarks can result in striking collider phenomenology~\cite{Kang:2008ea,Strassler:2006im,Han:2007ae}. 
If $m_d \sim \sqrt{s}$, dark fermions are frequently produced via Drell-Yan and other SM processes. 
If the dark quark mass is much larger than the dark confinement scale, $m_d \gg \Lambda_d$, the dark color strings do not fragment, and the dark fermions are bound by a dark color string for macroscopic distances. This results in exotic tracks, dependent on the SM charges of the dark fermions, which are unique and not producible by the SM alone~\cite{Kang:2008ea,Strassler:2006im}. 
Searches for such long-lived particles have been rapidly increasing at the LHC and beyond~\cite{Beacham:2019nyx,Alimena:2019zri}.

Lattice gauge theory calculations have been performed for dQCD models with several choices of $N_c$: $SU(2)$~\cite{Lewis:2011zb,Hietanen:2014xca,Hietanen:2013fya,Detmold:2014qqa,Detmold:2014kba}, $SU(3)$~\cite{LatticeStrongDynamicsLSD:2013elk}, $SU(4)$~\cite{LatticeStrongDynamicsLSD:2014osp}, and higher $N_c$~\cite{DeGrand:2012hd,DeGrand:2013nna,DeGrand:2016pur,DeGrand:2019vbx}, as well as other gauge groups including $SO(N_c)$ and $Sp(N_c)$~\cite{Lee:2022xbp,Hietanen:2012sz}. 
The primary challenge for using lattice gauge theory to explore dQCD is that there is a vast space of possibilities to explore depending on the gauge group and matter content~\cite{Kribs:2016cew}.
The utility of pNRQCD is that precise results can be obtained quickly with very modest computational resources, which enables scans over wide ranges of parameters such as $m_d$ and $N_c$.
The major downside of pNRQCD is its restriction to theories with dark quark masses $m_d \gg \Lambda_d$; however, there are phenomenologically viable dQCD models of DM that land firmly in this regime~\cite{Mitridate:2017oky,Asadi:2021yml,Asadi:2021pwo}.

\begin{figure*}[t!]
  \centering
  \subfigure{\includegraphics[width=.49\linewidth]{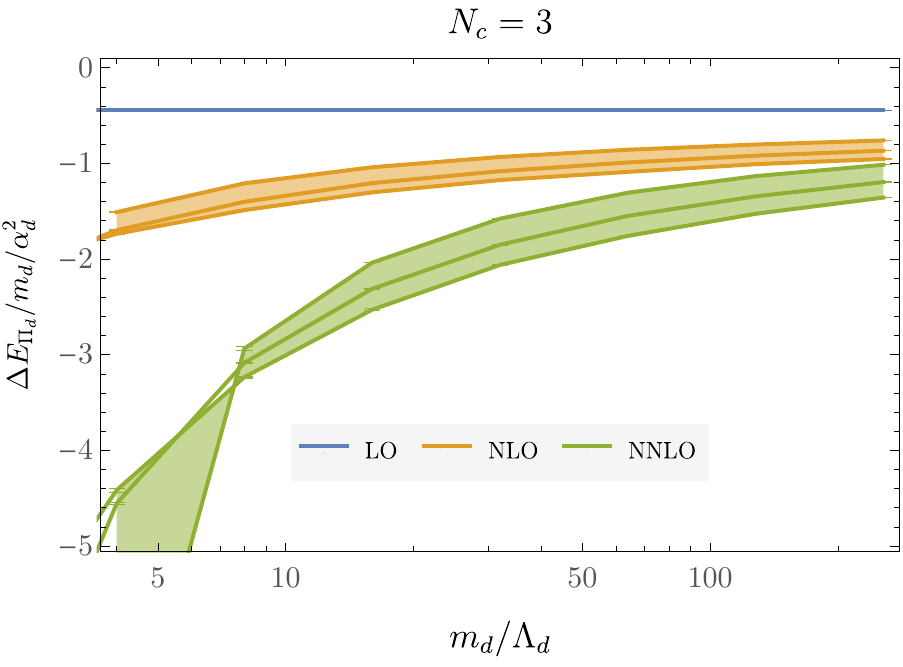}}
  \subfigure{\includegraphics[width=.49\linewidth]{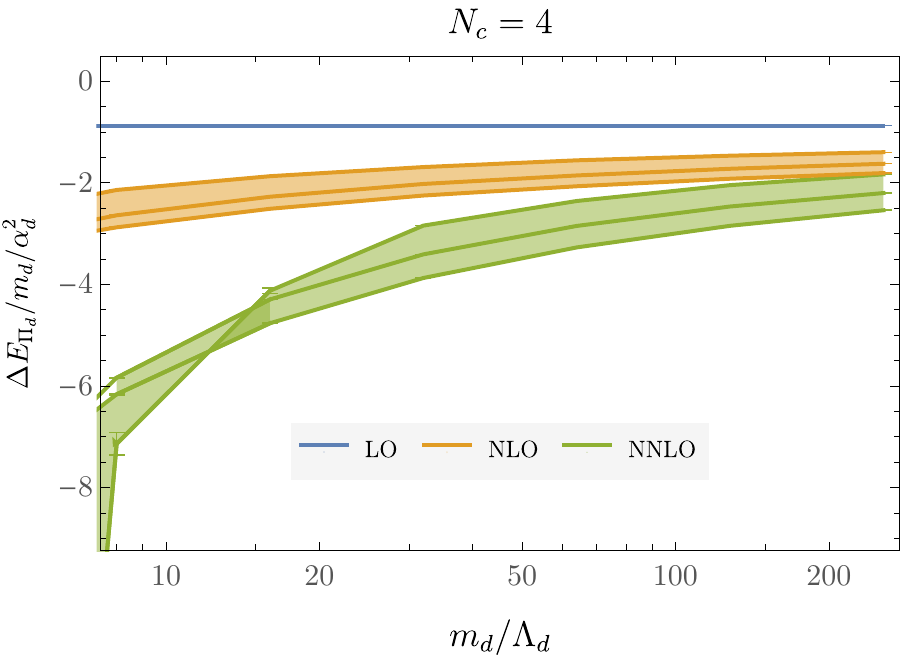}}\\
  \subfigure{\includegraphics[width=.49\linewidth]{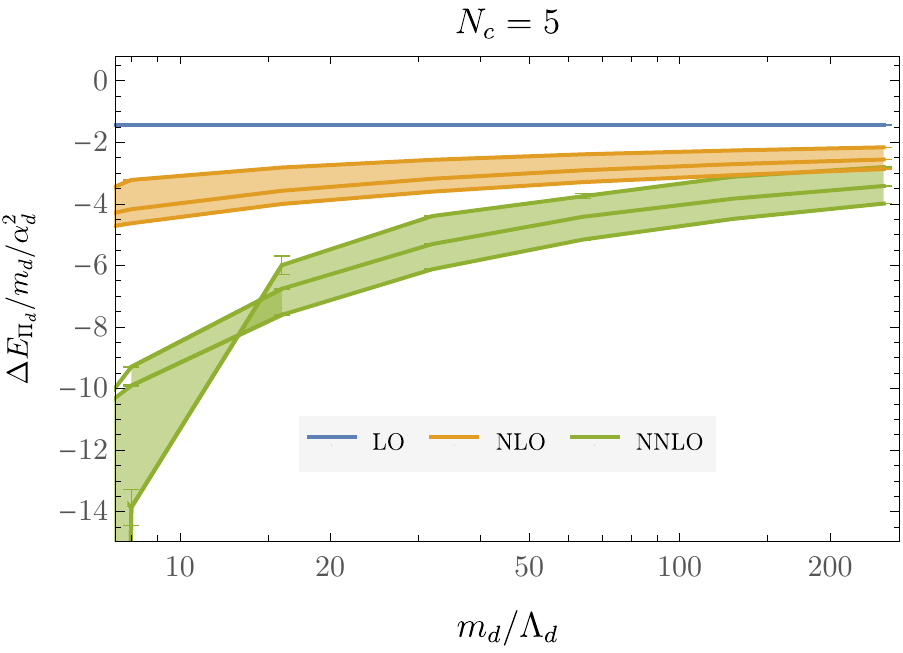}}
  \subfigure{\includegraphics[width=.49\linewidth]{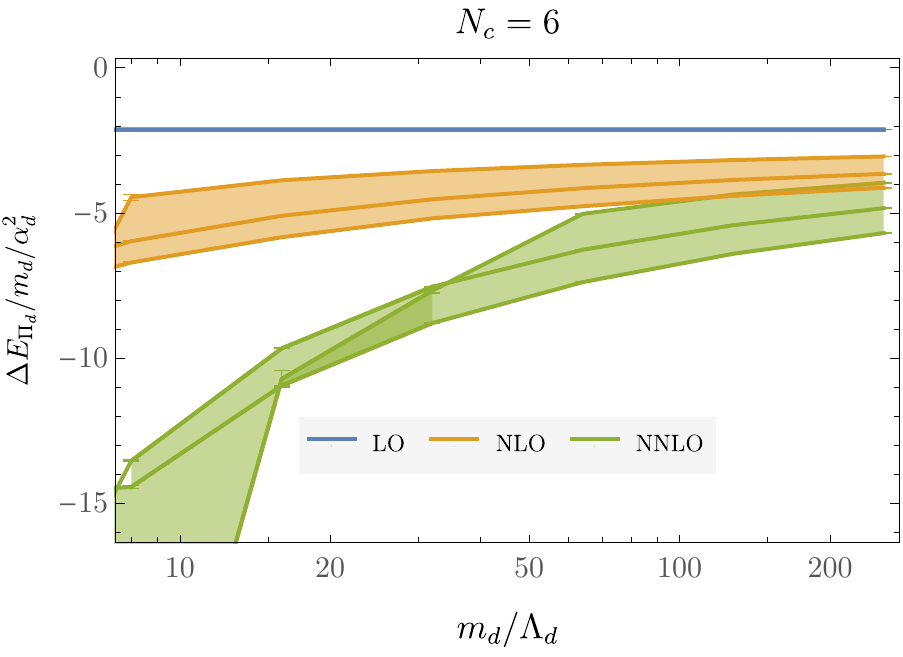}}
  \caption{Dark meson binding energy results as functions of $m_d/\Lambda_d$ with shaded bands connecting results with renormalization scale choices $\mu \in \{\mu_d, 2\mu_d, \mu_d/2$\} and $SU(N_c)$ gauge groups with $N_c \in \{3,\ldots,6\}$ as indicated. }
\label{fig:mesons_Nc3-6}
\end{figure*}

Models of composite DM with $m_d \ll \Lambda_d$ and models with $m_d \gg \Lambda_d$ have distinct phenomenological features.
In the regime $m_d\ll\Lambda_d$, if one assumes that all $B_d \overline{B}_d$ annihilation channels scale simply with the dark baryon mass $M_{B_d}$ as $\sigma v \sim 100/m_{B_d}^2$, then matching to thermal freezeout cross section~\cite{Steigman:2012nb,Antipin:2014qva}, requires cross-sections nearly as strong as allowed by unitarity and $m_{B_d}\sim 200\text{  TeV}$~\cite{Griest:1989wd,vonHarling:2014kha,Smirnov:2019ngs}.
In the heavy dark quark mass regime $m_d \gg \Lambda_d$, thermal freezeout occurs before the confinement transition in the dark sector. 
The confinement transition's subsequent dynamics involve trapping dark quarks inside pockets of the deconfined phase that significantly reduce the resulting DM relic abundance~\cite{Asadi:2021yml}.
Studies of the dynamics of this phase transition for the case of $N_c=3$ show that the correct relic abundance for dark baryons to account for all of DM can be achieved with $m_{d}/\Lambda_d \in [100,10^4]$ and in particular $m_{d} \in [1,100]$ PeV~\cite{Asadi:2021yml,Asadi:2021pwo}. 
This motivates more detailed studies of the dynamics and possible detection signatures of $SU(N_c)$ composite DM with $m_d \gg \Lambda_d$.

Dark baryon masses, $M_{B_d}$, and dark meson masses, $M_{\Pi_d}$, can be calculated for generic $SU(N_c)$ gauge theories in the $m_d \gg \Lambda_d$ regime using GFMC calculations of pNRQCD that are entirely analogous to the $SU(3)$ calculations above. 
These results can be used to relate these dark hadron observables to the dark-sector Lagrangian's fundamental parameters, particularly $m_d$ and $\alpha_d$.
Since the relation between the pole mass $m_d$ appearing in the pNRQCD Hamiltonian and observables such as hadron masses do not show good convergence in $\alpha_d$ as discussed for the QCD case above, these relations can then be used to replace dependence on $m_d$ with dependence on $M_{\Pi_d}$ in other dark hadron quantities and enable better-converging predictions relating different dark-sector observables.
Dependence on $\alpha_d$ can similarly be exchanged with dependence on $\Lambda_d$ using Eq.~\eqref{eqn:lamdark} and its higher-order analogs.
In particular, perturbative expansions for meson and baryon masses as functions of $N_c$ and $\alpha_d$ obtained by fitting to GFMC results are used below to predict the ratios of dark baryon and meson masses for $SU(N_c)$ dark sectors as a function of $N_c$ and $M_{\Pi_d}/\Lambda_d$ below.
Other observables, such as the dark-sector matching coefficients relating $N_c$ and $M_{\Pi_d} / \Lambda_d$ to interaction rates in dark matter direct detection experiments~\cite{Fitzpatrick:2012ix,Hill:2014yka,Cirigliano:2012pq,Hoferichter:2017olk,Aalbers:2022dzr}, can be studied in future pNRQCD calculations of dark-baryon matrix elements using the optimized wavefunctions obtained here.

\subsection{Dark Mesons}

The dark meson binding energy $\Delta E_{\Pi_d}$ and mass $M_{\Pi_d} = 2 m_d + \Delta E_{\Pi_d}$ can be calculated as functions of $m_d / \Lambda_d$ by applying GFMC methods to the pNRQCD Hamiltonian with the appropriate value of $N_c$ and the corresponding zero-flavor strong-coupling $\alpha_d$.
As above, calculations are performed for renormalization scales $\mu_d \equiv 4 m_d \alpha_s(\mu_d)$ as well as scales $\mu_d /2$ and $2 \mu_d$ in order to study scale dependence.
We considered a wide range of dark quark masses $m_d / \Lambda_d \in \{2, 4, 8, 16, 32, 64, 128, 256\}$ for $N_c \in \{3,4,5,6\}$.
Our GFMC calculations used\footnote{These bounds were saturated except that $\delta \tau = 0.4 m_Q$ was used for $N_c \in \{3,4\}$ and $N_\tau \delta \tau = 4/\alpha_d^2$ was used for $N_c = 3$.} $\delta \tau \leq 0.8 m_Q$ and $N_\tau \delta \tau \geq 2 / \alpha_d^2$ with statistical ensembles of size  $N_{\text{walkers}} = 5,000$.
The results for $\Delta E_{\Pi_d}$ with are shown as functions of $m_d/\Lambda_d$ for each $N_c \in \{3,\ldots,6\}$ in Fig.~\ref{fig:mesons_Nc3-6}.
Similar qualitative features arise as in the QCD results for $\Delta E_{Q\overline{Q}}$: significant scale dependence arises beyond LO, large order-by-order changes in dependence on $m_d / \Lambda_d$ are apparent, and for the smallest $m_d / \Lambda_d$ considered the results with $\mu = \mu_d / 2$ begin to show significant curvature arising from logarithmic effects in the potential.

\begin{figure}[t!]
  \centering
  \subfigure{\includegraphics[width=\linewidth]{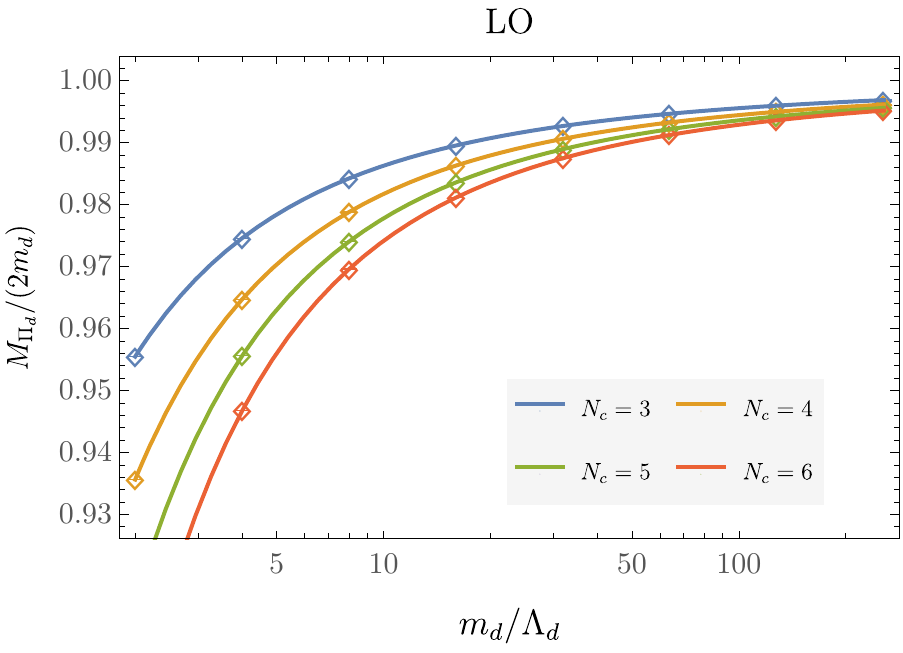}}
  \subfigure{\includegraphics[width=\linewidth]{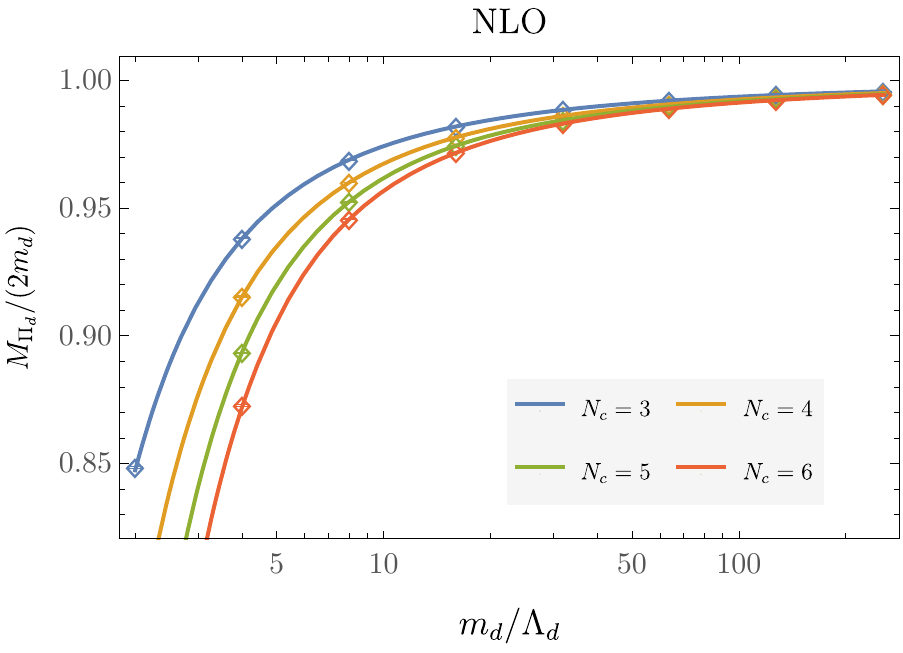}}
  \subfigure{\includegraphics[width=\linewidth]{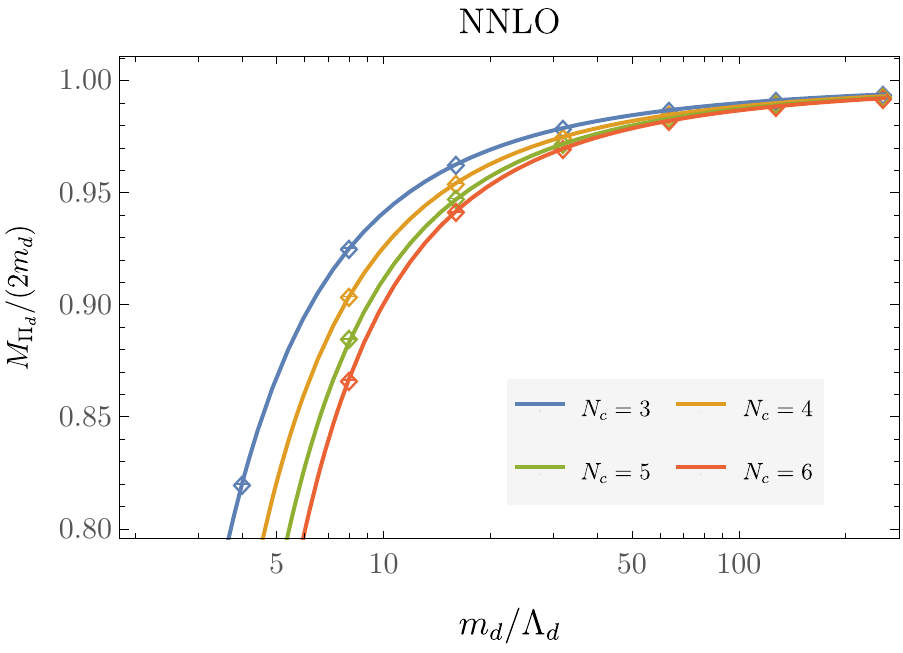}}
\caption{Dark meson binding energy GFMC results as functions of $m_d/\Lambda_d$ wth $\mu = \mu_d$ and $N_c \in \{3,\ldots,6\}$ are shown in comparison with the power series fit results described in the main text.}
\label{fig:qqbar_Nc_fits}
\end{figure}

\begin{figure*}[t!]
  \centering
  \subfigure{\includegraphics[width=.49\linewidth]{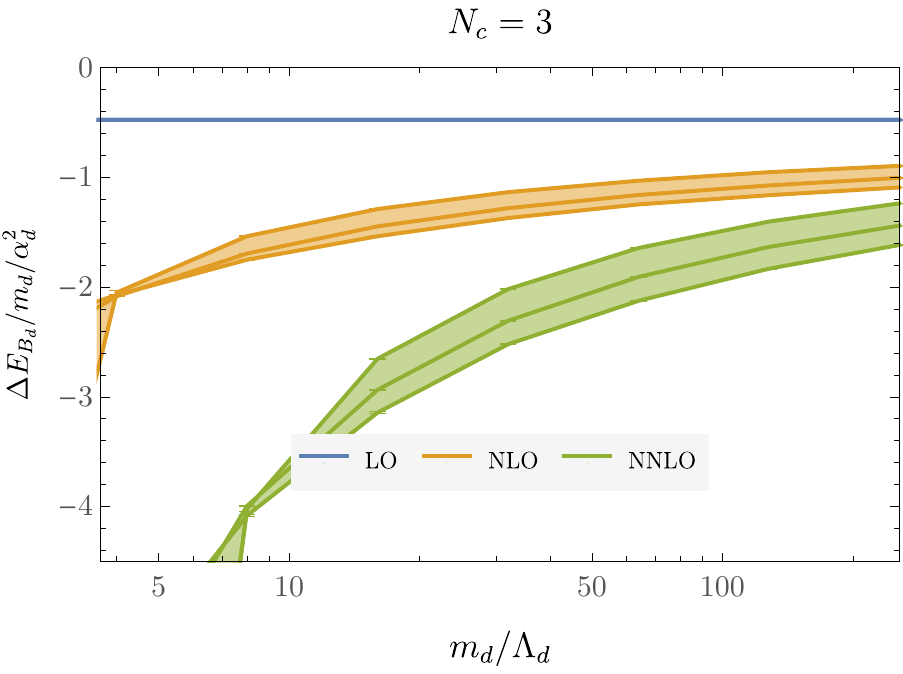}}
  \subfigure{\includegraphics[width=.49\linewidth]{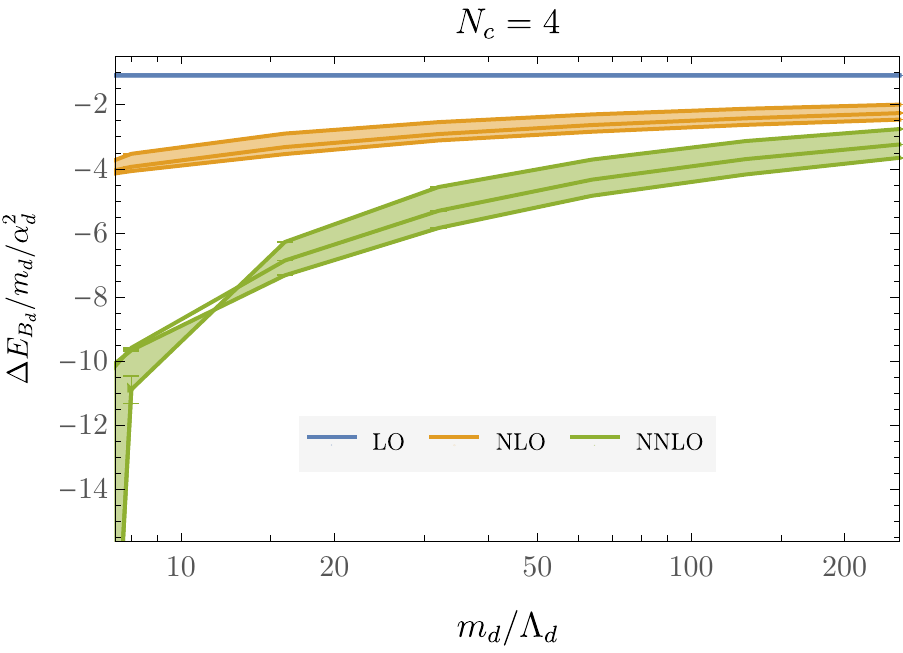}}\\
  \subfigure{\includegraphics[width=.49\linewidth]{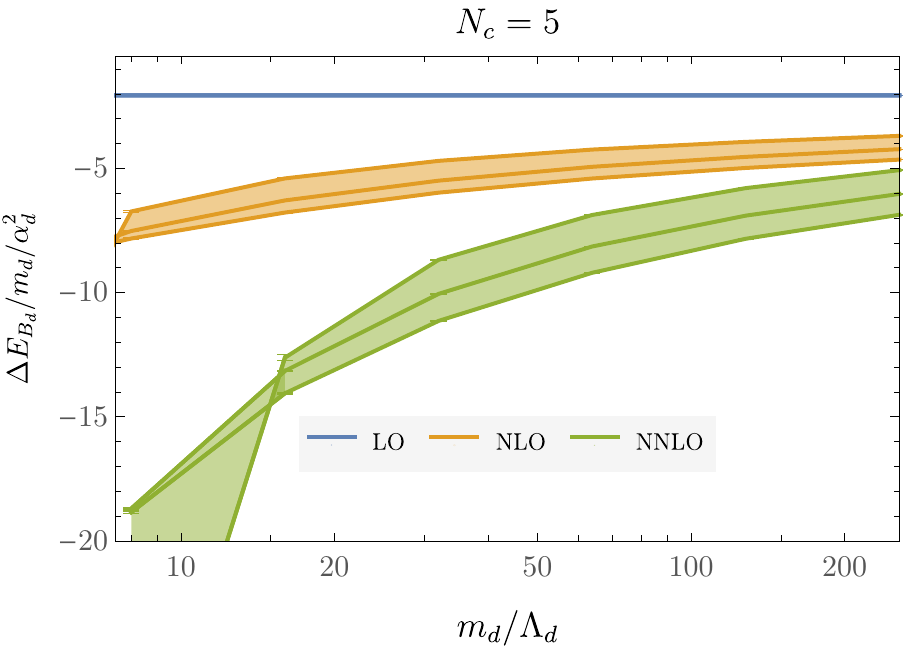}}
  \subfigure{\includegraphics[width=.49\linewidth]{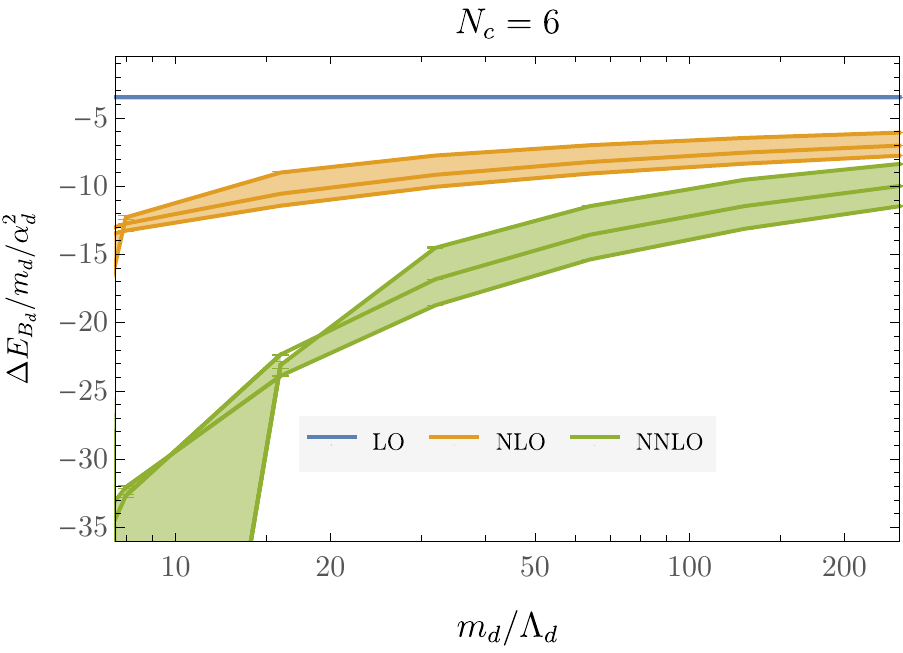}}
\caption{Dark baryon binding energy results as functions of $m_d/\Lambda_d$ with shaded bands connecting results with renormalization scale choices $\mu \in \{\mu_d, 2\mu_d, \mu_d/2$\} and $SU(N_c)$ gauge groups with $N_c \in \{3,\ldots,6\}$ as indicated. }
\label{fig:baryons_Nc3-6}
\end{figure*}

Although precise predictions for dark hadron observables with $m_d \gg \Lambda_d$ require pNRQCD calculations with particular choices of $m_d / \Lambda_d$, phenomenological estimates of the dependence of dark-sector observables on $m_d / \Lambda_d$ can be made more conveniently using analytic parameterizations that have been fit to pNRQCD results over the relevant range of $m_d / \Lambda_d$.
To provide such a parameterization, we perform fits to power series expansions in $\alpha_d$ and $1/N_c$ of these GFMC results for $\Delta E_{\Pi_d} / m_d / \alpha_d^2$.
At LO, the exact result
\begin{equation}
    \Delta E_{\Pi_d}^{(\text{LO})} = -\frac{ C_F^2 }{4} m_d \alpha_d^2,
\end{equation}
can be cast into this form by dividing by $N_c^2$ to remove the leading large $N_c$ dependence of $C_F$,
\begin{equation}
  \frac{ \Delta E_{\Pi_d}^{(\text{LO})} }{m_d \alpha_d^2 N_c^2} = \frac{C_F^2}{4N_c^2} = 0.0625 - \frac{0.0125}{N_c^{2}}  + \frac{0.0625}{N_c^4}.
\end{equation}
At NLO, $\mathcal{O}(\alpha_d)$ corrections can be expected to lead to $\mathcal{O}(\alpha_d)$ and $\mathcal{O}(\alpha_d^2)$ corrections to binding energies for an approximately Coulombic potential and we adopt the power series ansatz,
\begin{equation}
\begin{split}
  \frac{ \Delta E_{\Pi_d}^{(\text{NLO})} }{m_d \alpha_d^2 N_c^2} &\approx -\frac{ C_F^2 }{4 N_c^2} - \alpha_d  A^{(\text{NLO},1)}_{\Pi_d} -  \alpha_d^2 A^{(\text{NLO},2)}_{\Pi_d}. \label{eq:NLOQQbar}
  \end{split}
\end{equation}
The coefficients  $A^{(\text{NLO},1)}$ and $A^{(\text{NLO},2)}$ can be further expanded as power series in $1/N_c$ that are truncated to include at most three terms since calculations are only performed for four values of $N_c$.
Fits to GFMC results are performed using $\chi^2$-minimzation with results with all $m_d / \Lambda_d$ for $N_c=3$ and all $m_d / \Lambda_d \geq 4$ for $N_c \in \{4,5,6\}$, which corresponds to a total of 25 points.
Fit parameter uncertainties are determined using bootstrap resampling methods. The Akaike information criterion~\cite{AkaikeAIC} (AIC) is used to determine whether one, two, or three terms are included in the $1/N_c$ expansion for each coefficient.
This leads to the results
\begin{equation}
    \begin{split}
        A^{(\text{NLO},1)}_{\Pi_d} &\approx 1.1801(23) - \frac{3.051(25)}{N_c} + \frac{2.59(4)}{N_c^2}, \\
        A^{(\text{NLO},2)}_{\Pi_d} &\approx 0.487(6) - \frac{0.721(18)}{N_c},
    \end{split}
\end{equation}
with $\chi^2 / \text{dof} = 1.3$.

Analogous fits can be performed at NNLO using a series expansion, including two additional orders in $\alpha_d$,
\begin{equation}
\begin{split}
  \frac{ \Delta E_{\Pi_d}^{(\text{NNLO})} }{m_d \alpha_d^2 N_c^2} &\approx -\frac{ C_F^2 }{4 N_c^2} - \alpha_d  A^{(\text{NLO},1)}_{\Pi_d} -  \alpha_d^2 A^{(\text{NNLO},2)}_{\Pi_d}  \\
  &\hspace{10pt}  -  \alpha_d^3 A^{(\text{NNLO},3)}_{\Pi_d} -  \alpha_d^4 A^{(\text{NNLO},4)}_{\Pi_d}.
  \label{eq:NNLOQQbar}
  \end{split}
\end{equation}
The constant and $\mathcal{O}(\alpha_d)$ terms should be unaffected by NNLO corrections to the potential, and we, therefore, fix these terms to their lower order values as indicated in Eq.~\eqref{eq:NNLOQQbar}.
It is not possible to obtain a fit with $\chi^2 / \text{dof} \sim 1$ using $\mathcal{O}(1/N_c^3)$ power series expansions, and in particular an $\mathcal{O}(1/N_c^4)$ term in $A^{(\text{NNLO,2})}_{\Pi_d}$ is required to achieve $\chi^2 / \text{dof} \lesssim 2$.
Since such a term would lead to interpolation rather than fitting of $1/N_c$ dependence, we do not include such a term and take this to indicate that a simple power series ansatz is not able to describe the $N_c$ dependence of $\Delta E_{\Pi_d}^{(\text{NNLO})}$ in pNQRCD to the level of precision of our GFMC results.
We therefore multiply our GFMC uncertainties on $\Delta E_{\Pi_d}^{(\text{NNLO})}$ by a factor of 5 so that the best $\mathcal{O}(1/N_c^3)$ fit for $A^{(\text{NNLO},2)}_{\Pi_d}$ obtains a $\chi^2 / \text{dof} \sim 1$. This fit corresponds to
\begin{equation}
    \begin{split}
      A^{(\text{NNLO},2)}_{\Pi_d} &\approx 25.5(3) - \frac{126(2)}{N_c} + \frac{178(6)}{N_c^2}, \\
        A^{(\text{NNLO},3)}_{\Pi_d} &\approx 13.6(8) - \frac{32(9)}{N_c}, \\
        A^{(\text{NNLO},4)}_{\Pi_d} &\approx -1(5).
    \end{split}
\end{equation}
Comparisons of GFMC results with these fit results for each order are shown in Fig.~\ref{fig:qqbar_Nc_fits}.

The $N_c$ scaling behavior of meson masses has previously been studied using LQCD in Refs.~\cite{Bali:2013kia,DeGrand:2016pur}. 
However, without computing the relationship between either $\Lambda_d$ or the pole mass $m_d$ used here and another dimensionful observable such as the pion decay constant, it is not possible to compare results for $M_{\Pi_d} / \Lambda_d$ or $M_{\Pi_d} / m_d$ directly with the LQCD results of these works.
Such comparisons are therefore deferred to future studies, including dark meson matrix element calculations in pNRQCD.

\subsection{Dark Baryons}

Dark baryon binding energies $\Delta E_{B_d}$ and masses $M_{B_d}$ are computed by applying GFMC methods to $SU(N_c)$ baryon states with the pNRQCD Hamiltonian at LO, NLO, and NNLO with the same range of masses $m_d / \Lambda_d \in [2,256]$ and $N_c \in [3,6]$ as in the dark meson case discussed above.
The trial wavefunctions described in Sec.~\ref{sec:trial} are found to provide suitable initial states for GFMC evolution using the same relation between the Bohr radius and $\alpha_d$ as the QCD case shown in Eq.~\eqref{eq:psiTQQQSimp}.
Excited-state effects are found to increase only mildly with $N_c$ using this prescription.
Results for $\Delta E_{B_d}$ obtained from single-state fits as described above are shown for each $N_c$ as functions of $m_d / \Lambda_d$ in Fig.~\ref{fig:baryons_Nc3-6}.

\begin{figure}[t!]
  \centering
  \subfigure{\includegraphics[width=\linewidth]{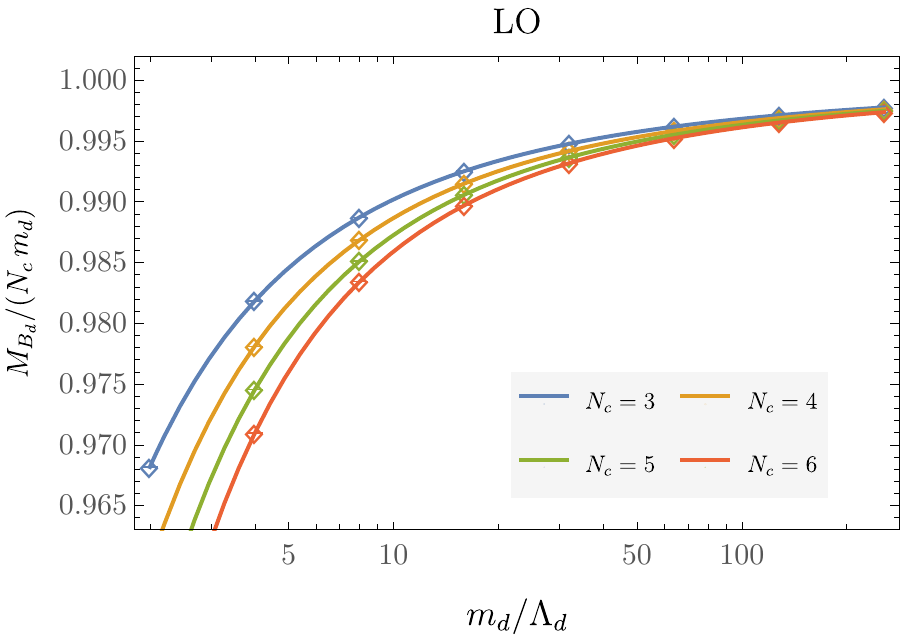}}
  \subfigure{\includegraphics[width=\linewidth]{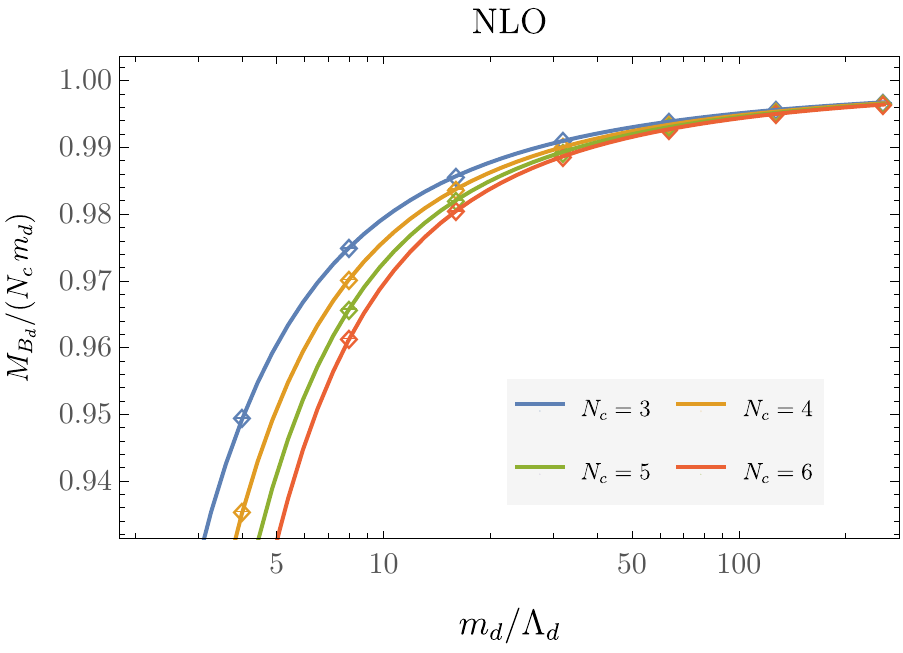}}
  \subfigure{\includegraphics[width=\linewidth]{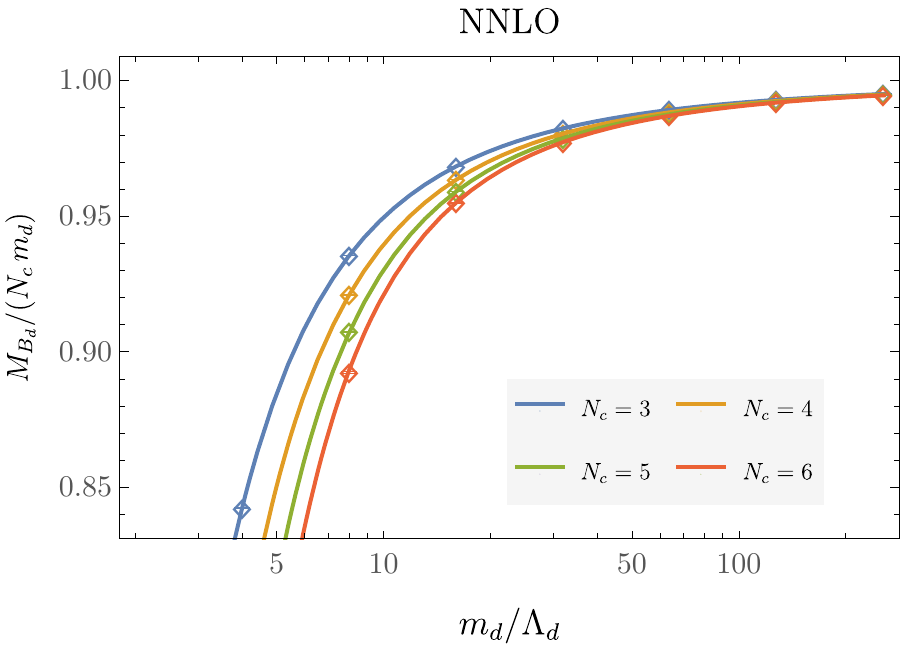}}
\caption{Dark baryon binding energy GFMC results as functions of $m_d/\Lambda_d$ wth $\mu = \mu_d$ and $N_c \in \{3,\ldots,6\}$ are shown in comparison with the power series fit results described in the main text.}
\label{fig:baryon_Nc_fits}
\end{figure}

As in the dark meson case above, we can analytically parameterize our GFMC dark baryon binding-energy results as a power series in $\alpha_d$ and $1/N_c$~\cite{tHooft:1973alw,Manohar:1998xv}.
These power series expressions cannot capture the complete non-analytical structure of pNRQCD, but they can provide convenient estimates and accurately describe our pNRQCD results to a relatively high level of precision over the range of quark masses, and $N_c$ studied.
At LO, it is sufficient to parameterize $\Delta E_{B_d} / m_d / \alpha_d^2$ as a constant that only depends on $N_c$,
\begin{equation}
\begin{split}
  \frac{ \Delta E_{B_d}^{(\text{LO})} }{ m_d \alpha_d^2 N_c^4 } &\approx  -A^{(\text{LO},0)}_{B_d}(N_c).  \label{eq:LORB}
  \end{split}
\end{equation}
The factor of $1/N_c^4$ is included to ensure that the result is finite as $N_c \rightarrow \infty$ and the following (naive) argument for the scaling of the binding energy with $N_c$: the quark-quark potential is proportional to $C_F / (N_c - 1) \sim N_c^0$ and the total potential, therefore, scales as $\sum_{I < J} \sim N_c^2$.
Since the binding energy for a Coulombic system is proportional to the { square of the prefactor of $1/r$ in the potential}, it can therefore be expected to scale as $N_c^4$.
However, fits to a constant plus $\mathcal{O}(1/N_c)$ and/or $\mathcal{O}(1/N_c^2)$ corrections lead to a vanishing constant term at LO.
Including two additional powers of $1/N_c$ and fitting to the same set of 25 GFMC results with varying $m_d / \Lambda_d$ and $N_c$ as in the dark meson case using the same $\chi^2$-minimization and bootstrap resampling techniques leads to
\begin{equation}
    \begin{split}
      A^{(\text{LO},0)}_{B_d} &\approx \frac{0.0132814(16)}{N_c} + \frac{0.020772(34)}{N_c^2} \\
      &\hspace{20pt} - \frac{0.02307(5)}{N_c^3},
    \end{split}
\end{equation}
with a $\chi^2 / \text{dof} = 1.4$.
This observed scaling $\Delta E_{B_d}/m_d~\sim~\alpha_d^2 N_c^3$ is consistent with Witten's large-$N_c$ arguments in Ref.~\cite{Witten:1979kh}.
Since the strong coupling is taken to scale as $\alpha_d \sim 1/N_c$~\cite{tHooft:1973alw} this leads to the usual result that $\Delta E_{B_d}/m_d~\sim~N_c$ while $\Delta E_{\Pi_d}/m_d~\sim~N_c^0$.

\begin{figure*}[t!]
  \centering
  \subfigure{\includegraphics[width=.49\linewidth]{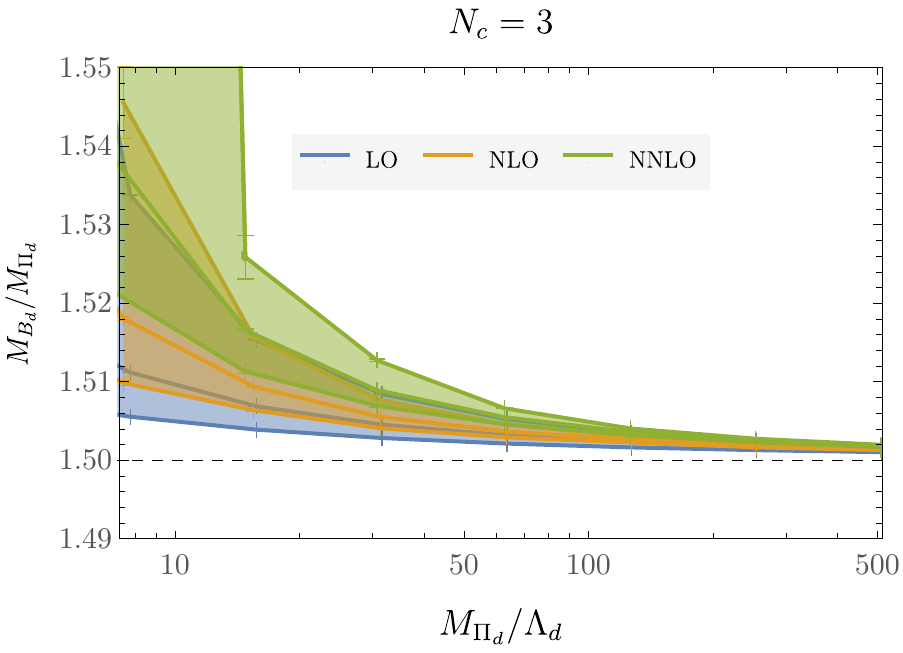}}
  \subfigure{\includegraphics[width=.49\linewidth]{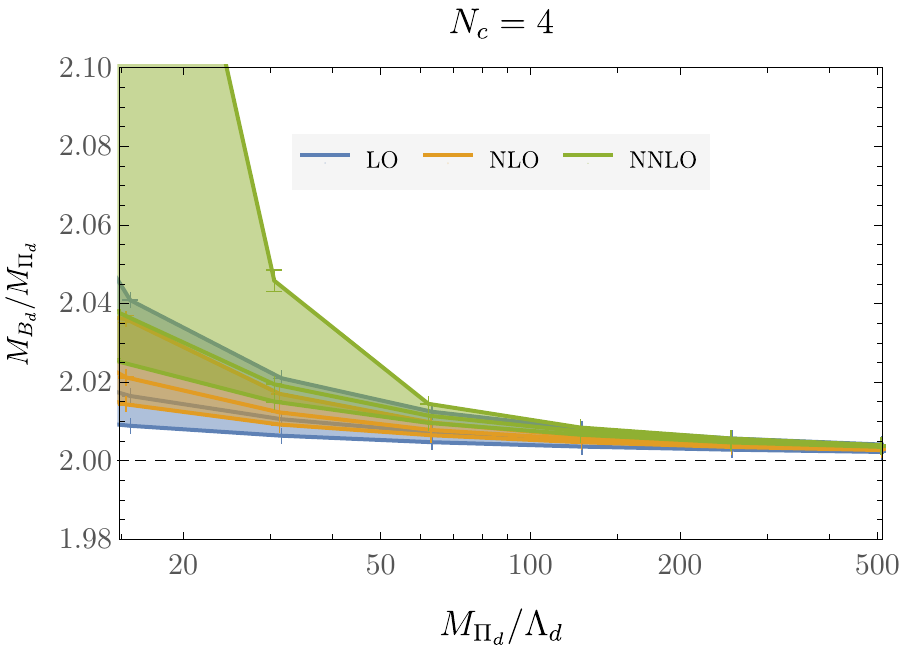}}\\
  \subfigure{\includegraphics[width=.49\linewidth]{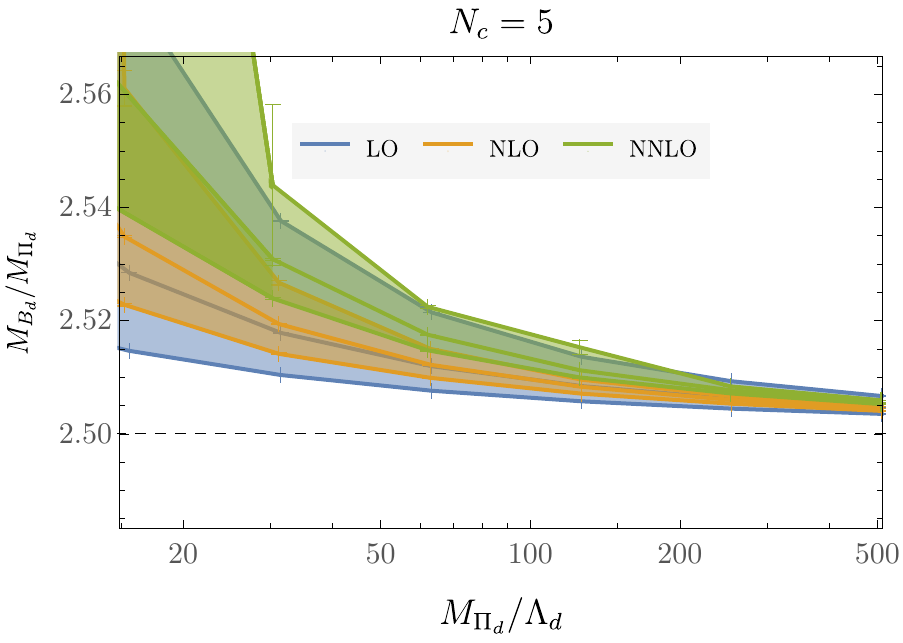}}
  \subfigure{\includegraphics[width=.49\linewidth]{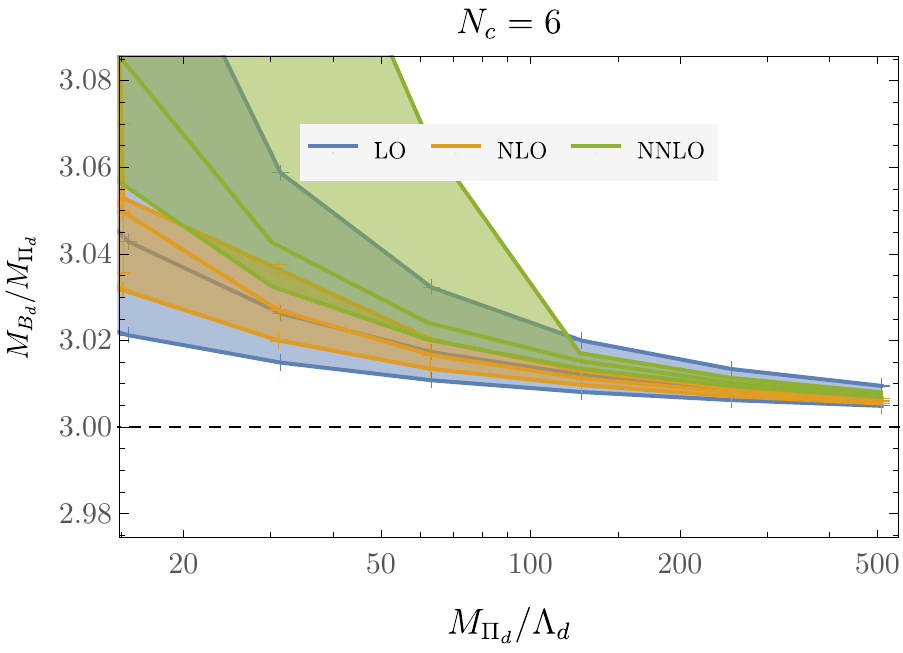}}
\caption{Ratios of dark baryon and meson masses as functions of the dark meson mass with shaded bands connecting results with renormalization scale choices $\mu \in \{\mu_d, 2\mu_d, \mu_d/2$\} and $SU(N_c)$ gauge groups with $N_c \in \{3,\ldots,6\}$ as indicated. }
\label{fig:baryons_Nc3-6_ratios}
\end{figure*}

At NLO, an $\mathcal{O}(\alpha_d^2)$ power series analogous to the one used in the dark meson case is given by
\begin{equation}
\begin{split}
  \frac{ \Delta E_{B_d}^{(\text{NLO})} }{m_d \alpha_d^2 N_c^4} &\approx -A^{(\text{LO},0)}_{B_d} -  \alpha_d  A^{(\text{NLO},1)}_{B_d} -  \alpha_d^2 A^{(\text{NLO},2)}_{B_d}, \label{eq:NLORB}
  \end{split}
\end{equation}
where $A^{(\text{LO},0)}_{B_d}$ is fixed to it's LO value. Expanding the $\mathcal{O}(\alpha_d)$ term to $\mathcal{O}(1/N_c^2)$ and the $\mathcal{O}(\alpha_d^2)$ term to $\mathcal{O}(1/N_c)$ gives
\begin{equation}
    \begin{split}
      A^{(\text{NLO},1)}_{B_d} &\approx 0.01917(10) + \frac{0.2073(8)}{N_c} - \frac{0.24181(5)}{N_c^2}, \\
        A^{(\text{NLO},2)}_{B_d} &\approx 0.0456(6) + \frac{0.002(1)}{N_c},
    \end{split}
\end{equation}
where GFMC uncertainties have been inflated by a factor of two before fitting in order to obtain a $\chi^2 / \text{dof} \sim 1$ since, as in the NNLO dark meson case, deviations from a simple power series ansatz can be seen at the high level of precision of our GFMC results.
In this case $N_c^4$ scaling is observed for fixed $\alpha_d$; however, since $\alpha_d \sim 1/N_c$ in the large $N_c$ scaling of Ref.~\cite{Witten:1979kh} the expected scaling $\Delta E_{B_d}/m_d \sim \alpha_d^2 N_c^3 \sim N_c$ is reproduced by pNRQCD at NLO.
The same arguments apply at higher orders since further powers of $\alpha_d$ contribute additional powers of $1/N_c$ and are, therefore, further subleading corrections in the large $N_c$ limit.

At NNLO, an analogous power series expansion to the dark meson case is used,
\begin{equation}
\begin{split}
  \frac{ \Delta E_{B_d}^{(\text{NNLO})} }{m_d \alpha_d^2 N_c^4} &\approx -A^{(\text{LO},0)}_{B_d} - \alpha_d  A^{(\text{NLO},1)}_{B_d} -  \alpha_d^2 A^{(\text{NNLO},2)}_{B_d}  \\
  &\hspace{10pt} -  \alpha_d^3 A^{(\text{NNLO},3)}_{B_d} -  \alpha_d^4 A^{(\text{NNLO},4)}_{B_d},
  \label{eq:NNLORB}
  \end{split}
\end{equation}
and fits to our GFMC results give
\begin{equation}
    \begin{split}
      A^{(\text{NNLO},2)}_{B_d} &\approx 0.985(4) - \frac{2.35(3)}{N_c} + \frac{2.41(10)}{N_c^2}, \\
        A^{(\text{NNLO},3)}_{B_d} &\approx 1.34(2) - \frac{1.34(17)}{N_c}, \\
        A^{(\text{NNLO},4)}_{B_d} &\approx -1.00(8),
    \end{split}
\end{equation}
where uncertainties have again been inflated by a factor of two to achieve $\chi^2 / \text{dof} \sim 1$.
Comparisons of these power series fit results with GFMC results for $N_c \in \{3,\ldots,6\}$ dark baryon masses at each perturbative order are shown in Fig.~\ref{fig:baryon_Nc_fits}.

\begin{figure}[t!]
  \centering
  \subfigure{\includegraphics[width=\linewidth]{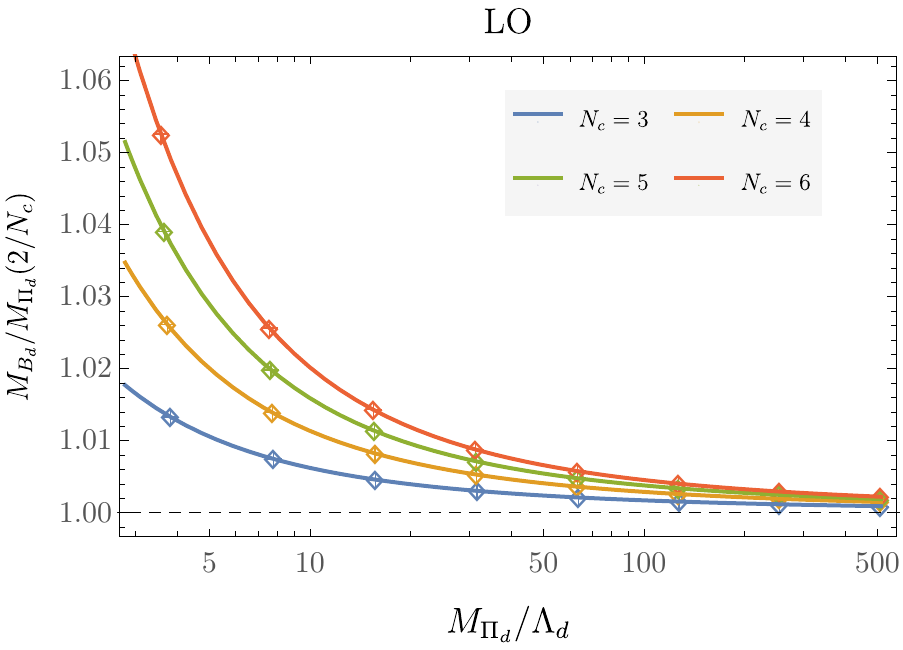}}
  \subfigure{\includegraphics[width=\linewidth]{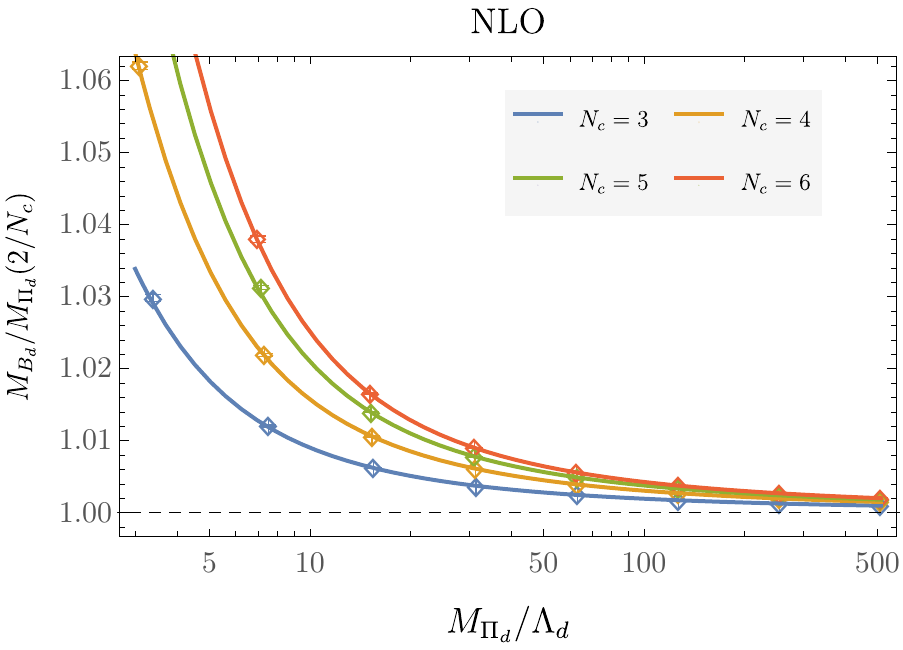}}
  \subfigure{\includegraphics[width=\linewidth]{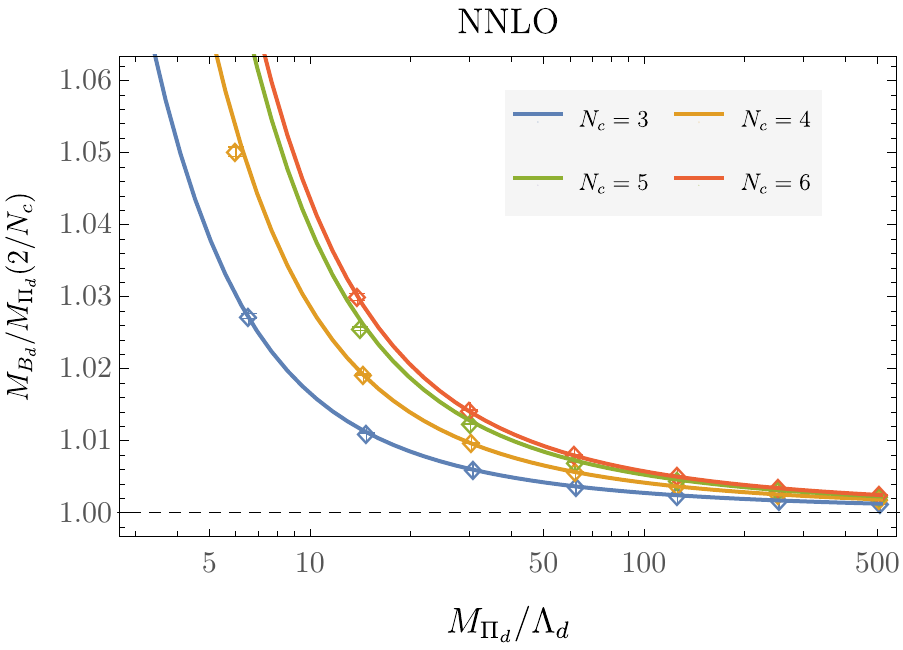}}
\caption{Ratios of dark baryon and meson masses as functions of the dark meson mass with $N_c \in \{3,\ldots,6\}$ computed using $\mu = \mu_d$ are shown in comparison with the power series fit results described in the main text.}
\label{fig:baryon_Nc_ratios}
\end{figure}

The ratio $M_{B_d} / M_{\Pi_d}$ is shown as a function of $M_{\Pi_d} /\Lambda_d$ for GFMC results in Fig.~\ref{fig:baryons_Nc3-6_ratios} and compared with power series fits in Fig.~\ref{fig:baryon_Nc_ratios}.
To obtain hadron mass ratios as functions of $M_{\Pi_d} / \Lambda_d$, the functions $M_{\Pi_d}(m_d) = m_d(2 - \alpha_d^2 C_F^2/4 - \ldots)$ defined at NLO and NNLO by the series expansions in Eq.~\eqref{eq:NLOQQbar} and Eq.~\eqref{eq:NNLOQQbar}, which implicitly depend on $m_d$ through $\alpha_d(\mu = 4\alpha_d m_d)$, are inverted numerically to obtain $m_d(M_{\Pi_d})$ and subsequently $\alpha_d(\mu = 4\alpha_d m_d(M_{\Pi_d}))$ at each order.
The $m_d$ and $\alpha_d$ determined in this way can be inserted in Eq.~\eqref{eq:LORB}-\eqref{eq:NNLORB} to obtain $M_{B_d}(M_{\Pi_d})$.
These results have the advantage of only depending on dark hadron masses and the $\overline{\text{MS}}$ Landau pole scale $\Lambda_d$ and are free from ambiguities in the scheme used to define $m_d$, apart from the renormalization scale dependence arising in fixed-order results from perturbative truncation effects.

These results can be compared with generalizations of the QCD inequalities discussed in Sec.~\ref{sec:qcd_baryons}.
The proof in Ref.~\cite{Detmold:2014iha} that there are no multi-meson bound states with maximal isospin is valid for $SU(N_c)$ gauge-theory with generic $N_c$, and if $1/m_Q^{2}$ effects are neglected are valid for heavy-quark hadrons in $SU(N_c)$ gauge-theory with generic $N_f$.
By the arguments in Section 10 of Ref~\cite{Nussinov:1999sx}, this is sufficient to establish that meson and baryon masses in $SU(N_c)$ gauge theory satisfy the inequality 
\begin{equation}
  M_{B_d} \geq \frac{N_c}{2} M_{\Pi_d}. \label{eq:DarkDetmold}
\end{equation}
This bound holds for the lightest meson and baryon constructed from quarks of a given flavor and, therefore, to generic $SU(N_c)$ dark sectors.
As discussed after Eq.~\eqref{eq:Detmold}, this leads to an equivalent bound on binding energies
\begin{equation}
  \Delta E_{B_d} \geq \frac{N_c}{2} \Delta E_{\Pi_d}. \label{eq:DarkDeltaDetmold}
\end{equation}
Both Eq.~\eqref{eq:DarkDetmold} and Eq.~\eqref{eq:DarkDeltaDetmold} are respected by all GFMC results of this work where $\Lambda_d / m_d$ corrections are expected to be perturbative,\footnote{For sufficiently small $m_d / \Lambda_d$, corrections to the static potential considered here from effects suppressed by $1/m_Q$ will be significant and pNRQCD results using only the static potential may not satisfy general features of the QCD. Indeed, our pNRQCD results with $N_c = 6$, $m_d / \Lambda_d = 8$, and $\mu = \mu_p / 2$ predict $\Delta E_{Q\overline{Q}} < -2m_Q$ and therefore lead to unphysical predictions of negative meson masses as well as unphysical violations of Eq.~\eqref{eq:DarkDetmold} and Eq.~\eqref{eq:DarkDeltaDetmold}. } as seen in Fig.~\ref{fig:baryon_Nc_ratios}.
It is noteworthy that pNRQCD results approximately saturate Eq.~\eqref{eq:DarkDetmold} with $M_{B_d} / M_{\Pi_d} / (N_c / 2)$ within 5\% of unity for $m_d / \Lambda_d \gtrsim 5$ for $N_c \in \{3,\ldots,6\}$.
As in the QCD case discussed above, $\Delta E_{B_d} / \Delta E_{\Pi_d}$ is approximately independent of $m_d$ and Eq.~\eqref{eq:DarkDeltaDetmold} is not saturated in the $m_d \rightarrow \infty$ limit, which means that $M_{B_d} / M_{\Pi_d}$ approaches $N_c/2$ logarithmically as $m_d \rightarrow \infty$.
The degree to which Eq.~\eqref{eq:DarkDetmold} is saturated for a given $m_d / \Lambda_d$ is further seen to decrease with increasing $N_c$. This behavior is unsurprising because for $N_c = 2$ meson and baryon masses are guaranteed to be identical and therefore saturate  Eq.~\eqref{eq:DarkDetmold}, while saturation is not exact for $N_c = 3$.

In the large $N_c$ limit, the NLO and NNLO results above provide subleading corrections, and the LO result above simplifies to 
\begin{equation}
  M_{B_d} = N_c m_d \left( 1 - 0.0132814(16) \alpha_d^2 N_c^2 \right)  + \mathcal{O}\left( \frac{1}{N_c} \right). \label{eq:NcBaryon}
\end{equation}
An analogous formula was derived using mean-field results in the joint large quark mass and large $N_c$ limit in Ref.~\cite{Cohen:2011cw}.
Identical scaling with quark mass, strong coupling, and $N_c$ is obtained here and in Ref.~\cite{Cohen:2011cw}; however, the numerical value of the coefficient obtained there is 0.05426, which is larger than our result by roughly a factor of four.
The corresponding LO meson result is known analytically,
\begin{equation}
  M_{\pi_d} = 2 m_d \left( 1 - \frac{C_F^2}{8} \alpha_d^2 \right)  + \mathcal{O}\left( \frac{1}{N_c} \right),
\end{equation}
and so the $SU(N_c)$ heavy-quark Detmold bound implies that the numerical coefficient in Eq.~\eqref{eq:NcBaryon} must be smaller in magnitude than $C_F^2 / (8 N_c^2) = 0.03125 + \mathcal{O}(1/N_c)$.
This bound is satisfied by Eq.~\eqref{eq:NcBaryon} but not by the results of Ref.~\cite{Cohen:2011cw}, which indicates that the discrepancy must arise from uncertainties in the mean-field approach used there.

The large-$N_c$ behavior of baryon masses has also been studied in lattice gauge theory calculations~\cite{Jenkins:2009wv,DeGrand:2012hd,DeGrand:2013nna,Cordon:2014sda,LatticeStrongDynamicsLSD:2014osp,DeGrand:2016pur}.
The baryon-to-meson mass ratio provides a well-defined dimensionless observable that can be matched to lattice gauge theory results for each $N_c$, allowing us to select the $m_q / \Lambda_d$ that reproduces lattice gauge theory results with any particular quark mass.
However, other observables must be calculated to make non-trivial predictions to compare with $SU(N_c)$ lattice gauge theory, which is left to future work.

\section{Outlook}
\label{sec:out}

We have presented a formulation of pNRQCD suitable for calculating binding energies and matrix elements of generic hadron and multi-hadron states made of heavy quarks in $SU(N_c)$ gauge theory using quantum Monte Carlo techniques.
The complete two- and three-quark potentials required for generic multi-hadron systems are constructed up to NNLO in the strong coupling.
The appearance of four-quark potentials arising at NNLO is pointed out, and a complete construction of these potentials should be pursued in future work.

We further employed VMC and GFMC to compute quarkonium and triply-heavy baryon binding energies in pNRQCD at $\mathcal{O}(m_Q^0)$.
Precise results are obtained with modest computational resources, but we underpredict the baryon masses computed using LQCD by 
 1-2\% for all baryons comprised of $b$ and $c$ quarks.
Differences between perturbative orders demonstrate good convergence for the $\alpha_s$ expansion of the pNRQCD potential.
The remaining differences between NNLO and LQCD results likely arise primarily from  $1/m_Q$ and $1/m_Q^2$ effects in the pNRQCD potential that are neglected in this work.
Extending this work by incorporating spin-dependent potentials and determining suitable trial wavefunctions with these potentials included will be an essential step toward improving the predictive power of this framework.
It will also be interesting to extend these studies towards heavy exotics such as tetraquarks and multi-baryon systems, as well as quarkonium and baryon excited states.

Applying quantum Monte Carlo methods to pNRQCD may be particularly useful for studies of composite dark matter.
A $SU(N_c)$ dark sector with one heavy dark quark provides a simple, UV-complete, phenomenological viable model of composite DM~\cite{Asadi:2021yml,Asadi:2021pwo}.
QMC calculations using pNRQCD can provide computationally simple predictions for composite DM observables that enable efficient scanning over a wide range of mass scales. This is particularly useful in the composite DM context, where the underlying theory's actual parameters are not yet known.
The works provide pNRQCD results and simple analytic parameterizations of the dark meson and dark baryon masses in $SU(N_c)$ gauge theory as functions of $N_c$ and the dark sector parameters $m_d$ and $\Lambda_d$.
The properties and interactions of these dark hadrons should be studied in future applications of QMC to pNRQCD.\\

\begin{acknowledgments}
We thank Matthew Baumgart, Elias Bernreuther, Nora Brambilla, William Detmold, Jacopo Ghiglieri, Florian Herren, Chia~Hsien-Shen, Gurtej Kanwar, Aneesh Manohar, Joan~Soto, Daniel Stolarski, and Antonio Vairo for helpful discussions and insightful comments. 
This manuscript has been authored by Fermi Research Alliance, LLC under Contract No. DE-AC02-07CH11359 with the U.S. Department of Energy, Office of Science, Office of High Energy Physics.
\end{acknowledgments}

\bibliography{pnrqcd}


\end{document}